\documentclass[acmsmall,screen,nonacm]{acmart}

\AtBeginDocument{%
  }

\setcopyright{none}
\copyrightyear{2024}
\acmYear{2024}
\acmDOI{XXXXXXX.XXXXXXX}

\acmJournal{JACM}
\acmVolume{37}
\acmNumber{4}
\acmArticle{111}
\acmMonth{8}

\usepackage[english]{babel}
\usepackage{subcaption}
\usepackage{amsmath}
\usepackage{graphicx}
\graphicspath{{figures/}}
\usepackage{mathtools}
\usepackage{array}
\usepackage[frozencache=true]{minted}
\usepackage{ifthen}
\usepackage{ebproof}
\ebproofset{label separation=0pt}
\usepackage{multirow}
\usepackage{wrapfig}

\makeatletter
\DeclareFontFamily{OMX}{MnSymbolE}{}
\DeclareSymbolFont{MnLargeSymbols}{OMX}{MnSymbolE}{m}{n}
\SetSymbolFont{MnLargeSymbols}{bold}{OMX}{MnSymbolE}{b}{n}
\DeclareFontShape{OMX}{MnSymbolE}{m}{n}{
    <-6>  MnSymbolE5
   <6-7>  MnSymbolE6
   <7-8>  MnSymbolE7
   <8-9>  MnSymbolE8
   <9-10> MnSymbolE9
  <10-12> MnSymbolE10
  <12->   MnSymbolE12
}{}
\DeclareFontShape{OMX}{MnSymbolE}{b}{n}{
    <-6>  MnSymbolE-Bold5
   <6-7>  MnSymbolE-Bold6
   <7-8>  MnSymbolE-Bold7
   <8-9>  MnSymbolE-Bold8
   <9-10> MnSymbolE-Bold9
  <10-12> MnSymbolE-Bold10
  <12->   MnSymbolE-Bold12
}{}

\let\llangle\@undefined
\let\rrangle\@undefined
\DeclareMathDelimiter{\llangle}{\mathopen}%
                     {MnLargeSymbols}{'164}{MnLargeSymbols}{'164}
\DeclareMathDelimiter{\rrangle}{\mathclose}%
                     {MnLargeSymbols}{'171}{MnLargeSymbols}{'171}
\makeatother

\setminted{
    style=colorful, 
    fontsize=\tiny,
    linenos=false, 
    breaklines, 
    breakanywhere, 
    autogobble,
}
\setmintedinline{fontsize=\small}

\ifthenelse{1=0}{
\newcommand{\TODO}[1]{\textcolor{red}{TODO: #1}}
\newcommand\help[1]{\textcolor{green}{Help Needed: #1}}
\newcommand\kyle[1]{\textcolor{olive}{Kyle: #1}}
\newcommand\teo[1]{\textcolor{red}{teo: #1}}
\newcommand\willow[1]{\textcolor{purple}{Willow: #1}}
\newcommand{\saman}[1]{\textcolor{orange}{Saman: #1}}

\newcommand\changwan[1]{\textcolor{blue}{CW: #1}}
}{
\newcommand{\TODO}[1]{}
\newcommand{\help}[1]{}
\newcommand{\saman}[1]{}
\newcommand{\kyle}[1]{}
\newcommand{\teo}[1]{}
\newcommand{\willow}[1]{}
\newcommand{\changwan}[1]{}

}

\newcommand{\finchextent}{\ensuremath{\text{\textbf{extent}}}}
\newcommand{\finchliteral}{\ensuremath{\text{\textbf{literal}}}}
\newcommand{\finchvalue}{\ensuremath{\text{\textbf{value}}}}

\newcommand{\finchmode}{\ensuremath{\text{\textbf{mode}}}}
\newcommand{\finchindex}{\ensuremath{\text{\textbf{index}}}}
\newcommand{\finchvar}{\ensuremath{\text{\textbf{variable}}}}
\newcommand{\finchcall}{\ensuremath{\text{\textbf{call}}}}
\newcommand{\finchaccess}{\ensuremath{\text{\textbf{access}}}}
\newcommand{\finchread}{\ensuremath{\text{\textbf{read}}}}
\newcommand{\finchassign}{\ensuremath{\text{\textbf{assign}}}}
\newcommand{\finchupdate}{\ensuremath{\text{\textbf{update}}}}
\newcommand{\finchloop}{\ensuremath{\text{\textbf{loop}}}}
\newcommand{\finchdefine}{\ensuremath{\text{\textbf{define}}}}
\newcommand{\finchsieve}{\ensuremath{\text{\textbf{sieve}}}}
\newcommand{\finchblock}{\ensuremath{\text{\textbf{block}}}}
\newcommand{\finchdeclare}{\ensuremath{\text{\textbf{declare}}}}
\newcommand{\finchfreeze}{\ensuremath{\text{\textbf{freeze}}}}
\newcommand{\finchthaw}{\ensuremath{\text{\textbf{thaw}}}}

\newcommand{\finchthunk}{\ensuremath{\text{\textbf{thunk}}}}
\newcommand{\finchphase}{\ensuremath{\text{\textbf{phase}}}}
\newcommand{\finchswitch}{\ensuremath{\text{\textbf{switch}}}}
\newcommand{\finchlookup}{\ensuremath{\text{\textbf{lookup}}}}
\newcommand{\finchrun}{\ensuremath{\text{\textbf{run}}}}
\newcommand{\finchspike}{\ensuremath{\text{\textbf{spike}}}}
\newcommand{\finchsequence}{\ensuremath{\text{\textbf{sequence}}}}
\newcommand{\finchstepper}{\ensuremath{\text{\textbf{stepper}}}}
\newcommand{\finchjumper}{\ensuremath{\text{\textbf{jumper}}}}

\newcommand{\HIDE}[1]{}

\newcommand{\rothead}[1]{\rotatebox{90}{\textbf{#1}}}

\newcommand\den[2][]{%
  \ensuremath{%
    \left\llbracket
    #2
    \right\rrbracket^{#1}}}

\usepackage{varwidth}

\begin{document}

\title{Finch: Sparse and Structured Tensor Programming with Control Flow}

\author{Willow Ahrens}
\affiliation{%
  \institution{MIT CSAIL}
  \city{Cambridge}
  \state{Massachusetts}
  \country{USA}}
\email{willow@csail.mit.edu}

\author{Teodoro Fields Collin}
\affiliation{%
  \institution{MIT CSAIL}
  \city{Cambridge}
  \state{Massachusetts}
  \country{USA}}
\email{teoc@mit.edu}

\author{Radha Patel}
\affiliation{%
  \institution{MIT CSAIL}
  \city{Cambridge}
  \state{Massachusetts}
  \country{USA}}
\email{rrpatel@alum.mit.edu}

\author{Kyle Deeds}
\affiliation{%
  \institution{University of Washington}
  \city{Seattle}
  \state{Washington}
  \country{USA}}
\email{kdeeds@cs.washington.edu}

\author{Changwan Hong}
\affiliation{%
  \institution{MIT CSAIL}
  \city{Cambridge}
  \state{Massachusetts}
  \country{USA}}
\email{changwan@mit.edu}

\author{Saman Amarasinghe}
\affiliation{%
  \institution{MIT CSAIL}
  \city{Cambridge}
  \state{Massachusetts}
  \country{USA}}
\email{saman@csail.mit.edu}

\renewcommand{\shortauthors}{Ahrens et al.}

\begin{abstract}
From FORTRAN to NumPy, tensors have revolutionized how we express computation. However, tensors in these, and almost all prominent systems, can only handle dense rectilinear integer grids.  Real world tensors often contain underlying structure, such as sparsity, runs of repeated values, or symmetry.  Support for structured data is fragmented and incomplete.  Existing frameworks limit the tensor structures and program control flow they support to better simplify the problem.

In this work, we propose a new programming language, Finch, which supports \textit{both} flexible control flow and diverse data structures. Finch facilitates a programming model which resolves the challenges of computing over structured tensors by combining control flow and data structures into a common representation where they can be co-optimized. Finch automatically specializes control flow to data so that performance engineers can focus on experimenting with many algorithms. Finch supports a familiar programming language of loops, statements, ifs, breaks, etc., over a wide variety of tensor structures, such as sparsity, run-length-encoding, symmetry, triangles, padding, or blocks. Finch reliably utilizes the key properties of structure, such as structural zeros, repeated values, or clustered non-zeros. We show that this leads to dramatic speedups in operations such as SpMV and SpGEMM, image processing, and graph analytics.
\end{abstract}

%

\received{20 February 2007}
\received[revised]{12 March 2009}
\received[accepted]{5 June 2009}

\maketitle

\section{Introduction}

Arrays are the most fundamental abstraction in computer science. Arrays and lists are often the first-taught datastructure
\cite[Chapter 2.2]{abelson_structure_1996}, \cite[Chapter 2.2]{knuth_art_1997}.
Arrays are also universal across programming languages, from their introduction
in Fortran in 1957 to present-day languages like Python
\cite{backus_fortran_1957}, keeping more-or-less the same semantics.
Modern array programming languages such as NumPy~\cite{harris_array_2020},
SciPy~\cite{virtanen_scipy_2020}, MatLab~\cite{moler_history_2020},
TensorFlow~\cite{abadi_tensorflow_2016}, PyTorch~\cite{paszke_pytorch_2019}, and
Halide~\cite{ragan-kelley_halide_2013} have pushed the limits of productive data
processing with arrays, fueling breakthroughs in machine learning, scientific
computing, image processing, and more.

The success and ubiquity of arrays is largely due to their simplicity. 
Since their introduction, multidimensional arrays have represented dense, rectilinear,
integer grids of points. 
By \textbf{dense}, we mean that indices are mapped to value via a simple formula relating multidimensional space to linear memory.
Consequently, dense arrays offer extensive compiler optimizations and many convenient interfaces.
%
%
Compilers understand dense computations across many
programming constructs, such as for and while loops, breaks, parallelism,
caching, prefetching, multiple outputs, scatters, gathers, vectorization,
loop-carry-dependencies, and more. A myriad of optimizations have been developed for
dense arrays, such as loop fusion, loop tiling, loop unrolling, and loop
interchange.
However, while dense arrays are the easiest way to program for performance, real world applications
often require more complex data structures to reach peak efficiency.

\begin{figure}
	\includegraphics[width=\linewidth]{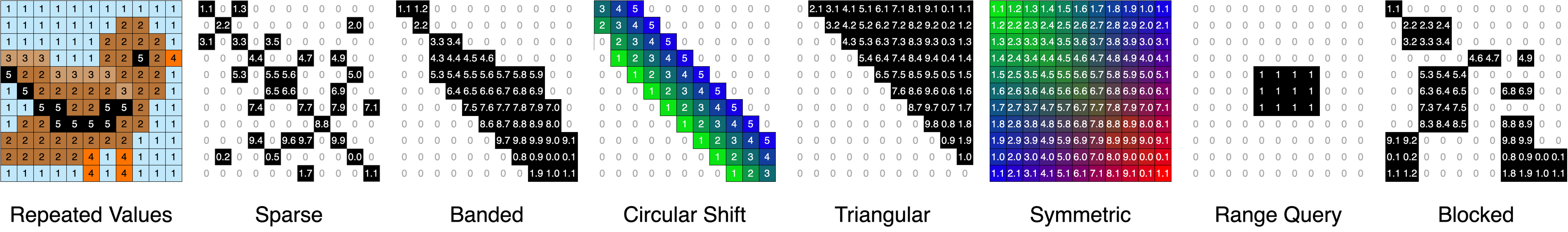}
    \vspace{-12pt}
    \caption{A few examples of matrix structures arising in practice}
    \vspace{-8pt}
\end{figure}

\textbf{Our world is full of structured data.}
In this work, we make the distinction between a \textbf{tensor}, which describes any multidimensional object which relates tuples of integer coordinates to values, such as vectors or matrices, and an \textbf{array}, the previously described classical data structure.
We say that a tensor is \textbf{structured} when it has patterns that allows us to optimize storage or computation of the tensor.
Sparse tensors (which store only nonzero elements) describe networks, databases, and simulations~\cite{abhyankar_petsc_nodate,bell_lessons_2007,mcauley_hidden_2013,balay_petsc_2020}.
Run-length encoding describes images, masks, geometry, and
databases (such as a list of transactions with the date field all the same)~\cite{shi_column_2020,golomb_run-length_1966}.
Symmetry, bands, padding, and blocks arise due to modeling choices in scientific computing (e.g., higher order FEMs) as well as in intermediate structures in many linear solvers (e.g., GMRES)~\cite{burns_dedalus_2020,saad_iterative_2003,oleary_scientific_2009}.
Combinations of sparse and blocked matrices are increasingly under consideration in machine learning~\cite{dao_monarch_2022}.
Even complex operators can be expressed as structured tensors.
For example, convolution can be expressed as a matrix multiplication
with the Toeplitz matrix of all the circular shifts of the filter~\cite{sze_efficient_2020}.

\textbf{Currently, support for structured data is fragmented and incomplete}.
Experts must hand write variations of even the simplest kernels, like matrix
multiply, for each data structure/data set and architecture to get performance.
Implementations must choose a small set of features to support well, resulting
in a compromise between \textbf{program flexibility} and \textbf{data structure
flexibility}.
Hand-written solutions are collected in diverse libraries like
MKL, OpenCV, LAPACK or SciPy~\cite{bradski_opencv_2000,anderson_lapack_1999,virtanen_scipy_2020,psarras_linear_2022}. 
However, libraries will only ever support a subset of
programs on a subset of data structure combinations.
Even the most advanced
libraries, such as GraphBLAS, which support a wide variety of sparse
operations over various semi-rings always lack support for other features, such
as $N$-D tensors, fused outputs, or runs of repeated values~\cite{buluc_design_2017,mattson_lagraph_2019}.
While dense tensor
compilers support an enormous variety of program constructs like early break and
multiple left hand sides, they only support dense tensors~\cite{ragan-kelley_halide_2013,grosser_pollyperforming_2012}.  
Special-purpose
compilers like TACO~\cite{kjolstad_tensor_2019}, Taichi~\cite{hu_taichi_2019}, StructTensor~\cite{ghorbani_compiling_2023}, or CoRa~\cite{fegade_cora_2022} which support a select subset of structured data
structures (only sparse, or only ragged tensors) must compromise by greatly
constraining the classes of programs which they support, such as tensor
contractions.
This trade-off is visualized in Tables \ref{tab:features} and \ref{tab:data_structures}.

\newcommand*\rot{\rotatebox{90}}

\begin{table}[b!]
\vspace{-12pt}
\noindent
\flushleft
\begin{minipage}[t]{.49\textwidth}
  \scriptsize
  \begin{tabular}{l|cccccc}
  \textbf{Feature / Tool} & \rothead{Halide} & \rothead{Taco} & \rothead{Cora} & \rothead{Taichi} & \rothead{Stur} & \rothead{Finch} \\
  \hline
  Einsums/Contractions & \checkmark & \checkmark & \checkmark & \checkmark & \checkmark & \checkmark \\
  Multiple LHS             & \checkmark &            & \checkmark & \checkmark &            & \checkmark \\
  Affine Indices           & \checkmark &            &            & \checkmark & \checkmark & \checkmark \\
  Recurrence               & \checkmark &            &            &            &            &           \\
  If-Conditions and Masks  & \checkmark & \checkmark &            & \checkmark &            & \checkmark \\
  Scatter Gather           & \checkmark &            &            & \checkmark &            &\checkmark \\
  Early Break              &            & \checkmark &            & \checkmark &            &\checkmark \\
  Unrestricted Read/Write              &    \checkmark        &  &            &  &            &  \\
  \end{tabular}
  \caption{Control flow support across various tools.}
  \label{tab:features}
\end{minipage} 
\begin{minipage}[t]{.49\textwidth}
  \flushright
  \scriptsize
  \begin{tabular}{l|cccccc}
  \textbf{Feature / Tool} & \rothead{Halide} & \rothead{Taco} & \rothead{Cora} & \rothead{Taichi} & \rothead{Stur} & \rothead{Finch} \\
  \hline
  Dense                    & \checkmark & \checkmark & \checkmark & \checkmark & \checkmark & \checkmark \\
  Padded                   & \checkmark &            &            &            &            & \checkmark \\
  One Sparse Operand              &            & \checkmark &            & \checkmark &            &\checkmark \\
  Multiple Sparse  Operands                 &            & \checkmark &            &            &            &\checkmark \\
  Run-length               &            &            &            &            &            & \checkmark \\
  Symmetric                &            &            &            &            & \checkmark & \checkmark \\
  Regular Sparse Blocks    &            & \checkmark &            &            &            & \checkmark \\
  Irregular Sparse Blocks  &            &            &            &            &            &\checkmark \\
  Ragged                   &            &            & \checkmark &            &            & \checkmark \\
  \end{tabular}
  \caption{Data structure support across various tools.  Finch supports \textbf{both} complex programs and complex data structures.}
  \label{tab:data_structures}
\end{minipage}
\vspace{-24pt}
\end{table}

Prior implementations are incomplete because the abstractions they use are tightly coupled with the specific data structures that they support.
For example, TACO merge lattices represent Boolean logic over sets of non-zero values on an integer grid~\cite{kjolstad_tensor_2017}.
The polyhedral model allows various compilers to represent dense computations on affine regions~\cite{grosser_pollyperforming_2012}.
Taichi enriches single static assignment form with a specialized instruction for accessing only a single sparse structure, but it supports more control flow ~\cite{hu_taichi_2019}.
These systems tightly couple their control flow to narrow classes of data structures to avoid the challenges that occur when we intersect complex control flow with structured data. There are two challenges:

\textbf{Optimizations are specific to the indirection and patterns in data structures}: 
These structures break the simple mapping between tensor elements and where they are stored in memory.
For example, sparse tensors store lists of which coordinates are nonzero, whereas run-length-encoded tensors map several pixels to the same color value. 
These zero regions or repeated regions are optimization opportunities, and we must adapt the program to avoid repetitive work on these regions by referencing the stored structure.

\textbf{Performance on structured data is highly algorithm dependent}: The landscape of implementation decisions is dramatically unpredictable. 
For example, the asymptotic performance of sparse matrix multiplication can be impacted by the distribution of nonzeros, the sparse format, and the loop order~\cite{ahrens_autoscheduling_2022,zhang_gamma_2021}. 
This means that performance engineering for such kernels requires the exploration of a large design space, changing the algorithm as well as the data structures.

\textbf{In this work, we propose a new programming language, Finch, which supports \textit{both} flexible control flow and diverse data structures.}
Finch facilitates a programming model which resolves the challenges of computing over structured tensors by \textbf{combining control flow and data structures into a common representation where they can be co-optimized}.
In particular, Finch automatically specializes the control flow to the data so that performance engineers can focus on experimenting with many algorithms.
Finch supports a familiar programming language of loops, statements, if conditions, breaks, etc., over a wide variety of tensor structures, such as sparsity, run-length-encoding, symmetry, triangles, padding, or blocks. 
This support would be useless without the appropriate level of structural specialization; Finch reliably utilizes the key properties of structure, such as structural zeros, repeated values, or clustered non-zeros.

As an example, a programmer might explore different ways to intersect only the even integers of two lists (represented as sparse vectors with sorted indices). The control flow here is only useful if the first example differs from the next two in that it actually selects only even indices as the two integer lists are merged and different from the last in that it does not require another tensor:

%
%
%

\begin{minipage}{0.25\linewidth}
\begin{minted}{julia}
     for i = _
         if i % 2 == 0
             c[i]=a[i]*b[i]
\end{minted}
\end{minipage}%
\begin{minipage}{0.25\linewidth}
\begin{minted}{julia}
     for i = _
         if i % 2 == 0
             ap[i] = a[i]
     for i = _
         c[i] = ap[i] * b[i]
\end{minted}
\end{minipage}%
\begin{minipage}{0.25\linewidth}
\begin{minted}{julia}
     for i = _
         cp[i] = a[i] * b[i]
     for i = _
         if i % 2 == 0
             c[i] = cp[i]
\end{minted}
\end{minipage}%
\begin{minipage}{0.25\linewidth}
\begin{minted}{julia}
     for i = _
        if i % 2 == 0
            f[i] = 1
     for i = _
        c[i] = a[i] * b[i] * f[i]
\end{minted}
\end{minipage}%

\subsection{Contributions}

\begin{enumerate}
\item More complex tensor structures than ever before. We are the first to extend level-by-level hierarchical descriptions to capture banded, triangular, run-length-encoded, or sparse datasets, and any combination thereof.
We have chosen a set of level formats that completely captures all combinations of relevant structural properties (zeros, repeated values, and/or blocks).
Although many systems (TACO, Taichi, SPF, Ebb)~\cite{chou_format_2018,  hu_taichi_2019, strout_sparse_2018, bernstein_ebb_2016} feature a flexible structure description, our level abstraction is more capable and extensible because it uses looplets \cite{ahrens_looplets_2023} to express the structure of each level. 
\item A rich structured tensor programming language with for-loops and complex control flow constructs at the same level of productivity of dense tensors. 
To our knowledge, the Finch programming language is the first to support if-conditions, early breaks, and multiple left hand sides over structured data, as well as complex accesses such as affine indexing or scatter/gather of sparse or structured operands.
\item A compiler that specializes programs to data structures automatically, facilitating an expressive language that makes it easier to search the complex space of algorithms and data structures. Finch reliably utilizes four key properties of structure, such as structural zeros, repeated values, clustered non-zeros, and singletons.
\item Our compiler is highly extensible, evidenced by the variety of level formats and control flow constructs that we implement in this work.
For example, Finch has been extended to support real-valued tensor indices with continuous tensors. 
Finch is also used as a compiler backend for the Python PyData/Sparse library \cite{abbasi_plans_2023}.
\item We evaluate the efficiency, flexibility, and expressiveness of our language in several case studies on a wide range of applications, demonstrating speedups over the state of the art in classic operations such as SpMV (geomean $1.26\times$, max $3.04\times$) and SpGEMM (geomean $1.30\times$, max $1.62\times$), to more complex applications such as graph analytics (geomean $2.47\times$ on Bellman-Ford, reducing lines of code by $4\times$ over GraphBLAS), and image processing ($19.5\times$ on the humansketches dataset \cite{eitz_how_2012}).

\end{enumerate}

\section{Background}

\subsection{Looplets}
Finch represents iteration patterns using looplets, a language that decomposes datastructure iterators hierarchically. 
Looplets represent the control-flow structures needed to iterate over any given datastructure, or multiple datastructures simultaneously. 
Because looplets are compiled with progressive lowering, structure-specific mathematical optimizations such as integrals, multiply by zero, etc. can be implemented using simple compiler passes like term rewriting and constant propagation during the intermediate lowering stages.

\begin{figure}[h]
  \footnotesize
  \begin{subfigure}[t]{0.23\linewidth}
    \centering
    \includegraphics[scale=0.25]{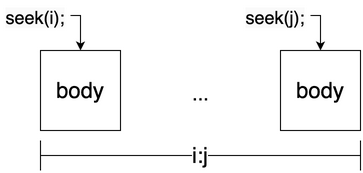}
    \caption{A \textbf{Lookup} looplet represents a randomly accessible sequence as a function of the index.}
  \end{subfigure}\hfill%
  \begin{subfigure}[t]{0.23\linewidth}
    \centering
    \includegraphics[scale=0.25]{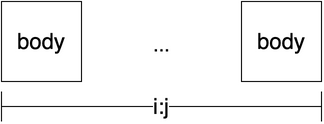}
    \caption{A \textbf{Run} looplet represents a sequence of many of the same value, usually stored once.}
  \end{subfigure}\hfill%
  \begin{subfigure}[t]{0.26\linewidth}
    \centering
    \includegraphics[scale=0.25]{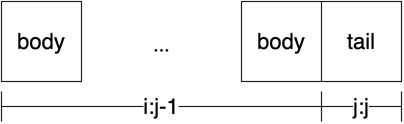}
    \caption{A \textbf{Spike} looplet represents a run followed by a single, different value at the end of the target region.}
  \end{subfigure}\hfill%
  \begin{subfigure}[t]{0.24\linewidth}
    \centering
    \includegraphics[scale=0.25]{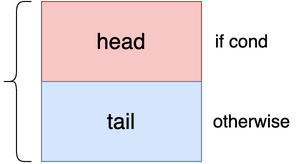}
    \caption{A \textbf{Switch} looplet represents a choice between different looplets under different conditions.}
  \end{subfigure}

  \begin{subfigure}[t]{0.32\linewidth}
    \centering
    \includegraphics[scale=0.25]{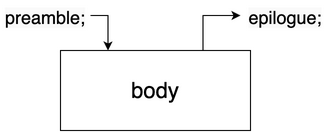}
    \caption{A \textbf{Thunk} looplet allows us to add side effects such as caching a value for sublooplets to use.}
  \end{subfigure}\hfill%
  \begin{subfigure}[t]{0.32\linewidth}
    \centering
    \includegraphics[scale=0.25]{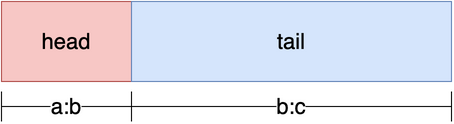}
    \caption{A \textbf{Sequence} looplet represents a sequence of a few different looplets, one after the other.}
  \end{subfigure}\hfill%
  \begin{subfigure}[t]{0.32\linewidth}
    \centering
    \includegraphics[scale=0.25]{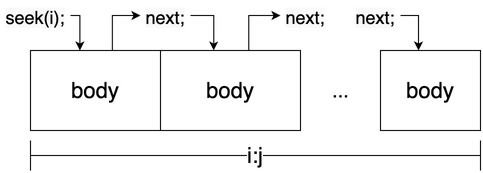}
    \caption{A \textbf{Stepper} looplet represents a sequence of an unbounded number of identical looplets.}
  \end{subfigure}\hfill%
  \caption[Visual representations of the looplet language]{The looplet language, as understood in a correct execution of a Finch program.}
  \label{fig:looplets}
\end{figure}

The looplets are described in Figure~\ref{fig:looplets}. We simplify the presentation to focus on the semantics, rather than precise implementation. 
For more background on looplets, we recommend the original work \cite{ahrens_looplets_2023}.
Several looplets introduce or modify variables in the scope of the target language.
This allows looplets to lift code to the highest possible loop level. 
It is assumed that if a looplet introduces a variable, the child looplet will not modify that variable.  

Finch advances the state-of-the-art over the looplets work \cite{ahrens_looplets_2023}. 
While looplets presented a way to merge iterators over single dimensional structures, Finch is the only framework to support such a broad range of multi-dimensional structured data in a programming language with fully-featured control flow. 
Looplets provide a powerful mechanism to simplify structured loops, but our paper shows how to make this functionality practical; Finch uses looplets as a symbolic loop simplification engine. 
The precise choice and implementation of tensor level structures, the lifecycle interface between levels and looplets, and the canonicalization of fancy indexing and masking all serve to utilize and recombine looplets to achieve efficient computation over structured tensors.


\begin{table}[H]
  \footnotesize
  \begin{tabular}{|p{8.1cm}|p{4.9cm}|}
  \hline
  \textbf{Description} & \textbf{Arguments} \\
  \hline
  
  \textbf{\finchlookup(body):} The Lookup looplet represents a randomly accessible region of an iterator, where the element at index $i$ is given by the expression returned by the function $body(i)$. While this expression is often a tensor access, it could also be a function call, like $f(i) = \sin(\pi i/7)$. Lookups are leaf looplets, and the body is a value, not a looplet. & 
  • $body(i)$: A function which returns an expression representing the value at index $i$ in the current program state. \\
  \hline
  
  \textbf{\finchrun(body):} The Run looplet represents a constant region of an iterator. Runs are leaf looplets, and the body of a run is a value, not a looplet, similar to a Lookup. Run looplets do not need to store any information about their region because it is specified by the enclosing loop. & 
  • $body$: An expression representing the value within the run in the current program state. \\
  \hline
  
  \textbf{\finchswitch(cond, head, tail):} The Switch looplet allows us to specialize the body of a looplet based on a condition, evaluated in the embedding context. If the condition is true, we use $head$, otherwise $tail$. Switch has a high lowering priority so we can see the looplets it contains and lower them appropriately. Lowering Switch looplets first also lifts the condition as high as possible into the loop nest. & 
  • $cond$: A function that returns a Boolean value. \newline
  • $head$: A looplet to execute if the condition is true. \newline
  • $tail$: A looplet to execute if the condition is false. \\
  \hline
  
  \textbf{\finchthunk(preamble, body, epilogue):} The Thunk looplet allows us to cache certain computations in program state. The computations can be used by the Thunk $body$, making Thunks useful for computing and caching the results of expensive computations. The $epilogue$ can be used to clean up any relevant side effects.& 
  • $preamble$: A program that executes before $body$, modifying program state. \newline
  • $body$: A looplet that can reference variables defined in $preamble$. \newline
  • $epilogue$: A program that cleans up any relevant side effects of $preamble$ or $body$. \\
  \hline
  
  \textbf{\finchsequence(bodies...):} The Sequence looplet represents the concatenation of two or more looplets. The arguments must be $phase$ objects which regions on which each body is defined. & 
  • $bodies...$: One or more phase objects, whose regions must be non-overlapping, covering, and ordered. \\
  \hline
  
  \textbf{\finchphase(ext, body):} The Phase object is not a looplet, but instead helpfully couples a sublooplet with the subregion of indices it is defined on in a larger compound looplet. & 
  • $ext$: An expression representing the absolute range on which the $body$ is defined. \newline
  • $body$: The looplet describing the sequence within the $range$. \\
  \hline
  
  \textbf{\finchspike(body, tail):} The Spike looplet represents a run followed by a single value. Spike can be considered a shorthand for $\finchsequence(\finchphase(i:j-1, \finchrun(body)), \finchphase(j:j, \finchrun(tail)))$. In the Finch compiler, spikes are handled with special care, since they are an opportunity to align the final run to the end of the root loop extent without using any special bounds inference. & 
  • $body$: An expression representing the value within the run. \newline
  • $tail$: An expression representing the value at the end of the spike. \\
  \hline
  
  \textbf{\finchstepper([seek], next, stride, body):} The Stepper looplet represents a variable number of looplets, concatenated. Since our looplets may be skipped over due to conditions or various rewrites, the $seek$ function allows us to fast-forward the state to the start of the root loop extent when it comes time to lower the stepper. \newline
  \textbf{\finchjumper(seek, next, body):} The Jumper looplet is identical to a stepper looplet, but when two jumpers interact, the largest stride between them is taken, and the jumper with the smaller stride is demoted to a stepper within that region. Jumpers allow us to request leader-follower strategies or mutual-lookahead coiteration.
  & 
  • $seek(j)$: A function that returns a program that advances state to the iteration of the stepper which processes the absolute coordinate $j$. \newline
  • $next$: A program that advances the state to the next iteration of the stepper. \newline
  • $stride$: The absolute endpoint of the current subregion of the stepper. \newline
  • $body$: The looplet to execute for the current iteration of the stepper.\\
  \hline
  \end{tabular}
  \caption{Detailed descriptions of looplet behavior. An example compilation is given later in Figures~\ref{fig:lowering_example1} and \ref{fig:lowering_example2}}
  \vspace{-24pt}
  \end{table}

\subsection{Fiber Trees}

Fiber-tree style tensor abstractions have been the subject of extensive study
\cite{sze_efficient_2020,chou_compilation_2022,chou_format_2018}.  
The underlying
idea is to represent a multi-dimensional tensor as a nested vector
datastructure, where each level of the nesting corresponds to a dimension of the
tensor. Thus, a matrix would be represented as a vector of vectors. This kind of
abstraction lends itself to representing sparse tensors if we vary the type of
vector used at each level in a tree. Thus, a sparse matrix might be represented
as a dense vector of sparse vectors. The vector of subtensors in this
abstraction is referred to as a \textbf{fiber}.

\begin{wrapfigure}{r}{0.45\linewidth}
  \centering
  \vspace{-14pt}
  \includegraphics[width=\linewidth]{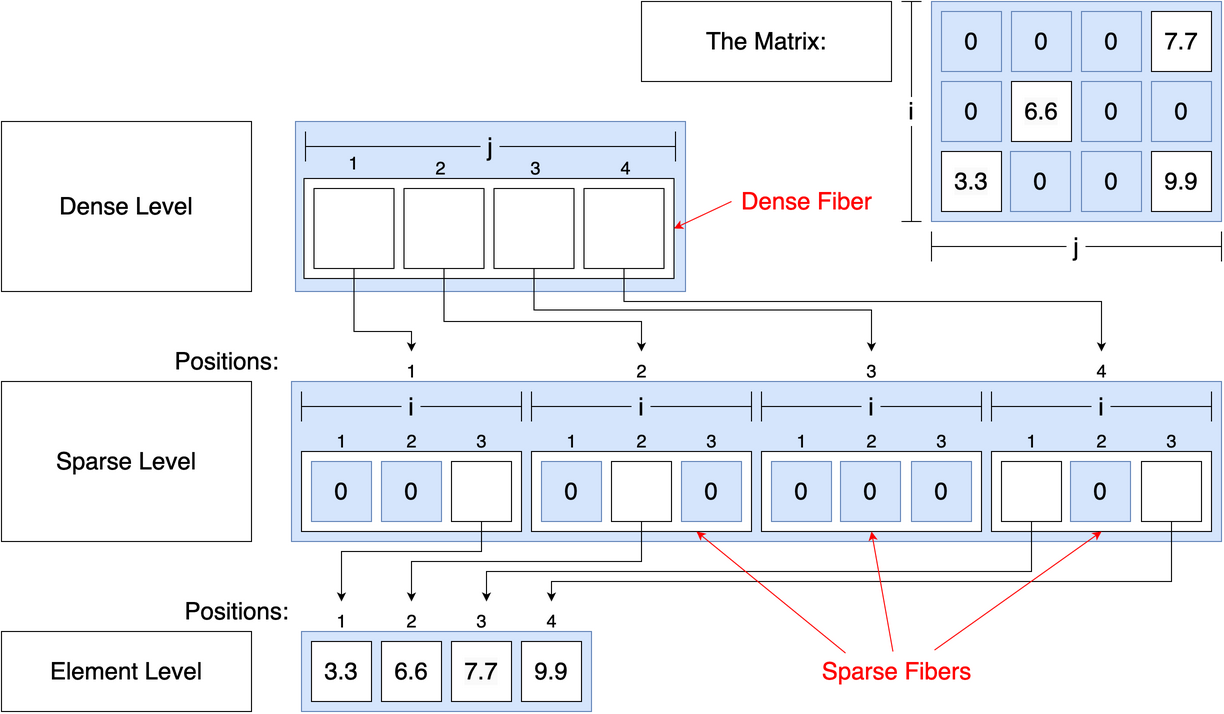}
  \vspace{-16pt}
  \caption{Levels in the fiber tree representation of a sparse matrix in CSC format, with a dense outer level and a sparse inner level. The element level holds the leaves of the tree.}
  \label{fig:levelsvsfibers}
  \vspace{-20pt}
\end{wrapfigure}

Instead of storing the data for each subfiber separately, most sparse tensor
formats such as CSR, DCSR, and COO usually store the data for all fibers in a
level contiguously. In this way, we can think of a level as a bulk allocator for
fibers. Continuing the analogy, we can think of each fiber as being
disambiguated by a \textbf{position}, or an index into the bulk pool of
subfibers. The mapping $f$ from indices to subfibers is thus a mapping from an
index and a position in a level to a subposition in a sublevel.
Figure~\ref{fig:levelsvsfibers} shows a simple example of a level as a pool of fibers.
When we need to refer to a particular fiber at position $p$ in the level $l$, we
may write $fiber(l, p)$. Note that the formation of fibers from levels is lazy,
and the data underlying each fiber is managed entirely by the level, so the
level may choose to overlap the storage between different fibers. Thus, the only
unique data associated with $fiber(l, p)$ is the position $p$.
\section{Bridging Looplets and Finch: The Tensor Interface}
%
%
%
%
Tensors use multiple dimensions to organize data with respect to orthogonal concepts.
Thus, the Finch language supports multi-dimensional tensors.
Unfortunately, the looplet abstraction is best suited towards iterators over a single dimension.
Our level abstraction provides a bridge between the single dimensional iterators created from looplets and the multi-dimensional fiber-tree abstractions common to tensor compilers.
This bridge must address three challenges.
First, while looplets represent an instance of an iterator over a tensor, we may access the same tensor twice with different indices.
Thus, the $unfurl$ function creates separate looplet nests for each iterator.
Next, since Finch programs go beyond just single Einsums, they may read and write to the same data at different times.
The $declare$, $freeze$, and $thaw$ functions provide machinery to manage transition between these states.
Finally, we must be able to write looplet nests that modify tensors, as well as reading them.
The $assemble$ function manages the allocation of new data in the tensor.

Additionally, prior fiber-tree representations focus on sparsity (where only the nonzero elements are represented) and treat sparse vectors as sets of represented points. Since our fiber-tree representation must handle other kinds of structure, such as diagonal, repeated, or constant values, we must generalize our fiber abstraction to allow arbitrary mappings from indices into a space of subfibers.

In the rest of this section, we discuss how these 5 core functions ($declare$, $freeze$, $thaw$, $unfurl$, and $assemble$) function as part of a life cycle abstraction that defines a level in Finch.
These interfaces add to the level abstraction, expanding the types of data that they can express via mapping to looplets and expanding the contexts in which they can be used.
We then identify a taxonomy of four key structural properties exhibited in data.
We implement several levels in this abstraction that capture all combinations of these structures, 
including specializations to zero dimensional tensors (scalars) and level structures that support different access patterns.

\subsection{Tensor Lifecycle, Declare, Freeze, Thaw, Unfurl}

Our simplified view of a level is enabled by our use of looplets to represent
the structure within each fiber.
In fact, our level interface requires only 5
highly general operations, described below.

The first three of these functions, $declare$, $freeze$, and $thaw$, have to do with managing when tensors can be assumed mutable or immutable.
As we use looplets to represent iteration over a tensor, we must restrict the mutability of tensors in the region of code which iterates over them. 
For example, if a tensor declares it has a constant
region from $i = 2:5$, but some other part of the computation modifies the
tensor at $i = 3$, this would result in incorrect behavior.
It is much easier to
write correct looplet code if we can assume that the tensor is immutable while
it is being read from.
Thus, we introduce the notion that a tensor can be in
read-only mode or update-only mode.  
In read-only mode, the tensor may only
appear in the right-hand side of assignments. 
In update-only mode, the tensor
may only appear in the left-hand side of an assignment, either being overwritten
or incremented by some operator. 
We can switch between these modes using freeze
and thaw functions.
The $declare$ function is used to allocate a tensor,
initialize it to some specified size and value, and leave it in update-only
mode. 

\begin{table}[h]
    \footnotesize
    \begin{tabular}{|p{7cm}|p{6.2cm}|}
        \hline
        \textbf{Description} & \textbf{Arguments} \\
        \hline
        
        \textbf{$\mathit{declare}(\mathit{tns}, \mathit{init}, \mathit{dims}...):$} Returns a program that declares a tensor of size $\mathit{dims}$ and an initial value of $\mathit{init}$. Requires the level to be in read-only mode. & 
        • $tns$: The tensor object to declare. \newline
        • $init$: An expression for the initial value. \newline
        • $dims...$: Several expressions for the tensor dimensions. \\
        \hline
        
        \textbf{$\mathit{freeze}(\mathit{tns}):$} Returns a program that finalizes the updates in the tensor, and readies the tensor for reading. Requires the tensor to be in update-only mode. & 
        • $tns$: The tensor object to freeze. \\
        \hline
        
        \textbf{$\mathit{thaw}(\mathit{tns}):$} Returns a program that prepares the level to accept updates. Requires the tensor to be in read-only mode. & 
        • $tns$: The tensor object to thaw. \\
        \hline
        
        \textbf{$\mathit{unfurl}(\mathit{tns}, \mathit{ext}, \mathit{mode}):$} Returns a looplet that iterates over subtensors within the tensor along the extent $\mathit{ext}$. When $\mathit{mode} = \finchread$, returns a looplet nest over the values in the read-only fiber. When $\mathit{mode} = \finchupdate$, returns a looplet nest over mutable subfibers in the update-only fiber. The compiler calls $\mathit{unfurl}$ directly before iterating over the corresponding loop, so it has access to any state variables introduced by freezing or thawing the tensor.& 
        • $tns$: The tensor or subtensor to unfurl. \newline
        • $ext$: An expression representing the range to unfurl over. \newline
        • $mode$: An enum representing whether to unfurl in read-only or update-only mode. \\
        \hline
        
        \textbf{$\mathit{unwrap}(\mathit{tns}, \mathit{mode}, [\mathit{op}], [\mathit{rhs}]):$} Returns code to read or update the scalar value of a scalar or leaf node $\mathit{tns}$, using $\mathit{op}$ and $\mathit{rhs}$ in the case of update. Parent fibers may ask their children to use this function to set a dirty bit in $\mathit{tns}$, indicating a non-fill value has been written and that the child fiber needs to be stored. & 
        • $tns$: The tensor object to increment, possibly a fiber. \newline
        • $mode$: An enum representing whether to unwrap in read-only or update-only mode. \newline
        • $op$: An expression representing the operation to apply to the scalar value. \newline
        • $rhs$: An expression representing the second argument to $op$. \\
        \hline
        
        \textbf{$\mathit{assemble}(\mathit{lvl}, \mathit{pos}_{\mathit{start}}, \mathit{pos}_{\mathit{stop}}):$} Returns a program that allocates subfibers in the level from positions $\mathit{pos}_{\mathit{start}}$ to $\mathit{pos}_{\mathit{stop}}$. In looplet nests which modify the output, this function is often called to construct the output tensor. For example, to handle the case where a new nonzero is discovered, the compiler might call $\textit{assemble}$ to obtain a location in memory to which the nonzero may be written. & 
        • $lvl$: The level object in which subfibers are allocated. \newline
        • $pos_{start}$: The first subfiber position to assemble. \newline
        • $pos_{stop}$: The last subfiber position to assemble. \\
        \hline
        \end{tabular}
        \caption{The five functions that define a level.}
        \vspace{-24pt}
\end{table}

The $unfurl$ function is used to manage iteration over a subfiber.
When it comes time to iterate over a tensor, be in on the left or right hand
side of an assignment, the compiler calls $unfurl$ to return a looplet nest that describes the hierarchical structure of the outermost dimension of the tensor.
The compiler calls $unfurl$ directly before
compiling the corresponding loop, so the called has access to any state variables introduced
by freezing or thawing the tensor. Looplets were chosen for this purpose as a symbolic engine to ensure certain simplifications take place, but another symbolic system could have been used (e.g. polyhedral\cite{zhao_polyhedral_2022} or e-graph search \cite{shaikhha_functional_2022}). We chose looplets because they reliably process structured iterators, predictably eliminating zero regions, using faster lookups when available, and utilizing repeated work.

Our view of a level as a fiber allocator implies an allocation function
$assemble(tns, pos_{start}:pos_{stop})$, which allocates fibers at positions
$pos_{start}:pos_{stop}$ in the level. We don't specify a de-allocation
function, instead relying on initialization to reset the fiber if it needs to be
reused. While all of the previous functions are used to manage the lifecycle and
iteration over a general tensor, $assemble$ is quite specific to the
level abstraction, and the notion of positions within sublevels. Note: it was an
intentional choice to hold the parent level responsible for managing the
data of the sublevels, which positions they allocate, etc. This allows the parent
level to reuse allocation logic from internal index datastructures. For example,
a sparse level might use a list of indices to store which nonzeros are present,
and when it comes time to resize that list, it could also call $assemble$ to resize the
sublevel, reducing the number of branches in the code. The $assemble$ function
lends itself particularly to a "vector doubling" allocation approach, which we
have found to be effective and flexible when managing the allocation
of sparse left hand sides. This benefit is made clear in our case studies,
where prior systems like TACO do not support all possible loop
orderings and format combinations for sparse matrix multiply because they do
not have a flexible enough allocation strategy, instead using a two-phase approach
which requires computing a complicated closed-form kernel to iterate over the
data twice to determine the number of required output nonzeros.

\subsection{The 4 Key Structures}

\begin{wrapfigure}{r}{.35\textwidth}
    \vspace{-26pt}
    \centering
    \scriptsize
    \begin{tabular}{|c|c|c|c|l|}
        \hline
        \rothead{Sparse} & \rothead{Blocked} & \rothead{Runs} & \rothead{Singular } & \begin{tabular}{@{}l@{}}\textbf{Corresponding} \\ \textbf{Format}\end{tabular} \\ \hline
        &  &  &  & Dense \\ \hline
        &  &  & $\checkmark$ & n/a \\ \hline
        &  & $\checkmark$ &  & RunList \\ \hline
        &  & $\checkmark$ & $\checkmark$ & n/a \\ \hline
        & $\checkmark$ &  &  & n/a \\ \hline
        & $\checkmark$ &  & $\checkmark$ & n/a \\ \hline
        & $\checkmark$ & $\checkmark$ &  & n/a \\ \hline
        & $\checkmark$ & $\checkmark$ & $\checkmark$ & n/a \\ \hline
       $\checkmark$ &  &  &  & SparseList \\ \hline
       $\checkmark$ &  &  & $\checkmark$ & SparsePinpoint \\ \hline
       $\checkmark$ &  & $\checkmark$ &  & SparseRunList \\ \hline
       $\checkmark$ &  & $\checkmark$ & $\checkmark$ & SparseInterval \\ \hline
       $\checkmark$ & $\checkmark$ &  &  & SparseBlockList \\ \hline
       $\checkmark$ & $\checkmark$ &  & $\checkmark$ & SparseBand \\ \hline
       $\checkmark$ & $\checkmark$ & $\checkmark$ &  & n/a \\ \hline
       $\checkmark$ & $\checkmark$ & $\checkmark$ & $\checkmark$ & n/a \\ \hline
    \end{tabular}
    \caption{All combinations of our 4 structural properties and the
    corresponding formats we have chosen to represent them. Not all combinations
    are relevant. Note that blocks and runs need not be considered together
    because we must store a run length for each run, so there isn't a
    significant storage benefit to combining them. Blocks and singletons only
    make sense in the context of sparsity.}
    \vspace{-24pt}
    \label{tab:TypesOfStructure}
\end{wrapfigure}

    In the Finch programming model, the programmer relies on the Finch compiler to specialize to the sequential properties of the data. In our experience, the main benefits of specializing to structure come from the following properties of the data:

    \begin{itemize}
        \item \textbf{Sparsity} Sparse data is data that is mostly zero, or some other
        fill value. When we specialize on this data, we can use annihilation ($x
        * 0 = 0$), identity ($x * 1 = 1$), or other constant propagation
        properties ($ifelse(false, x, y) = y$) to simplify the computation and avoid
        redundant work.
        
        \item \textbf{Blocks} Blocked data is a subset of sparse data where the nonzeros
        are clustered and occur adjacent to one another. This provides us with
        two opportunities: We can avoid storing the locations of the nonzeros
        individually, and we can use more efficient randomly accessible
        iterators within the block. \cite{im_optimizing_2001, vuduc_performance_2002, ahrens_looplets_2023}.

        \item \textbf{Runs} Runs of repeated values may occur in dense or sparse code,
        cutting down on storage and allowing us to use integration rules such as 
        \mintinline{julia}{for i = 1:n; s += x end} $\rightarrow$
        \mintinline{julia}{s += n * x} or code motion to lift operations out of loops \cite{donenfeld_unified_2022,ahrens_looplets_2023}.

        \item \textbf{Singular} When we have only one non-fill region in sparse data,
        we can avoid a loop entirely and reduce the complexity of iteration \cite{ghorbani_compiling_2023, ahrens_looplets_2023}.
    \end{itemize}
    In the following section, we consider a set of concrete implementations of levels that expose all combinations of these structures, paying some attention to a few important special cases: random access, scalars, and leaf levels. We summarize the structures in Table~\ref{table:formats} and Table~\ref{tab:TypesOfStructure}.
    
    \begin{figure}[htbp]
        \centering
        \includegraphics[width=\linewidth]{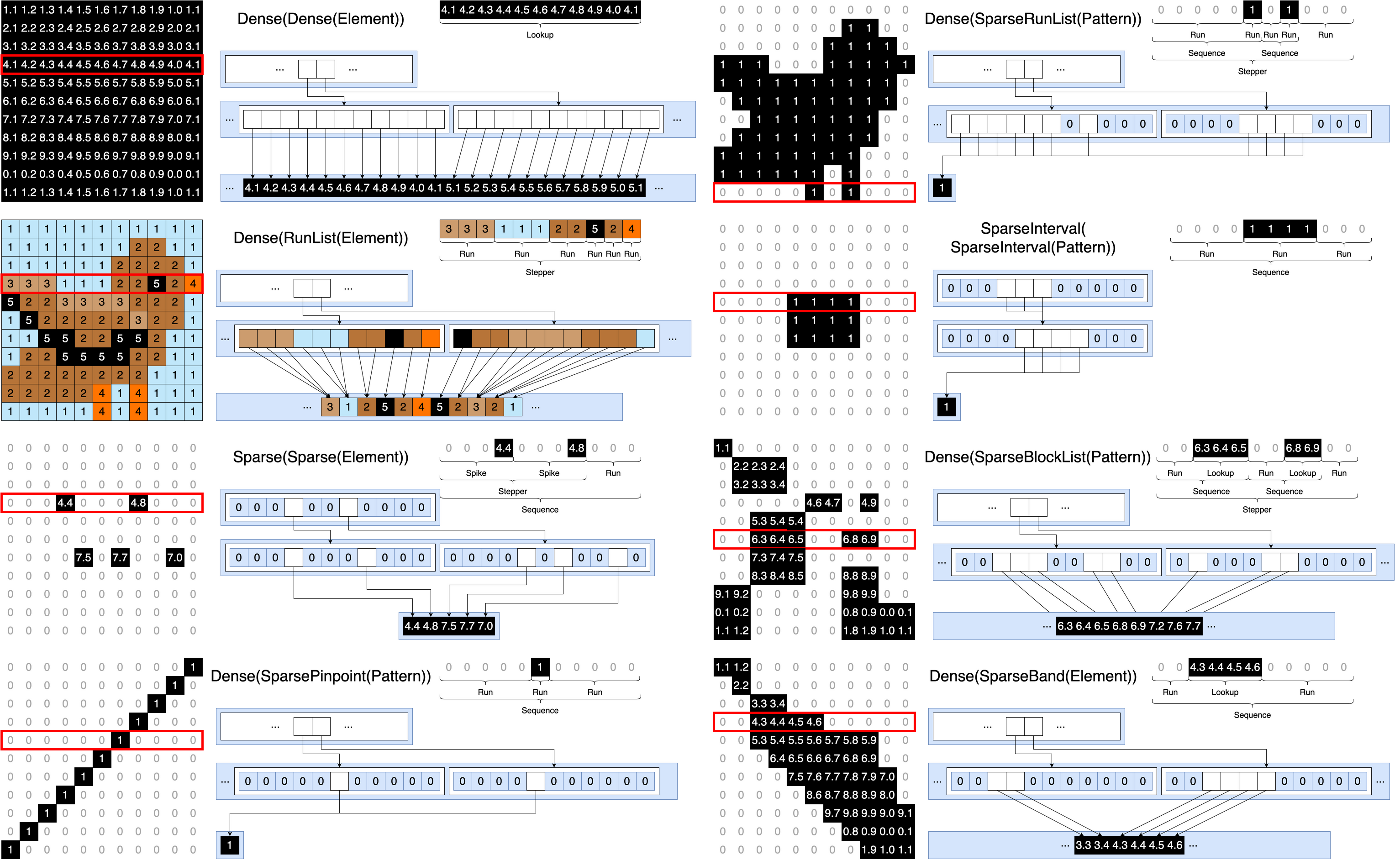}\hfill%
        \vspace{-8pt}
        \caption{Several examples of matrix structures represented using the
        level structures identified in Table~\ref{tab:TypesOfStructure}.
        Comparing this figure to \cite[Figure 3]{ahrens_looplets_2023}, we see
        that a level-by-level structural decomposition is diagrammed together
        with the looplets.}
        \label{fig:structuraldiversity}
        \vspace{-18pt}
    \end{figure}

\subsection{Implementations of Structures}\label{sec:formats}
\subsubsection{Sequentially Constructed Levels}
   We consider all combinations of these structural properties in Table
    \ref{tab:TypesOfStructure}, resulting in 8 key level formats that correspond
    to the 8 resulting situations. 
    While it is impossible to write code which
    precisely addresses every possible structure, our level
    formats can be combined to express a wide variety of hierarchical structures to a
    sufficient granularity that we can generate code which utilizes the four
    properties.  For example, though banded tensors are a superset of ragged
    tensors, a ragged matrix can be stored in our banded format where the band
    in each row starts at the first column. In practice, the overhead of storing a 
    1 for the start of each band is negligible.
    The structures we consider are exhaustive in the sense that they
    address all combinations of sparsity, blocks, runs, and singletons in each
    level. We can represent a wide variety of hierarchical tensor structures by
    combining these level structures in a tree, as shown in
    Figure~\ref{fig:structuraldiversity}.

\subsubsection{Non-sequentially Constructed Levels}
To reduce the implementation burden and improve efficiency in the common case, the levels we described in the previous section only support bulk,
sequential construction of formats. However, when users want to be able to write out
of order (which is a common requirement arising from loop order or from the
problem itself, it occurs in our SpGEMM algorithms and our histogram example in
the evaluation section), we must use more complicated datastructures like hash
tables and trees to support the random writes so that in-order levels can be constructed later. Because these datastructures are more complex and have a higher implementation burden and performance overhead,
we only support random access construction of sparse or dense structures.  We
can use these two more general structures as intermediates to convert to our
more specialized structures later.



\begin{table}[h]
    \centering
    \scriptsize
    \renewcommand{\arraystretch}{1.3}
    \begin{tabular}{p{14cm}}

    \hline
    \textbf{Sequentially Constructed Levels} \\
    \hline
    \textbf{Dense}:
    The dense format is the simplest format, mapping $fiber(l, p)[i] \rightarrow fiber(l.lvl, p \times l.shape + i)$.
    This format is used to store dense data and is often a convenient format for the root level of a tensor.
    Due to its simplicity, freezing and thawing the level are no-ops. \\
    \textbf{RunList}:
    Used to represent runs of repeated values, storing two vectors, $right$ and $ptr$, with $q^{th}$ run in the $p^{th}$ subfiber starting and ending at $right[ptr[p] + q]$ and $right[ptr[p] + q + 1] - 1$, respectively.
    A challenge arises for this level: it is difficult to merge duplicate runs.
    Such a scenario might arise when merging runs of subfibers of length 3, representing colors in an image.
    Ideally, we would be able to detect duplicate subfibers and merge them on the fly, but we cannot determine which subfibers are equal because the sublevel cannot be read in update-only mode.
    Instead, the duplicates are merged during the freeze phase.
    The compiler $freezes$ the sublevel, $declares$ a separate sublevel $buf$ as a buffer to store the deduplicated subfibers, and then compares each of the subfibers in the main level, copying the deduplicated subfibers into the buffer. \\
    %
    \textbf{SparseList}:
    The simplest sparse format, used to construct popular formats like CSR, CSC, DCSR, DCSC, and CSF.
    It stores two vectors, $idx$ and $ptr$, such that $idx[ptr[p] + q]$ is the index of the $q^{th}$ nonzero in the subfiber at position $p$. \\
    \textbf{SparsePinpoint}:
    Similar to SparseList, but only one nonzero in each subfiber, eliminating the need for the $ptr$ field.
    It stores a vector $idx$, such that $idx[p]$ is the nonzero index in the subfiber at position $p$. \\
    \textbf{SparseRunList}:
    Similar to RunList level, but because runs are sparse, we must also store the start of each run.
    It stores three vectors $left$, $right$, and $ptr$, such that the $q^{th}$ run in the $p^{th}$ subfiber begins and ends at $left[ptr[p] + q]$ and $right[ptr[p] + q]$, respectively.
    Like RunList, it also stores a duplicate sublevel, $buf$, for deduplication. \\
    \textbf{SparseInterval}:
    Similar to SparseRunList, but only stores one run per subfiber, eliminating the need for the $ptr$ field.
    This level does not deduplicate as it cannot store intermediate results with more than one run.
    It stores two vectors, such that the run in subfiber $p$ begins and ends at $left[p]$ and $right[p]$ respectively. \\
    \textbf{SparseBlockList}: Used to represent blocked data. It stores three vectors, $idx$, $ptr$, and $ofs$, such that $ofs[ptr[p] + q]:ofs[ptr[p] + q + 1] - 1$ are the subpositions of block $q$ ending at index $idx[ptr[p] + q]$ in the subfiber at position $p$. \\
    \textbf{SparseBand}: Similar to SparseBlockList, but stores only one block per subfiber, eliminating the need for the $ptr$ field. It stores two vectors $idx$ and $ofs$, such that $ofs[p]:ofs[p + 1] - 1$ are the subpositions of the block ending at $idx[p]$ in subfiber $p$. \\

    \hline
    \textbf{Nonsequentially Constructed Levels} \\
    \hline
    \textbf{SparseHash}:
    The sparse hash format uses a hash table to store the locations of nonzeros, and sorts the unique indices for iteration during the freeze phase.
    This allows for efficient random access, but not incremental construction, as the freeze phase runs in time proportional to the number of nonzeros in the entire level.
    It stores two vectors, $idx$ and $ptr$, such that $idx[ptr[p] + q]$ is the index of the $q^{th}$ nonzero in the subfiber at position $p$. 
    Also stores a hash table $tbl$ for construction and random access in the level. \\
    \textbf{SparseBytemap}
    The SparseBytemap format uses a bytemap to store which locations have been written to.
    Unlike the SparseHash format, the bytemap assembles the entire space of possible subfibers.
    This accelerates random access in the format, but requires a high memory overhead.
    Because we don't want to reallocate all of the memory in each iteration, the declaration of this format instead re-assembles only the dirty locations in the tensor.
    This format is analogous to the default workspace format used by TACO.
    It stores two vectors, $idx$ and $ptr$, such that $idx[ptr[p] + q]$ is the index of the $q^{th}$ nonzero in the subfiber at position $p$.
    These vectors are used to collect dirty locations.
    It also stores $tbl$, a dense array of Booleans such that $tbl[shape * p + i]$ is true when there is a nonzero at index $i$ in the subfiber at position $p$. \\
    \hline
    \textbf{Leaf Levels} \\
    \hline
    \textbf{Element}:
    The element level uses an array $val$ to store a value for each position $p$. The zero (fill) value is configurable.\\
    \textbf{Pattern}:
    The pattern level statically represents a leaf level with a fill value of $false$ and whose stored values are all $true$. \\
    \hline
    \textbf{Scalars} \\
    \hline
    \textbf{Scalar}:
    A dense scalar that, unlike a variable, supports reduction.  \\
    \textbf{SparseScalar}:
    A scalar with a dirty bit which specializes on the fill value when it occurs.\\
    \textbf{ShortCircuitScalar}:
    A scalar which triggers early breaks in reductions whenever an annihilator is encountered.\\
    \hline
    \end{tabular}
    \vspace{8pt}
    \caption{The main level formats supported by Finch. Note that all non-leaf
    levels store a the dimension of the subfibers and a child level. Since we
    must be able to handle the case where a sublevel is not stored because a
    parent level is sparse, all of Finch's sparse formats use a dirty bit during
    writing to determine whether the sublevel has been modified from it's
    default fill value and thus, whether it needs to be stored.}
    \label{table:formats}
    \vspace{-24pt}
\end{table}

\subsubsection{Scalars}
Because leaf levels are geared towards representing multiple leaves, we also introduce a much simpler Scalar format to represent 0-dimensional tensors.
Scalars don't have as much structure because they only concern one value.
However, we allow the programmer to declare that a scalar might be sparse, or that it might be used in a reduction which can be exited early.
Through these structures, scalars can also affect other tensors in crucial ways.

When a scalar is sparse, this means that it might be equal to the fill value and the user has requested for the compiler to simplify subsequent computations accordingly.
Constant propagation through tensors is known to be a complex compiler pass \cite{mcmichen_representing_2024}.
We provide sparse scalars as an alternative, which allow for similar semantics by specializing reads for the possible fill value.

\mintinline{julia}|ShortCircuitScalars| trigger stepper looplets to
re-specialize the loop whenever a reduction into the scalar hits an
annihilator, removing the reduction from the specialized case since it has hit the annihilator value and can no longer change. Short-circuiting conditions are lowered by inserting a branch into the loop body which checks for the short circuit condition. The branch contains the (hopefully simplified) remainder of the loop, followed by a \mintinline{julia}|break|.
Re-specialization of other looplets is not required because the stepper
is the only one which repeats a non-constant number of times.

\begin{wrapfigure}{r}{.23\textwidth}
    \vspace{-4pt}
    \begin{minted}{julia}
        p = ShortCircuitScalar{0}()
        @finch begin
            p .= 0
            for j=_
                p[] *= A[j]
    \end{minted}
    \vspace{-12pt}
    \footnotesize
    \caption{Using a ShortCircuitScalar to find the product of values in A.}
    \label{fig:shortcircuit}
    \vspace{-12pt}
\end{wrapfigure}
We also provide ShortCircuitScalars, which signal that the compiler should check for an opportunity to early break out of a reduction loop when the loop hits an annihilator value. For example, Figure \ref{fig:shortcircuit} computes the product of vector elements, exiting the loop when one of them is zero.

We represent early break as a structural property rather than a program node
because it allows us to represent the tail of a loop where one scalar has hit an
annihilator but another scalar hasn't.
The effects of a \mintinline{julia}{break} statement would affect the value of all other statements in the loop, and violate some of the  dataflow assumptions lifecycle constraints would otherwise permit.
Representing breaks as structural properties allows us to more elegantly compose break statements with other structures in the language.
To our knowledge, sparse and ShortCircuitScalars are novel contributions of this
work; other systems don't include them, limiting the impact of sparsity.

\subsubsection{Leaf Levels}

The leaf level stores the actual entries of the tensor. In most cases, it is
sufficient to store each entry at a separate position in a vector. This is accomplished by the \textbf{ElementLevel}. However, when all of the
values are the same, an additional optimization can be made by storing the
identical value only once. In this work, we introduce the concept of a \textbf{PatternLevel}
 to handle this binary case. The PatternLevel has a fill value of $false$, and returning $true$ for all ``stored'' values. The PatternLevel allows us to easily represent unweighted graphs or other Boolean matrices.
\section{The Finch Language}

\begin{figure}[H]
    \centering
    \begin{minipage}[t]{0.5\textwidth}
        \centering
         \begin{minted}{julia}
     EXPR := LITERAL|VALUE|INDEX|VARIABLE|EXTENT|CALL|ACCESSS
     STMT := ASSIGN|LOOP|DEFINE|SIEVE|BLOCK|DECLARE|FREEZE|THAW

      DECLARE := TENSOR .= EXPR(EXPR...)   #V is the set of all values
       FREEZE := @freeze(TENSOR)           #S is the set of all Symbols
         THAW := @thaw(TENSOR)             #T is the set of all types                   
       TENSOR := TENSORNAME :: WRAPPER(TENSOR, EXPR...)                    
       ASSIGN := ACCESS <<EXPR>>= EXPR         TENSORNAME := S
         LOOP := for INDEX = EXPR              LITERAL := V                           
                   STMT                          VALUE := S::T                        
                 end                              WRAPPER := S
       DEFINE := let VARIABLE = EXPR             INDEX := S                            
                   STMT                       VARIABLE := S                            
                 end                            EXTENT := EXPR : EXPR                  
        SIEVE := if EXPR                          CALL := EXPR(EXPR...)                
                   STMT                         ACCESS := TENSOR[EXPR...]              
                 end                              MODE := @mode(TENSOR)                
        BLOCK := begin                            
                   STMT...                 
                 end                        
    \end{minted}
    \vspace{-6pt}
    \end{minipage}%
    \begin{minipage}[t]{0.5\textwidth}
        \centering
        
    $\scriptscriptstyle
\den[F]{\finchloop(i, \finchextent(a, b), \finchblock)} =  \cup^{F}_{iv\in \mathbb{Z}\cap [\den[F]{a},\den[F]{b}]}\den[{F, i\mapsto iv}]{\finchblock}
$

$\scriptscriptstyle
\den[F]{\finchaccess(tensor, exprs...)} = \den[F]{tensor}(\den[F]{exprs}...)
$
$\scriptscriptstyle
\den[F]{tensorname} = F(tensorname)
$

$\scriptscriptstyle
\den[F]{wrapper(tensor, exprs)} = (\den[W]{wrapper}(exprs))(\den[F]{tensor})
$

$\scriptscriptstyle
\den[F]{\finchblock(stmt_{1}, stmts...)} = \den[F]{stmt_{1}}\cup^{F}\den[F]{\finchblock(stmts...)}
$

$\scriptscriptstyle
\den[F]{\finchblock()} = \{\}
$

$\scriptscriptstyle
 \den[F]{\finchsieve(expr, stmt)} = \begin{cases}\scriptscriptstyle\den[F]{stmt}  & \scriptscriptstyle\den[F]{expr} \\
 \scriptscriptstyle\{\} & \\
 \end{cases}
$

$\scriptscriptstyle
\den[F]{\finchdeclare(var, expr, stmt)} = \den[{F, var\mapsto \den[F]{expr}}]{stmt}
$

$\scriptscriptstyle
\den[F]{\finchassign(\finchaccess(tensor, idxExpr), op, expr)} = F\cup^{F} \{tensor\mapsto \den[F]{tensor}\cup \{ \den[F]{idxExprs}... \mapsto \den[F]{op}(\den[F]{tensor}(\den[F]{idxExprs}...), \den[F]{expr} \} \}
$
\end{minipage}
\begin{subfigure}[t]{0.49\textwidth}
\caption{The syntax of the Finch language. Compare this grammar to the Concrete
Index Notation of TACO \cite[Figure~3]{kjolstad_tensor_2019}, noting the
addition of multiple left-hand sides through code blocks, access with arbitrary expressions, and explicit declaration, as well as freeze and thaw.}\label{fig:syntax}
\end{subfigure}\hfill%
\begin{subfigure}[t]{0.49\textwidth}
  \caption{Semantics of Finch: The semantic domains are $F$, an assignment of tensor names to functions ( $\mathbb{Z}^{N}\to V$) as well as $W$, an assignment of wrappers to functions with type $V^{M}\mapsto ((\mathbb{Z}^{N}\mapsto V)\mapsto \mathbb{Z}^{N^{\prime}}\mapsto V)$, representing transformations of tensors. The dimension $a:b$ of an index $i$ or a declaration is computed via the rules laid out in Section \ref{sec:dim}. 
  }
   \label{fig:semantics_new}
    \end{subfigure}
    \vspace{-10pt}
    \caption{Syntax and Semantics for Finch}
    \vspace{-1em}
\end{figure}

The syntax of Finch is displayed in Figure \ref{fig:syntax}, and a denotational semantics is displayed in Figure \ref{fig:semantics_new}. The Finch language
mirrors most imperative languages such as C with for-loops and control flow. Notable
statements that have been added to the language include \mintinline{julia}{for},
\mintinline{julia}{let}, blocks of code with \mintinline{julia}{if}, wrappers of tensors, and 
the lifecycle functions that let us declare, freeze, and thaw tensors.

The denotational semantics of our language concern large dense iteration spaces, but the implementation eliminates many of these unnecessary iterations through aggressive optimizations, carefully using life cycles, dimensions, sparsity via looplets, and control flow as a form of sparsity. Section~\ref{sec:compiler} details the specifics of how we compile our syntax to efficient code over structured data.

Our expressions support a wide variety of scalar operations on literals, indices, extents, wrappers, and calls to externally defined functions.
Wrappers are static higher order functions on tensors that serve to implement complex indexing logic such as $i+j$ or $i <= j$; an initial pass in the compiler converts indexing logic to a wrapper function when possible.
Since the wrappers are rather simpler higher order functions, we can implement them as transformations on looplets or other properties of the tensor format, which will mean careful implementations of wrapper will allow a more efficient, lazy implementation of complex tensor accesses as opposed to naive look ups or naively rebuilding a tensor at each iteration. For examples of wrappers, see Table \ref{tab:wrappers}.
Finally, as detailed in the previous section, tensors are defined externally via an interface that supports the $\mathit{declare}$, $\mathit{freeze}$, $\mathit{thaw}$, and $\mathit{unfurl}$ functions.
The first three are supported directly in the syntax whereas the fourth will be introduced through evaluation of loops and accesses,
in the next section.
We do not intend the user to insert freeze or thaw manually, but we include them in the language because it allows us to handle tensor lifecycles with a separate, simpler, compiler pass, rather than all at once. Tensors can only change between read and write mode in the scope in which they were defined, so we can insert freeze/thaw automatically by checking whether the tensor is being read or written to in each child scope. We error if a tensor appears on both the left hand and right hand sides within the same child scope. This algorithm is described later in Section\ref{sec:compiler}.

Our syntax is highly permissive: by allowing blocks of code with multiple statements, we implicitly support many features gained through complicated scheduling commands in other frameworks, such as multiple outputs, masking to avoid work, temporary tensors, and arbitrary loop fusion and nesting.
These features are seen most prominently in our implementation of Gustavson's algorithm for sparse-sparse matrix multiply, which simply writes to a temporary tensor in an inner loop and then reuses it; or in our breadth-first search, which uses an \mintinline{julia}{if} statement to avoid operating on vertices outside the frontier. 
The only restriction we impose on our syntax is that it must respect tensor life cycles.
As discussed in Section~\ref{sec:formats}, our language does not include a \mintinline{julia}{break} statement. We instead represent breaks as a structural property, so that they can more elegantly compose with other structures in the language.

\subsection{Dimensionalization Rules}\label{sec:dim}

    Looplets typically require the dimension of the loop extent to match the dimensions of the tensor. 
    However, it is cumbersome to write the dimensions
    in loop programs, and most tensor compilers have a means of specifying the dimensions automatically.
    In many pure Einsum languages like TACO, determining dimensions
    is not needed because any tensor dimensions that share an index are assumed to be the same~\cite{kjolstad_tensor_2017}.
    Other languages, such as Halide, perform bounds inference where
    known bounds are symbolically propagated to fill in unknown bounds, often from output/input sizes to intermediates via some approximation such as interval analysis or polyhedral methods~\cite{ragan-kelley_halide_2013,grosser_pollyperforming_2012}.
    We refer to the process of discovering suitable
    dimensions as \textbf{dimensionalization}.

Loop bounds in Finch are computed automatically via a few simple rules. There are currently two kinds of dimensions in Finch: \texttt{\_} represents a dimensionless quantity, and \texttt{a:b} represents an integer dimension.
Dimensions can be joined with the \texttt{meet} operation, which returns the dimension that is not \texttt{\_} or else asserts that the two extents match.
\begin{itemize}
  \item The dimension of an index is defined as the \texttt{meet} of the loop bound and the tensor dimension corresponding to any right-hand-side accesses with that index.
  \item The $n^{th}$ dimension of a tensor declaration is defined as the \texttt{meet} of all index dimensions in the $n^{th}$ mode of left-hand-side accesses to that tensor, from its declaration to its first read.
  \item The dimension of \texttt{i + c}, where \texttt{c} is a constant, is the dimension of \texttt{i} shifted by \texttt{c}.
  \item The dimension of \verb|~(x)|, or any other unrecognized function, is \texttt{\_}.
  \item  More rules may be added as Finch is extended to recognize more indexing syntax.
\end{itemize}

\section{The Finch Compiler}\label{sec:compiler}
\begin{figure}[H]
  \vspace{-12pt}
  \includegraphics[width=\textwidth]{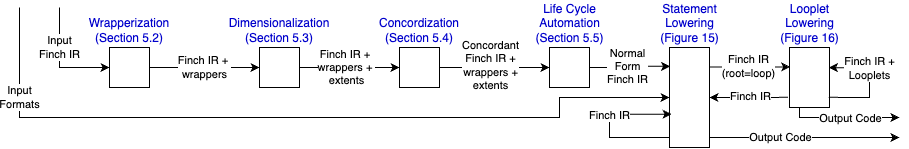}
  \vspace{-24pt}
  \caption{Stages of the Finch Compiler.}\label{fig:compiler_flow}
  \vspace{-12pt}
\end{figure}

The Finch compiler takes a Finch program together with a program state defining the formats of tensors, and produces efficient structure aware code.
The compiler operates in several stages. The first stages normalize the program to make it easier to process. The final stages lower a normalized program recursively, one loop at a time. For each loop, all tensors that are indexed by the loop index are transformed into looplets based on their structure, and these looplets are lowered to executable code. The overall flow is summarized in Figure~\ref{fig:compiler_flow}.

\subsection{Finch Normal Form}
Our core recursive lowering compiler described in Figure~\ref{fig:semantics_core} and
Figure~\ref{fig:semantics_looplets} is designed to handle a particular class of programs we refer to as \textbf{Finch Normal Form}. This section defines the properties of Finch Normal Form. Later sections will describe how to normalize all Finch programs which are well-defined under the semantics in Figure \ref{fig:semantics_new}.

The properties of such a program are as follows:
\begin{itemize}
    \item \textbf{Access with Indices:} Though Finch allows general expressions
    (including affine expressions and general function calls) in an access (i.e.
    \mintinline{julia}{A[i + j]} or \mintinline{julia}{A[I[i]]}), the normal
    form restricts to allow only indices in accesses (i.e.
    \mintinline{julia}{A[i]}), rather than more general expressions.
    \item \textbf{Evaluable Dimensions:} Loop dimensions and declaration dimensions must
    be evaluable at the time we compile them, so we restrict the normal form to
    programs whose loop dimensions and declaration dimensions are extents with
    limits defined in the scope of the corresponding loop or declaration
    statement.
    \item \textbf{Concordant:} Finch is column-major by default to match Fortran\cite{backus_fortran_1957} and Julia\cite{bezanson_julia_2012}. A Finch program is \textbf{concordant}
    when the order of indices in each access match the order in which loops
    are nested around it.  For example,
    \mintinline{julia}{for j = _; for i = _; s[] += A[i, j] end end}
    is concordant but
    \mintinline{julia}{for i = _; for j = _; s[] += A[i, j] end end} is not.
    \item \textbf{Lifecycle Constraints:} Tensors in read mode may appear on the right
    hand side only. Tensors in update mode may appear on the left hand side
    only. To make it easier to analyze lifecycle constraints statically, we
    restrict tensors to only change modes in the same scopes in which they were
    defined.
\end{itemize}

The subsequent sections will explain how programs that violate each of these
constraints can be rewritten to programs that satisfy them and thus how we can
support such a wide variety of programs. For example, we can write nonconcordant
programs like  \mintinline{julia}{for i = _; for j = _; s[] += A[i, j] end end}
by inserting a loop to randomly access \mintinline{julia}{A}.
      
\subsection{Wrapperization}
    Many fancy operations on indices can be resolved by introducing equivalent
    \textbf{wrapper tensors} which modify the behavior of the tensors they wrap,
    or by introducing \textbf{mask tensors} which replace index expressions like
    \mintinline{julia}{i <= j} with their equivalent masks (in this case, a
    triangular mask tensor).  Wrappers and masks are summarized in Table \ref{tab:wrappers}.

    \begin{figure}[H]
    \vspace{-12pt}
    \begin{minipage}{0.25\linewidth}
       \begin{minted}{julia}
          for i=_, j=_
            if i <= j
              s[] += A[i - 1, j]
        \end{minted}
    \end{minipage}%
    $\rightarrow$%
    \begin{minipage}{0.35\linewidth}
        \begin{minted}{julia}
          for i=_, j=_
            if UpTriMask()[i, j]
              s[] += OffsetTensor(A, (-1, 0))[i, j]
        \end{minted}
    \end{minipage}%
    $\rightarrow$%
    \begin{minipage}{0.3\linewidth}
        \begin{minted}{julia}
          for i = 1:n
            for j = 1:i
              s[] += A.val[(i - 1) + j * n]
        \end{minted} 
    \end{minipage}
    \vspace{-10pt}
    \caption{Wrapperization. While \mintinline{julia}{i <= j} is only an expression, \mintinline{julia}{UpTriMask()[i, j]} can use looplets to end each loop over \mintinline{julia}{j} at \mintinline{julia}{i}.}\label{fig:wrapperization}
        \vspace{-20pt}
\end{figure}

    \begin{table}[H]
      \footnotesize
      \centering
      \begin{tabular}{|p{13.4cm}|}
      \hline
      \textbf{OffsetTensor} shifts tensors such that \mintinline[fontsize=\footnotesize]{julia}{offset(tns, delta...)[i...] == tns[i + delta...]}. The shifting is achieved by modifying the ranges returned by the looplets in the wrapped tensor.
      
      \begin{minipage}[t]{\linewidth}
        \begin{minted}[fontsize=\footnotesize]{julia}
        A[i..., j + c, k...] -> OffsetTensor(A, (0..., c, 0...))[i..., j, k...]
        \end{minted}
      \end{minipage}
      \\ \hline
      
      \textbf{ToeplitzTensor} adds a dimension that shifts another dimension of the original tensor. The added dimensions are produced during a call to \mintinline[fontsize=\footnotesize]{julia}{Unfurl}, when a lookup looplet is emitted for the first dimension.
      
      \begin{minipage}[t]{\linewidth}
      \begin{minted}[fontsize=\footnotesize]{julia}
      A[i_1, ..., i_n, j + k, l...] -> ToeplitzTensor(A, n)[i_1, ..., i_n, j, k, l...]
      \end{minted}
      \end{minipage}
      \\ \hline
      
      \textbf{PermissiveTensor} allows for out-of-bounds access or padding. PermissiveTensor returns a dimensionless value for any permissive indices. PermissiveTensor returns \mintinline[fontsize=\footnotesize]{julia}{missing} as the out-of-bounds value, where the \mintinline[fontsize=\footnotesize]{julia}{coalesce} function can be used to return the first nonmissing value.
      
      \begin{minipage}[t]{\linewidth}
      \begin{minted}[fontsize=\footnotesize]{julia}
      A[i..., ~j, k...] -> PermissiveTensor(A, (false..., true, false...))[i..., j, k...]
      \end{minted}
      \end{minipage}
      \\ \hline
      
      \textbf{ProtocolizedTensor} allows for advanced iteration protocols. The ProtocolizedTensor selects between several different implementations of \mintinline[fontsize=\footnotesize]{julia}{unfurl} that a level may support.
      
      \begin{minted}[fontsize=\footnotesize]{julia}
      A[i..., p(j), k...] -> ProtocolizedTensor(A, (nothing..., p, nothing...))[i..., j, k...]
      \end{minted}
      
      Finch recognizes several protocols:

      \begin{minipage}[t]{\linewidth}
      \begin{itemize}
          \item The \mintinline[fontsize=\footnotesize]{julia}{follow} protocol indicates the structure should be ignored and random access used for each element.
          \item The \mintinline[fontsize=\footnotesize]{julia}{walk} protocol declares that the structure of the iterator should be used in the computation.
          \item The \mintinline[fontsize=\footnotesize]{julia}{gallop} protocol declares that the structure of a tensor should lead an iteration and the compiler should specialize to that structure with a higher priority than others. A galloping protocol over two SparseList levels produces a mutual-binary-search merge algorithm popularized in the case of worst-case-optimal join queries~\cite{barbay_experimental_2010, veldhuizen_triejoin_2014, ngo_worst-case_2018}.
      \end{itemize}
      \end{minipage}
      \\ \hline
      
      \textbf{SwizzleTensor} is a lazily transposed tensor that changes the interpretation of the order of modes in the tensor. Unlike other wrappers, a SwizzleTensor is compiled during the wrapperization pass rather than introduced by it.
      
      \begin{minipage}[t]{\linewidth}
      \begin{minted}[fontsize=\footnotesize]{julia}
      swizzle(A, perm)[idx...] -> A[idx[perm]...]
      \end{minted}
      \end{minipage}
      \\ \hline
      
    \begin{minipage}[t]{0.58\linewidth}
      \textbf{UpTriMask} is a mask tensor that represents Boolean triangular matrices. \mintinline[fontsize=\footnotesize]{julia}{UpTriMask()[i, j]} is true when \mintinline[fontsize=\footnotesize]{julia}{i <= j}.
      
      It is introduced via several rewrite rules, such as:
      
      \begin{minted}[fontsize=\footnotesize]{julia}
      i < j  ->  UpTriMask()[i, j - 1]
      i > j  ->  !UpTriMask()[i, j - 1]
      i <= j ->  UpTriMask()[i, j]
      i >= j ->  !UpTriMask()[i, j]
      \end{minted}
      
      When \mintinline[fontsize=\footnotesize]{julia}{i} would be bound at a higher loop depth than \mintinline[fontsize=\footnotesize]{julia}{j}, care is taken to reverse the index order and emit the mask in column-major order.
    \end{minipage}\hfill%
    \begin{minipage}[t]{0.4\linewidth} 
      \begin{minted}[fontsize=\footnotesize]{julia}
      unfurl(UpTriMask(), ext, reader) =
        Lookup(body(j) = UpTriMaskCol(j))
      unfurl(UpTriMaskCol(j), ext, reader) =
        Sequence(
          Phase(
            stop = j,
            body = Run(true)),
          Phase(
            body = Run(false)))
      \end{minted}
    \end{minipage}
      \\ \hline
    \begin{minipage}[t]{0.58\linewidth}
      \textbf{DiagMask} is a mask tensor that represents Boolean diagonal matrices. \mintinline[fontsize=\footnotesize]{julia}{DiagMask()[i, j]} is true when \mintinline[fontsize=\footnotesize]{julia}{i == j}.
      
      It is introduced via several rewrite rules, such as:
      
      \begin{minted}[fontsize=\footnotesize]{julia}
      i == j -> DiagMask()[i, j]
      i != j -> !DiagMask()[i, j]
      \end{minted}
      
      When \mintinline[fontsize=\footnotesize]{julia}{i} would be bound at a higher loop depth than \mintinline[fontsize=\footnotesize]{julia}{j}, care is taken to reverse the loop order and emit the mask in column-major order.
      
      \begin{minted}[fontsize=\footnotesize]{julia}
      unfurl(DiagMask(), ext, reader) =
        Lookup(body(j) = DiagMaskCol(j))
      \end{minted}
    \end{minipage}\hfill%
    \begin{minipage}[t]{0.4\linewidth}
      \begin{minted}[fontsize=\footnotesize]{julia}
      unfurl(DiagMaskCol(j), ext, reader) =
        Sequence(
          Phase(
            stop = j - 1,
            body = Run(false)),
          Phase(
            stop = j,
            body = Run(true)),
          Phase(
            body = Run(false)))
      \end{minted}
    \end{minipage}
      \\ \hline
      \end{tabular}
      \caption{Wrapper tensors}
      \label{tab:wrappers}
      \vspace{-24pt}
      \end{table}

    More formally, a \textbf{wrapper tensor} is any tensor that wraps a tensor
    variable in an access, and can overload the behavior of $\mathit{unfurl}$, $\mathit{unwrap}$,
    and $size$, as well as modify the ranges declared by any looplets the
    wrapper contains. For example, the \mintinline{julia}{offset} wrapper tensor shifts looplets to shift the tensor with respect to the loop index. Wrappers are implemented either through a program rewrite during the wrapperization procedure, or by overloading format and looplet APIs \ref{fig:semantics_looplets} with some minor modifications. For example, OffsetTensor shifts the declared ranges of contained looplets.
    
    A \textbf{mask tensor} is simply a Boolean Finch tensor
    with implicit structure that uses a predefined looplet nest, rather
    than the level abstraction. For example, the \mintinline{julia}{UpTriMask} tensor
    uses looplets to represent the structure of a Boolean upper triangular matrix. Mask tensors are implemented using static looplets
    that are constructed during the unfurl step. Mask tensors
    allow us to lift computations with masks to the level of the loop, without modifying the loop directly.

\subsection{Dimensionalization}
    \begin{wrapfigure}{r}{0.26\textwidth}
        \vspace{-30pt}
        \begin{minted}{julia}
          #A is 3 x 4
          #B is 4 x 5
          C .= 0
          for i = 1:3
            for j = _
              for k = _
                C[i, j] += A[i, k] * B[k, j]
        \end{minted}
        $\downarrow$
        \begin{minted}{julia}
          C .= 0
          for i = 1:3
            for j = 1:5
              for k = 1:4
                C[i, j] += A[i, k] * B[k, j]
        \end{minted}
        \vspace{-12pt}
        \caption{Dimensionalization}\label{fig:dimensionalization}
        \vspace{-12pt}
    \end{wrapfigure}

    In Section \ref{sec:dim}, we described a simple set of rules to calculate
    dimensions.
    In Finch, these rules are implemented through a straightforward dimensionalization algorithm which operates on loops and declaration statements (output tensors).
    Finch determines the dimension of a loop index i from all of the tensors using i in an access, as well as the bounds in the loop itself, and operates similarly for declarations.

%
%

    Finch can compute these dimensions in a single pass over the program. When the compiler reaches a {\finchread} access, the tensor must be dimensionalized and those dimensions are used to compute the loop index dimension. When the compiler reaches an {\finchupdate} access, it saves the indices for later. Because the {\finchfreeze} statement must occur outside of any loops which access a tensor, we can assume that those stored indices are dimensionalized and use them to compute the dimensions of the corresponding tensor declaration.

    For example, in Figure~\ref{fig:dimensionalization}, the second dimension of A must match the first dimension of B. Also, the first dimension of A must match the i loop dimension, 1:3. Finch will resize declared tensors to match indices used in writes, so C is resized to (1:3, 1:5). If no dimensions are specified elsewhere, then Finch will use the dimension of the declared tensor.
    Dimensionalization occurs after wrapper tensors are de-sugared, so wrapper tensors can be used to pass dimensions through more complex index expressions. The user can exempt an index from dimensionalization by wrapping it in \mintinline{julia}{~} to produce a ``PermissiveTensor'' (e.g. \mintinline{julia}{A[~i]}) which has dimension \texttt{\_}.

    \begin{wrapfigure}{r}{0.32\linewidth}
      \vspace{-30pt}
      \centering 
      \begin{minipage}{0.16\textwidth}
      \begin{minted}{julia}
        for i = _
          for j = _
              s[] += A[i, j]
      \end{minted}
      $\downarrow$
      \begin{minted}{julia}
        for i = _
          for j = _
            for k = i:i
              s[] += A[k, j]
      \end{minted}
      \end{minipage}\hfill%
      \begin{minipage}{0.14\textwidth}
          \begin{minted}{julia}
            for i = _
              A[I[i]] += 1
          \end{minted}
          $\downarrow$
          \begin{minted}{julia}
            for i = _
              for j = I[i]:I[i]
                A[j] += 1
          \end{minted}
      \end{minipage}\hfill
      \vspace{-12pt}
      \caption{Examples of concordization, transforming accesses to normal column major.}\label{fig:concordization}
      \vspace{-24pt}
  \end{wrapfigure} 
\subsection{Concordization}
After wrapperization, Finch runs a pass over the code to make the program concordant. Finch makes expressions concordant by inserting single-iteration loops. Examples are given in Figure~\ref{fig:concordization}. \kyle{After concordization, the structure of A affects the inserted loop, and no longer affects the original one.} The algorithm works by applying the following rewrite rule in any scope where the expression \mintinline{julia}{x} is bound:
\begin{minipage}{\linewidth}
\centering
\begin{minipage}{0.4\linewidth}
  \begin{minted}{julia}
    for i = _
        ...
        ... A[x, i] ...
        ...
  \end{minted}
\end{minipage}\hspace{-30pt}$\rightarrow$
\begin{minipage}{0.4\linewidth}
  \begin{minted}{julia}
    for i = _
      ...
      for j = x:x
        ... A[j, i] ...
      ...
  \end{minted}
\end{minipage}
\vspace{-5pt}
\end{minipage}
The variable \mintinline{julia}{x} might be bound by a for-loop or an earlier definition, or it may be a constant.

\subsection{Life Cycle Automation}
    \begin{wrapfigure}{R}{0.34\textwidth}
        \vspace{-30pt}
        \begin{minipage}{0.15\textwidth}
        \begin{minted}{julia}
            y .= 0
            for i = _
                y[i] = x[i] + 1
            for i = _
                x[i] += 1
                y[i] += 1
            for i = _
                x[i] += y[i]
        \end{minted}
        \end{minipage}
        $\rightarrow$
        \begin{minipage}{0.15\textwidth}
        \begin{minted}{julia}
        y .= 0
        for i = _
            y[i] = x[i] + 1
        @thaw(x)
        for i = _
            x[i] += 1
            y[i] += 1
        @freeze(y)
        for i = _
            x[i] += y[i]
        @freeze(x)
        \end{minted}
        \end{minipage}

        \vspace{-12pt}
        \caption{Life cycle automation.} \label{fig:lifecycles}
        \vspace{-18pt}
    \end{wrapfigure}
The last normalization pass inserts the \mintinline{julia}{@freeze} or
\\\mintinline{julia}{@thaw} macros automatically.
Tensors are only allowed to change mode within the scope in which they were declared.
If they have not been inserted already, this pass automatically inserts these statements in the program, easing the programmer's burden and bridging between structured and dense languages.

The pass walks the program and tracks the current mode of each tensor, depending on whether the tensor is read or updated in each statement within the tensor's declared scope block.

\begin{figure}[b]
    \centering
    \small
    \vspace{-12pt}

    \begin{prooftree}\scriptsize
    \hypo{\langle val, (e, t, d) \rangle
    \rightarrow val'}
    \hypo{var\not\in d}
    \infer2[$\mathit{Define}$]{\splitfrac{\langle\finchdefine(var, val, body), (e, t, d)\rangle}{\rightarrow \langle body, (e[var \mapsto val'], t, \{\}) \rangle}}
    \end{prooftree}
    \hfill
    \begin{prooftree}\scriptsize
      \hypo{}
      \infer1[$\mathit{Literal}$]{\langle\finchliteral(val), (e, t, d)\rangle \rightarrow val}
    \end{prooftree}
    \hfill
    \begin{prooftree}\scriptsize
        \infer0[$\mathit{Variable}$]{\splitfrac{\langle\finchvar(name), (e, t, d)\rangle}{\rightarrow e(\finchvar(name))}}
    \end{prooftree}
    \vspace{8pt}

    \begin{prooftree}\scriptsize
        \hypo{\langle args_i, (e, t) \rangle \Rightarrow vals_i}
        \hypo{\langle f, (e, t) \rangle \Rightarrow g}
        \infer2[$\mathit{Call}$]{\langle\finchcall(f, args...), (e, t)\rangle \rightarrow \llangle g(vals...), t \rrangle}
    \end{prooftree}
    \vspace{8pt}
    \hfill
    \begin{prooftree}\scriptsize
        \infer0[$\mathit{Index}$]{\splitfrac{\langle\finchindex(name), (e, t, d)\rangle}{\rightarrow e(\finchindex(name))}}
    \end{prooftree}
    \hfill
    \begin{prooftree}\scriptsize
        \hypo{\langle node, algebra \rangle \rightarrow node'}
        \infer1[$\mathit{Simplify}$]{\langle E[node], s\rangle \rightarrow \langle E[node'], s\rangle}
    \end{prooftree}
    \vspace{8pt}

    \begin{prooftree}\scriptsize
    \hypo{\langle body, s \rangle \rightarrow s'}
    \infer1[$\mathit{Block}$]{\langle\finchblock(body, tail...), s \rangle \rightarrow \langle \finchblock(tail...), s' \rangle }
    \end{prooftree}
    \hfill
    \begin{prooftree}\scriptsize
        \hypo{\langle cond, (e, t, d) \rangle \Rightarrow true}
        \infer1[$\mathit{SieveTrue}$]{\langle\finchsieve(cond, body), (e, t, d)\rangle \rightarrow \langle body, (e, t, \{\}) \rangle}
    \end{prooftree}%

    \vspace{8pt} 
    \begin{prooftree}\scriptsize
      \hypo{e(tns) \mapsto tns'}
      \hypo{e(\finchmode(tns)) \mapsto \finchread}
      \hypo{\llangle unwrap(tns', \finchread), t\rrangle \rightarrow tns''}
      \infer3[$\mathit{Access}$]{\langle E[\finchaccess(tns)], s\rangle \rightarrow \langle E[tns''], s\rangle} 
    \end{prooftree}
  \hfill
    \begin{prooftree}\scriptsize
        \hypo{\langle cond, s \rangle \Rightarrow false}
        \infer1[$\mathit{SieveFalse}$]{\langle\finchsieve(cond, body), s\rangle \rightarrow s}
    \end{prooftree}

    \vspace{8pt}

    \begin{prooftree}\scriptsize
      \hypo{e(tns) = tns'}
      \hypo{\langle op, (e, t, d)\rangle \rightarrow op'}
      \hypo{\langle rhs, (e, t, d)\rangle \rightarrow rhs'}
      \infer[no rule]3{
          \begin{prooftree}
      \hypo{e(\finchmode(tns)) = \finchupdate}
      \hypo{\llangle unwrap(tns', \finchupdate, op', rhs'), t\rrangle \rightarrow  t'}
      \infer[simple]2[$\mathit{Assign}$]{\langle E[\finchassign(\finchaccess(tns), op, rhs)], (e, t, d)\rangle \rightarrow (e, t', d)}
      \end{prooftree}}
  \end{prooftree}
    \hfill
    \begin{prooftree}\scriptsize
        \hypo{}
        \infer1[$\mathit{Value}$]{\langle\finchvalue(ex, type), (e, t, d)\rangle \Rightarrow \llangle ex, t \rrangle}
    \end{prooftree}

    \vspace{8pt}
    
    \begin{prooftree}  \scriptsize
    \hypo{s = (e, t, d)}
    \hypo{tns\not\in d}
    \hypo{e(tns) = tns'}
    \infer[no rule]3{
        \begin{prooftree}
    \hypo{\langle init, s \rangle \Rightarrow init'}
    \hypo{\forall i \langle init, dims_i \rangle \Rightarrow dims'_i}
    \hypo{\llangle declare(tns', init', dims'...), t \rrangle \rightarrow t'}
    \infer[simple]3[$\mathit{Declare}$]{\langle \finchdeclare(tns, init, dims), s\rangle \rightarrow (e [\finchmode(tns) \mapsto \finchupdate], t', d\cup\{ tns\})}
    \end{prooftree}}
    \end{prooftree}
    \vspace{8pt}
    
    \begin{prooftree}  \scriptsize
    \hypo{s = (e, t,d )}
    \hypo{e(\finchmode(tns)) = \finchupdate}
    \hypo{tns\in d}
    \hypo{e(tns) = tns'}
    \infer4[$\mathit{Freeze}$]{\langle\finchfreeze(tns), s\rangle \rightarrow (e [\finchmode(tns) \mapsto \finchread], \llangle freeze(tns'), t \rrangle, d)}
    \end{prooftree}
    \vspace{8pt}

    \begin{prooftree}  \scriptsize
    \hypo{s = (e, t, d)}
    \hypo{e(\finchmode(tns)) = \finchread}
    \hypo{tns\in d}
    \hypo{e(tns) = tns'}
    \infer4[$\mathit{Thaw}$]{\langle\finchthaw(tns), s\rangle \rightarrow (e [\finchmode(tns) \mapsto \finchupdate], \llangle thaw(tns'), t \rrangle, d)}
    \end{prooftree}

    \caption{Basic evaluation semantics, roughly defining most of these language
    constructs to function similarly to their classical definitions.
    The state, $s$, of the program is a tuple $(e, t, d)$ of a variable value
    environment, another state $t$ corresponding to the state in the
    host language, and finally the set of tensors defined within the current scope, $d$. 
    We evolve tensor state with $\langle \rangle$ and host state with $\llangle\rrangle$.
    Several looplets introduce variables into the host state, which may be read when evaluating the {\finchvalue} node.
    Lifecycle functions are designed to be implemented and executed in the host language, but these semantics enforce that each function may update state in the host language and flip the mode of the tensor between {\finchread} and {\finchupdate}.}
    \label{fig:semantics_core}
\end{figure}

\begin{figure}
    \centering
    \footnotesize

\begin{align*}\scriptsize
T :=& \texttt{EXPR} | \texttt{STMT} \\
E :=& [\cdot] |
\finchloop(T, T, E) |
\finchblock(E, T...) |
\finchblock(T, E, T...) |
\finchsieve(E, T) |
\finchassign(E, T, T) | 
\finchassign(T, T, E) |\\
&\finchdeclare(T, E, T) |
\finchdeclare(T, T, E) |
\finchcall(E, T...) |
\finchcall(T, E, T...) |
\finchaccess(T, E, T...) |
\finchaccess(T, T, E...)
\end{align*}

\begin{prooftree}\scriptsize
    \hypo{e(tns) \mapsto tns'}
    \hypo{e(\finchmode(tns)) \mapsto m}
    \hypo{\llangle unfurl(tns', ext, m), t\rrangle \Rightarrow tns''}
    \infer3[$\mathit{Unfurl}$]{\langle \finchloop(i, ext, E[\finchaccess(tns, j..., i)]), s\rangle \rightarrow \langle \finchloop(i, ext, E[\finchaccess(tns'', j..., i)]), s\rangle}
\end{prooftree}
\vspace{6pt}

\begin{prooftree}\scriptsize
    \infer0[$\mathit{Run}$]{\splitfrac{\langle \finchloop(i, ext, E[\finchaccess(\finchrun(body), j..., i)]), s\rangle}{\rightarrow \langle \finchloop(i, ext, E[\finchaccess(body, j...)]), s\rangle}}
\end{prooftree}
\hfill
\begin{prooftree}\scriptsize
    \hypo{e(i) = i'}
    \hypo{\llangle seek(i'), t \rrangle \rightarrow t'}
    \infer2[$\mathit{Lookup}$]{\splitfrac{\langle E[\finchaccess(\finchlookup(seek, body), j..., i)], (e, t, d)\rangle}{\rightarrow \langle E[\finchaccess(body, j...)], (e, t', d)\rangle}}
\end{prooftree}
\vspace{6pt}

\begin{prooftree}\scriptsize
    \infer0[$\mathit{AcceptRun}$]{\splitfrac{\langle \finchloop(i, \finchextent(a, b), E[\finchassign(\finchaccess(\finchrun(body), j..., i), op, rhs)]), s\rangle}{\rightarrow \langle \finchloop(i, \finchextent(a, b), E[\finchsieve(i == a, \finchassign(\finchaccess(body, j...), op, rhs))]), s\rangle}}
\end{prooftree}
\vspace{6pt}

\begin{prooftree}\scriptsize
    \hypo{\llangle cond, t \rrangle \Rightarrow true}
    \infer1[$\mathit{SwitchTrue}$]{\splitfrac{\langle E[\finchaccess(switch(cond, head, tail), i...)], s\rangle}{\rightarrow \langle E[\finchaccess(head, i...)], s\rangle}}
\end{prooftree}
\hfill
\begin{prooftree}\scriptsize
    \hypo{\llangle cond, t \rrangle \Rightarrow false}
    \infer1[$\mathit{SwitchFalse}$]{\splitfrac{\langle E[\finchaccess(switch(cond, head, tail), i...)], s\rangle}{\rightarrow \langle E[\finchaccess(tail, i...)], s\rangle}}
\end{prooftree}
\vspace{6pt}

\begin{prooftree}\scriptsize
    \infer0[$\mathit{Phase}$]{\splitfrac{\langle \finchloop(i, \finchextent(a, b), E[\finchaccess(\finchphase(\finchextent(c, d), body), j..., i)]), s\rangle}{\rightarrow \langle \text{\finchloop}(i, \finchextent(max(a, c), min(b, d)), E[\text{\finchaccess}(body, j..., i)]), s\rangle}}
\end{prooftree}
\vspace{6pt}

\begin{prooftree}\scriptsize
    \hypo{\llangle preamble, t \rrangle \rightarrow t'}
    \hypo{\langle E[body], (e, t', d) \rangle \rightarrow (e', t'', d)}
    \hypo{\llangle epilogue, t'' \rrangle \rightarrow t'''}
    \infer3[$\mathit{Thunk}$]{\langle E[thunk(preamble, body, epilogue)], (e, t, d)\rangle \rightarrow (e', t''', d)}
\end{prooftree}
\vspace{6pt}

\begin{prooftree}\scriptsize
    \hypo{\langle \finchloop(i, ext, E[\finchaccess(head, j..., i)]), s\rangle \rightarrow s'}
    \infer1[$\mathit{Sequence}$]{\splitfrac{\langle \finchloop(i, ext, E[\finchaccess(sequence(head, tail), j..., i)]), s\rangle}{\rightarrow \langle \finchloop(i, ext, E[\finchaccess(tail, j..., i)]), s'\rangle}}
\end{prooftree}
\hfill
\begin{prooftree}\scriptsize
    \hypo{\langle node, algebra \rangle \rightarrow node'}
    \infer1[$Simplify$]{\langle E[node], s\rangle \rightarrow \langle E[node'], s\rangle}
\end{prooftree}
\vspace{6pt}

\begin{prooftree}\scriptsize
    \hypo{\llangle seek(a), t \rrangle \rightarrow t'}
    \infer1[$\mathit{StepperSeek}$]{\splitfrac{\langle \finchloop(i, \finchextent(a, b), E[\finchaccess(stepper(seek, body, next), j..., i)]), (e, t, d)\rangle}{\rightarrow \langle \finchloop(i, \finchextent(a, b), E[\finchaccess(stepper(body, next), j..., i)]), (e, t', d)\rangle}}
\end{prooftree}
\vspace{6pt}

\begin{prooftree}\scriptsize
    \hypo{\langle \finchloop(i, ext, E[\finchaccess(body, j..., i)]), (e, t, d) \rangle \rightarrow (e', t', d)}
    \hypo{\llangle next, t' \rrangle \rightarrow t''}
    \infer2[$\mathit{StepperNext}$]{\splitfrac{\langle \finchloop(i, ext, E[\finchaccess(stepper(body, next), j..., i)]), s\rangle}{\rightarrow \langle \finchloop(i, ext, E[\finchaccess(stepper(body, next), j..., i)]), (e', t'', d)\rangle}}
\end{prooftree}

\vspace{8pt}
\begin{prooftree}\scriptsize
    \infer0[$\mathit{Loop}$]{\splitfrac{\langle\finchloop(i, \finchextent(a, b), body), s\rangle}{\rightarrow \langle \finchblock(\finchdefine(i, a, body), \finchsieve(a < b, \finchloop(i, \finchextent(a + 1, b), body))), s\rangle}}
\end{prooftree}

    \vspace{6pt}
    

    \caption{Looplet evaluation semantics. 
    %
%
%
%
    %
    The state $s$ of the program is a tuple $(e, t, d)$ of a variable value environment, host language state $t$, and the current tensor scope, $d$. 
Note that $E$ is an evaluation context that applies anywhere in
the syntax tree.
    The nonlocal evaluations of looplets are what allow looplets to
    hoist conditions and subranges out of loops.
    However, this also means we must specify
    the priority in which we apply looplet rules, which is as follows: $\mathit{Thunk} > \mathit{Phase} > \mathit{Switch} > \mathit{Simplify} > \mathit{Run} > \mathit{Spike} > \mathit{Sequence} > \mathit{StepperSeek} > \mathit{StepperNext} > \mathit{Lookup} > \mathit{AcceptRun} > \mathit{Unfurl} > \mathit{Loop} > \mathit{Access}$.
    Many looplets, most notably the thunk looplet, introduce variables into the
    host language environment. 
    While looplets may modify variables they introduce themselves (steppers often increment some state
    variables), we forbid child looplets from modifying state variables that
    they didn't introduce. 
    This allows us to treat the {\finchvalue} node as a
    constant.
    The $Simplify$ rule references $algebra$, which is our variable
    defining a set of straightforward simplification rules. 
    These rules include
    simple properties like $x * 0 \rightarrow 0$ to more complicated ones such
    as constant propagation. 
    We omit the full set of rules for brevity and refer to \cite[Figure 5]{ahrens_looplets_2023} for examples.}
    \label{fig:semantics_looplets}
    \vspace{-20pt}
\end{figure}

\subsection{Recursive Lowering}

Finally, after normalization, the program is lowered recursively, node by node.
This phase is presented as a staged execution of a small step operational semantics (SOS) for Finch Normal Norm programs. Figure~\ref{fig:semantics_core} evolves Finch control flow towards loops. Figure~\ref{fig:semantics_looplets} lowers loops with looplets.

Though are semantics are phrased as an interpreter, we stress that what goes into the compiler is a program and some formats, and what comes out is code. In Figure \ref{fig:semantics_new}, we offer a denotational semantics which described the format-agnostic mathematical behavior of Finch programs as if tensors were functions. In Figures~\ref{fig:semantics_core} and \ref{fig:semantics_looplets}, we offer a structural operational semantics which succinctly describes the format-specific behavior of a hypothetical Finch interpreter. Our semantics can formally answer questions such as "which expressions will be annihilated by zero?", or "how many steps would be required to traverse a certain combination of formats?".

Our evaluation rules in SOS are closely related to the lowering rules used to define a compiler. Rules for lowering the program would be similar to Figures 13 and 14, with a few key differences. First, any changes to variable values in the "target environment" would simply be lowered to variable assignments. Second, instead of evaluating expressions when we apply a rule, we lower the expressions to code, using variables to reference the results. Finally, anywhere runtime information is used to determine which rule to use, we instead emit a branch and lower both rules with a runtime check to decide between them.

For example, though there are two rules to lower $\finchsieve$ depending on
whether the condition is true ($\mathit{SieveTrue}$ and $\mathit{SieveFalse}$)
both branches are lowered with an if to decide between them.

This stage of the compiler carefully intermixes our control flow and tensors into looplets so the combination can be successfully symbolic simplified together.
The crux of this is that loops enter into looplets via the $\mathit{Unfurl}$ rule. Unfurl is defined in Section \ref{sec:formats}.
In this system, repeated structures and constants are slowly uncovered as accesses are lowered in various points in the program (e.g. $\mathit{Run}$  and $\mathit{Switch}$, respectively).
In this process, we are able to use rewrite rules in $\mathit{Simplify}$ to eliminate cases, unnecessary iterations, and so forth based on the information provided via looplets and via the control flow (loops, sieve, definitions).
Because the systems above this reliably transforms complex index accesses and control flow to control flow and wrappers via concordization and wrapperization, this stage of the compiler can use looplets to simplify the combination of tensor structures and control flow to eliminate unneeded work.
This also crucially relies on the tensor life cycles as otherwise arbitrary mixes of reads and writes would disrupt the simplifications of looplets, which is why we include life cycles in these semantics. 

Finch lowers loops from the outside to the inside, focusing on a single outer loop at a time. The lowering of a single loop rewrites the entire loop body, even when the body contains multiple inner loops. It is true that a complex loop body may require a fairly invasive rewrite, but the rewrite is broken into many manageable pieces. The unfurl operation applies to all tensor access expressions involving the outer loop index, simply substituting each tensor with a corresponding looplet nest expression. The Looplet lowering rules in Figure \ref{fig:semantics_looplets} specify more granular rewrites that affect the entire loop body and involve the interaction between multiple looplets in different accesses. Consider the following example of $\mathit{Unfurl}$ in action, where we unfurl all accesses on \mintinline{julia}|k|:

\begin{minipage}{0.5\linewidth}
\begin{minted}[]{julia}
  A = Tensor(Dense(SparseList(Element(0.0))), m)
  B = Tensor(SparseList(SparseList(Element(0.0))), m, l)
  C = Tensor(Dense(SparseList(Element(0.0))), n, l)
  D = Tensor(SparseList(Element(0.0)), l)
  @finch begin
    for k = _
        A .= 0
        for i = _
            A[i] = B[i, k] * 2
        for j = _
            C[j, k] = A[i]^2 + D[k]
\end{minted}
\end{minipage}\hfill%
\begin{minipage}{0.5\linewidth}
\begin{minted}[]{julia}
@finch begin
  for k = _
      A .= 0
      for i = _
          A[i] = Stepper(...)[i, k] * 2
      for j = _
          Lookup(...)[j, k] = A[i]^2 + Stepper(...)[k]
\end{minted}
\end{minipage}\hfill%

We chose this level-by-level design to avoid combinatorial explosions handling different formats across two or more levels. Each level format describes one dimension of a tensor at a time, and Finch only lowers one loop at a time. The $\mathit{Unfurl}$ function substitutes each level format with a looplet expression composed from a fixed set of looplets. Then, we need only consider the relationships between each looplet, and not each format.

\section{Example Lowering}
Though Finch programs look as if they are written for dense loops, Finch
specializes the code during lowering so that only the necessary elements of
structure need to be processed. In Figures \ref{fig:lowering_example1}-\ref{fig:lowering_example2}, we illustrate the lowering of a program that
sums the upper triangle of an \mintinline{julia}{m} by \mintinline{julia}{n} matrix, \mintinline{julia}{A}.

\begin{figure}[H]
  \footnotesize
  \begin{minipage}[t]{0.49\linewidth}
    \textbf{Input Program:}
    \begin{minted}[]{julia}
      A = Tensor(Dense(SparseList(Element(0.0))), m, n)
      s = Tensor(Element(0.0))
      @finch begin
        s .= 0.0
        for j = _
          for i = _
            if i <= j
              s[] += A[i, j]
    \end{minted}

    \textbf{Step 1: Normalization}
    To begin, the wrapperization pass replaces \mintinline[fontsize=\footnotesize]{julia}{i <= j} with \\\mintinline[fontsize=\footnotesize]{julia}{UpTriMask()[i, j]}. The dimensionalization pass assigns \mintinline[fontsize=\footnotesize]{julia}{i}, and \mintinline[fontsize=\footnotesize]{julia}{j} the dimensions \mintinline[fontsize=\footnotesize]{julia}{m} and \mintinline[fontsize=\footnotesize]{julia}{n}, respectively. All accesses are concordant, so, after adding lifecycle statements, we have:

    \begin{minted}[]{julia}
      T = UpTriMask()
      A = Tensor(Dense(SparseList(Element(0.0))), m, n)
      s = Tensor(Element(0.0))
      @finch begin
        @declare(s, 0.0)
        for j = 1:n
          for i = 1:m
            if T[i, j]
              s[] += A[i, j]
        @freeze(s)
    \end{minted}

    \textbf{Step 2: Declaring \texttt{s}}
    We then begin lowering the program. The \mintinline[fontsize=\footnotesize]{julia}{@declare(s, 0.0)} statement results in the initialization of the \mintinline[fontsize=\footnotesize]{julia}{val} tensor, \mintinline[fontsize=\footnotesize]{julia}{s.lvl.val[1] = 0.0}.
    \begin{minted}[]{julia}
      s.lvl.val[1] = 0.0
      @finch begin
        for j = 1:n
          for i = 1:m
            if UpTriMask()[i, j]
              s[] += A[i, j]
        @freeze(s)
    \end{minted}

    \textbf{Step 3: Unfurling \texttt{j}}
    To process the \mintinline[fontsize=\footnotesize]{julia}{j} loop, we unfurl both tensors that access \mintinline[fontsize=\footnotesize]{julia}{j}:

    \begin{minipage}{0.42\linewidth}
      \begin{minted}[]{julia}
        s.lvl.val[1] = 0.0
        @finch begin
          for j = 1:n
            for i = 1:m
              if (t[j])[i]
                s[] += (a[j])[i]
          @freeze(s)
      \end{minted}
    \end{minipage}\hfill%
    \begin{minipage}{0.56\linewidth}
      \begin{minted}[]{julia}
        t = unfurl(UpTriMask()) =
          Lookup(
            body(j) = UpTriMaskCol(j))
        a = unfurl(A::DenseLevel) =
          Lookup(
            body(j) = SubFiber(A.lvl.lvl, j))
      \end{minted}
    \end{minipage}

    \textbf{Step 4: Lowering Lookups}
    The Lookup pass inserts a for-loop:

    \begin{minted}[]{julia}
      s.lvl.val[1] = 0.0
      for j = 1:n
        @finch begin
          for i = 1:m
            if UpTriMaskCol(j)[i]
              s[] += SubFiber(A.lvl.lvl, j)[i]
      @finch @freeze(s)
    \end{minted}
    
    \textbf{Step 5: Unfurling \texttt{i}}
    Next, we process the \mintinline[fontsize=\footnotesize]{julia}{i} loop. Again, we unfurl both tensors:

    \begin{minipage}{0.34\linewidth}
      \begin{minted}[]{julia}
        s.lvl.val[1] = 0.0
        for j = 1:n
          @finch begin
            for i = 1:m
              if t[i]
                s[] += a[i]
        @finch @freeze(s)
      \end{minted}
    \end{minipage}\hfill%
    \begin{minipage}{0.68\linewidth}
      \begin{minted}[]{julia}
        t = unfurl(UpTriMaskCol(j)) = 
          Sequence(
            Phase(stop = j, Run(true)),
            Phase(Run(false)))
        a = unfurl(SubFiber(
            A.lvl.lvl::SparseListLevel, j)) =
          Thunk(
            preamble = (q = A.lvl.lvl.ptr[j]),
            Stepper(
              seek = (i) -> (
                q = binarysearch(A.lvl.lvl.idx, i)),
              stop = A.lvl.lvl.idx[q],
              body = Spike(
                body = 0,
                tail = A.lvl.lvl.val[q]),
              next = (q += 1)))
      \end{minted}
    \end{minipage}

  \end{minipage}\hfill%
  \begin{minipage}[t]{0.49\linewidth}

    \textbf{Step 6: Lowering Thunk}
    The Thunk pass moves the preamble out of the Thunks and unwraps it:

    \begin{minipage}{0.34\linewidth}
      \begin{minted}[]{julia}
        s.lvl.val[1] = 0.0
        for j = 1:n
          q = A.lvl.lvl.ptr[j]
          @finch for i = 1:j
            if t[i]
              s[] += a[i]
        @finch @freeze(s)
      \end{minted}
    \end{minipage}\hfill%
    \begin{minipage}{0.63\linewidth}
      \begin{minted}[]{julia}
        t = Sequence(
          Phase(stop = j, Run(true)),
          Phase(Run(false)))
        a = Stepper(
          seek = (i) -> (
            q = binarysearch(A.lvl.lvl.idx, i)),
          stop = A.lvl.lvl.idx[q],
          body = Spike(
            body = 0,
            tail = A.lvl.lvl.val[q]),
          next = (q += 1))
      \end{minted}
    \end{minipage}
    
\textbf{Step 7: Lowering Sequence}
    The Sequence pass introduces separate loops for each phase:

    \begin{minipage}{0.34\linewidth}
      \begin{minted}[]{julia}
        s.lvl.val[1] = 0.0
        for j = 1:n
          q = A.lvl.lvl.ptr[j]
          @finch for i = 1:j
            if t_1[i]
              s[] += a[i]
          @finch for i = j+1:m
            if t_2[i]
              s[] += a[i]
        @finch @freeze(s)
      \end{minted}
    \end{minipage}\hfill%
    \begin{minipage}{0.63\linewidth}
      \begin{minted}[]{julia}
        t_1 = Run(true)
        t_2 = Run(false)
        a = Stepper(
          seek = (i) -> (
            q = binarysearch(A.lvl.lvl.idx, i)),
          stop = A.lvl.lvl.idx[q],
          body = Spike(
            body = 0,
            tail = A.lvl.lvl.val[q]),
          next = (q += 1))
      \end{minted}
    \end{minipage}
    
    \textbf{Step 8: Lowering Run}
    The Run pass simply replaces runs with their value:

    \begin{minipage}{0.36\linewidth}
      \begin{minted}[]{julia}
        s.lvl.val[1] = 0.0
        for j = 1:n
          q = A.lvl.lvl.ptr[j]
          @finch for i = 1:j
            if true
              s[] += a[i]
          @finch for i = j+1:m
            if false
              s[] += a[i]
      @finch @freeze(s)
      \end{minted}
    \end{minipage}\hfill%
    \begin{minipage}{0.64\linewidth}
      \begin{minted}[]{julia}
        a = Stepper(
          seek = (i) -> (
            q = binarysearch(A.lvl.lvl.idx, i)),
          stop = A.lvl.lvl.idx[q],
          body = Spike(
            body = 0,
            tail = A.lvl.lvl.val[q]),
          next = (q += 1))
      \end{minted}
    \end{minipage}

    \textbf{Step 9: Simplify}
    The simplification pass removes the if statement in the first loop and removes the second loop:

    \begin{minipage}{0.34\linewidth}
      \begin{minted}[]{julia}
      s.lvl.val[1] = 0.0
      for j = 1:n
        q = A.lvl.lvl.ptr[j]
        @finch for i = 1:j
            s[] += a[i]
      @finch @freeze(s)
      \end{minted}
    \end{minipage}\hfill%
    \begin{minipage}{0.63\linewidth}
      \begin{minted}[]{julia}
      a = Stepper(
        seek = (i) -> (
          q = binarysearch(A.lvl.lvl.idx, i)),
        stop = A.lvl.lvl.idx[q],
        body = Spike(
          body = 0,
          tail = A.lvl.lvl.val[q]),
        next = (q += 1))
      \end{minted}
    \end{minipage}

    \textbf{Step 10: Lowering Steppers}
    The Stepper pass introduces a while loop:

    \begin{minipage}{0.48\linewidth}
      \begin{minted}[]{julia}
        s.lvl.val[1] = 0.0
        for j = 1:n
          q = A.lvl.lvl.ptr[j]
          k = 1
          while k < j
            k_2 = A.lvl.lvl.idx[q]
            @finch for i = k:min(k_2, j)
              s[] += a[i]
            q += 1
        @finch @freeze(s)
      \end{minted}
    \end{minipage}\hfill%
    \begin{minipage}{0.48\linewidth}
      \begin{minted}[]{julia}
        a = Spike(
            body = 0,
            tail = A.lvl.lvl.val[q])
      \end{minted}
    \end{minipage}

    \textbf{Step 11: Lowering Spike}
    The Spike pass acts like Sequence combined with a Run:

    \begin{minted}[]{julia}
    s.lvl.val[1] = 0.0
    for j = 1:n
      q = A.lvl.lvl.ptr[j]
      k = 1
      while k < j
        k_2 = A.lvl.lvl.idx[q]
        @finch for i = k:min(k_2, j)
          s[] += 0
        i = k_2
        @finch if i < j
          s[] += A.lvl.lvl.val[q]
        q += 1
    @finch @freeze(s)
    \end{minted}
  \end{minipage}\hfill%
  \begin{minipage}[t]{0.32\linewidth}

  \end{minipage}
  \caption{Example lowering of a Finch program, continued in Figure \ref{fig:lowering_example2}.}\label{fig:lowering_example1}
\end{figure}

\begin{figure}[h]
  \footnotesize
  \begin{minipage}[t]{0.49\linewidth}
    \textbf{Step 12: Simplify}
    The simplify pass recognizes that addition of \mintinline[fontsize=\footnotesize]{julia}{0} is a no-op:

    \begin{minted}[]{julia}
    s.lvl.val[1] = 0.0
    for j = 1:n
      q = A.lvl.lvl.ptr[j]
      k = 1
      while k < j
        k_2 = A.lvl.lvl.idx[q]
          i = k_2
          if i < j
            s[] += A.lvl.lvl.val[q]
        q += 1
      @finch @freeze(s)
    \end{minted}

  \end{minipage}\hfill%
  \begin{minipage}[t]{0.49\linewidth}

    \textbf{Step 13: Lower Freeze}
    Finally, the final \mintinline[fontsize=\footnotesize]{julia}{freeze} is a no-op and we obtain:

    \begin{minted}[]{julia}
    s.lvl.val[1] = 0.0
    for j = 1:n
      q = A.lvl.lvl.ptr[j]
      k = 1
      while k < j
        k_2 = A.lvl.lvl.idx[q]
          i = k_2
          if i < j
            s[] += A.lvl.lvl.val[q]
        q += 1
    \end{minted}

  \end{minipage}\hfill%

  \caption{Example lowering of a Finch program, continued from \ref{fig:lowering_example1}. The final program accesses only the upper triangle of A, though the original code looks as though it loops over all \mintinline[fontsize=\footnotesize]{julia}{i} and \mintinline[fontsize=\footnotesize]{julia}{j}.}\label{fig:lowering_example2}
\end{figure}

\section{Case Studies}

We evaluate Finch on a broad set of applications to showcase it's efficiency,
flexibility, and expressiveness. All of our implementations highlight the
benefits of data structure and algorithm co-design.  Our implementation of
sparse-sparse-matrix multiply (SpGEMM) translates classical lessons from sparse
performance engineering into the language of Finch, using temporaries and
randomly-accessible workspace formats to efficiently implement the three main
approaches. Our study of sparse-matrix-dense-vector multiply (SpMV) highlights
the benefits of precise structural specialization. Our studies of image
morphology and graph applications show how Finch's programming model can express more complex
real-world kernels. 

All experiments were run on a single core of a 12-core 2-socket Intel Xeon CPU
E5-2680 v3 running at 2.50GHz with 128GB of memory. Finch is implemented in
Julia v1.9, targeting LLVM through Julia. All timings are the minimum of 10,000
runs or 5s of measurement, whichever happens first.

\begin{wrapfigure}{r}{0.41\linewidth}
    \vspace{-36pt}
    \begin{minipage}[t]{0.18\textwidth}
        \begin{minted}{julia}
        
            y .= 0
            for j = _, i = _
              y[i] += A[i, j] * x[j]
        \end{minted}
        \vspace{24pt} 
        \begin{minted}{julia}
            y .= 0
            for j = _, i = _
              y[j] += A[i, j] * x[i]
        \end{minted}
    \end{minipage}\hfill%
    \begin{minipage}[t]{0.22\textwidth}
        \vspace{0pt} 
        \begin{minted}{julia}
            y .= 0
            for j = _
              let x_j = x[j]
                y_j .= 0
                for i = _
                  let A_ij = A[i, j]
                    y[i] += x_j * A_ij
                    y_j[] += A_ij * x[i]
                #D is the diagonal
                y[j] += y_j[] + D[j] * x_j
        \end{minted}
    \end{minipage}
    \vspace{-12pt}
    \caption{Finch row-major, column-major and symmetric SpMV Programs. Note
    that the upper triangle of the input is pre-computed for the symmetric
    program. Reads to the canonical triangle are reused with a $\finchdefine$
    statement, and the results are written to both relevant locations using
    multiple outputs.}
    \label{spmv_programs}
    \vspace{-12pt}
  \end{wrapfigure}
  \subsection{Sparse Matrix-Vector Multiply (SpMV)}
  Sparse matrix-vector multiply (SpMV) has a wide range of applications and has been thoroughly studied~\cite{liu_csr5_2015,
  zhou_enabling_2020}. 
  Because SpMV is bandwidth bound, many formats have
  been proposed to reduce the footprint~\cite{langr_evaluation_2016}. 
  The wide range of applications
  results in a wide range of tensor structures, making it an effective kernel to
  demonstrate the utility of our programming model. 

\begin{figure}[!h]
    \vspace{-6pt}
    \includegraphics[width=\linewidth]{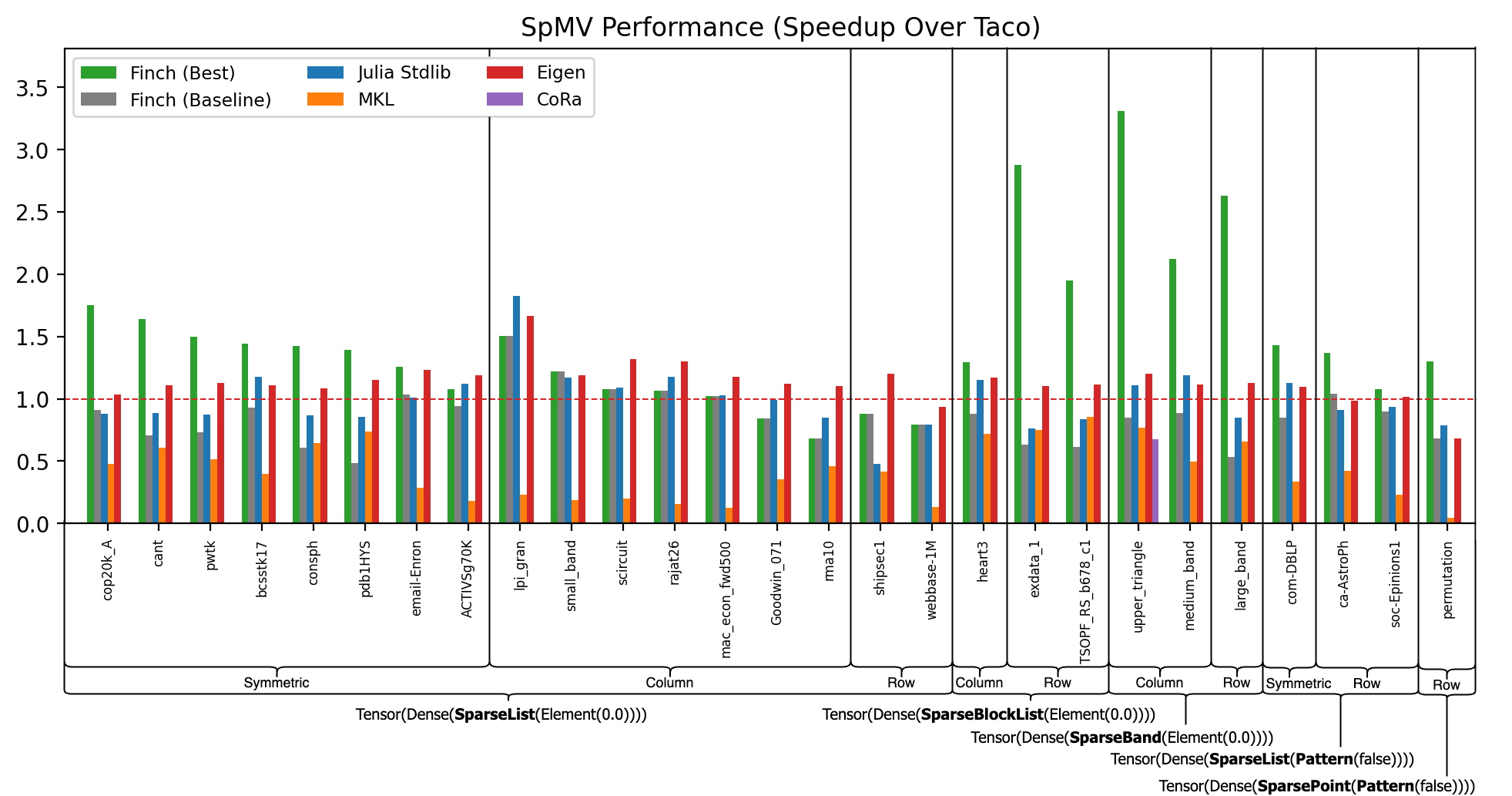}
    \vspace{-12pt}
    \caption{\footnotesize Performance of SpMV algorithms, organized by the best performing Finch format. The displayed Finch performance is the fastest among the formats we tested. Programs are from Figure~\ref{spmv_programs}. ``finch\_baseline'' is the faster of row major or column major ``SparseList''.}
    \vspace{-6pt}
    \label{fig:spmv_grouped}
  \end{figure}

  Figure~\ref{fig:spmv_grouped} displays the performance of SpMV measured relative to TACO.
  We varied both the data formats and the SpMV algorithm. 
  We display the best Finch format and algorithm among all the formats and algorithms listed in Figure~\ref{spmv_programs}, wherever each format and algorithm are applicable. 
  Precisely which Finch format performed best on which matrices is shown in the figure.
  We compare against TACO (best of row or column-major), Julia’s standard library (column major), baseline Finch (best of row or column-major with CSC format), Eigen (row-major) \cite{guennebaud_eigen_2010}, MKL (row-major) \cite{noauthor_developer_2024}, and CORA (unscheduled, row-major) \cite{fegade_cora_2022-1}.
  Our test suite is the union of datasets from three previous papers: the matrices used by Ahrens et al. to test a variable block row format partitioning strategy \cite{ahrens_optimal_2021}, Kjolstad et al.  to test the TACO library \cite{kjolstad_tensor_2017}, and Leskovec et al. to evaluate graph clustering algorithms \cite{leskovec_empirical_2010}. 
  We left out two very large matrices (\texttt{Janna/Emilia\_923} and \texttt{Janna/Geo\_1438}); the remaining matrices in our dataset had a maximum of 12 million nonzeros.
  We also added some synthetic matrices, $10,000 \times 10,000$ banded matrices with bandwidth 5, 30, and 100, a $1024 \times 1024$ upper triangular matrix, and a $1,000,000 \times 1,000,000$ reverse permutation matrix.

  %
  %

  Finch commonly introduces tradeoffs between branching and more complicated loop structures and the benefits of such specialization.
  Specialization is better in cases where the specialized routine is much faster and the common case (such as the zero region of a sparse tensor, which becomes a no-op).
  However, specialization introduces many different branches, and complicates the bounds of loops.
  For example, in our symmetric kernel, we found it was faster to pre-compute the upper triangle of the input matrix, rather than calculate it on the fly using a mask expression such as \mintinline{julia}{i < j}, which would change the exit condition of the inner loop.
  The option to de-specialize certain conditional expressions is another example of how Finch can widen the design space for structured operators.
   

  Our result shows that different formats perform better on different matrices, and that Finch can be used to exploit these formats effectively.
  We found that the SpMV performance was superior for the level format that best paralleled the structure of the tensor.
  The best Finch format had a geomean speedup of 1.27 over TACO.
  Matrices with a clear blocked structure like exdata\_1, TSOPF\_RS\_b678\_c1, and heart3 performed notably well with the SparseBlockList format with speedups of 2.75, 1.80, and 1.20 relative to TACO, while the baseline format was slightly slower than TACO.
  Furthermore, the synthetic banded matrices we constructed performed the best with the SparseBand matrix, in particular with the large\_band and the medium\_band matrices having a speedup of 2.50 and 2.02 relative to TACO, while the baseline format had minor slowdowns relative to TACO. 
  On our triangular matrix, Finch had a speedup of 3.04 over TACO, outperforming even CORA, which was designed for ragged tensors but whose optimizations were targeted more towards cache blocking than to the specific structure of the tensor.
  Similarly, using a SparsePoint format obtained a speedup of 1.30 by avoiding a loop over nonzeros (since there is only ever one nonzero in the SparsePoint level).
  The Pattern leaf level performed better than the Element leaf level for representing Boolean graph matrices.
  For example, the SparseList-Pattern format for \texttt{ca-AstroPh} resulted in a speedup of 1.17, while the baseline Finch format resulted only achieved 1.04. Our results clearly demonstrate the utility of being able to vary both the algorithm and the format to match the structure of the tensor.

  Though MKL is closed-source, using \texttt{perf}, we found that the MKL implementation had noticeably higher branch mispredictions than expected (23\%, as compared to TACO's 1\%), and that the code was vectorized (23\% of the instructions were AVX, as compared to TACO's 0.03\%), indicating a vectorized row-major traversal strategy, with a gather operation in each step and a horizontal sum at the end of each row. This would not be a good strategy when the matrix has only 6 nonzeros per row and the SIMD registers have 4 elements, as the inner loop would iterate only once between loop setup and loop cleanup, which proved expensive on this architecture. Taco and Eigen fared similarly, both lemmiting simple loops with loop variables that correspond to the nonzero position, which can sometimes have a slight advantage over Finch, which uses the coordinate as the loop variable. Still, Finch's structural specification showed a clear advantage on  our test inputs.

\subsection{Sparse-Sparse Matrix Multiply (SpGEMM)}

We compute the $M \times N$ sparse matrix $C$ as the product of $M \times K$ and $K \times N$ sparse matrices $A$ and $B$.
There are three main approaches to SpGEMM \cite[Section 2.2]{zhang_gamma_2021}.
The inner-products algorithm takes dot products of corresponding rows and columns, while the outer-products algorithm sums the outer products of corresponding columns and rows.
Gustavson's algorithm sums the rows of $B$ scaled by the corresponding nonzero columns in each row of $A$.
Inner-products is known to be asymptotically less efficient than the others, as we must do a merge operation to compute each of the $O(MN)$ entries in the output \cite{ahrens_autoscheduling_2022}.
We will show that our ability to implement these latter methods exceeds that of TACO, translating to asymptotic benefits. 

\begin{figure}[h]
    \begin{minipage}[t]{0.32\linewidth}
      \vspace{0pt}
    \begin{minted}{julia}
    @finch begin
      C .= 0
      for j=_
        for i=_
          for k=_
            C[i, j] += AT[k, i] * B[k, j]
      return C
    \end{minted}
    \end{minipage}%
    \begin{minipage}[t]{0.33\linewidth}
      \vspace{0pt}
    \begin{minted}{julia}
    w = Tensor(SparseByteMap(Element(0)))
    @finch begin
      C .= 0
      for j=_
        w .= 0
        for k=_
          for i=_
            w[i] += A[i, k] * B[k, j]
        for i=_
          C[i, j] = w[i]
    \end{minted}
    \end{minipage}%
    \begin{minipage}[t]{0.35\linewidth}
      \vspace{0pt}
    \begin{minted}{julia}
    w = Tensor(SparseHash(SparseHash(Element(0))))
    @finch begin
      w .= 0
      for k=_
        for j=_
          for i=_
            w[i, j] += A[i, k] * BT[j, k]
      C .= 0
      for j=_, i=_
        C[i, j] = w[i, j]
    \end{minted}
    \end{minipage}
    \caption{Inner Products, Gustavson's, and Outer Products matrix multiply in Finch}\label{fig:spgemm_listing}
\end{figure}

\begin{figure}[h]
    \vspace{-12pt}
    \begin{minipage}[t]{0.5\linewidth}
	\includegraphics[width=\linewidth]{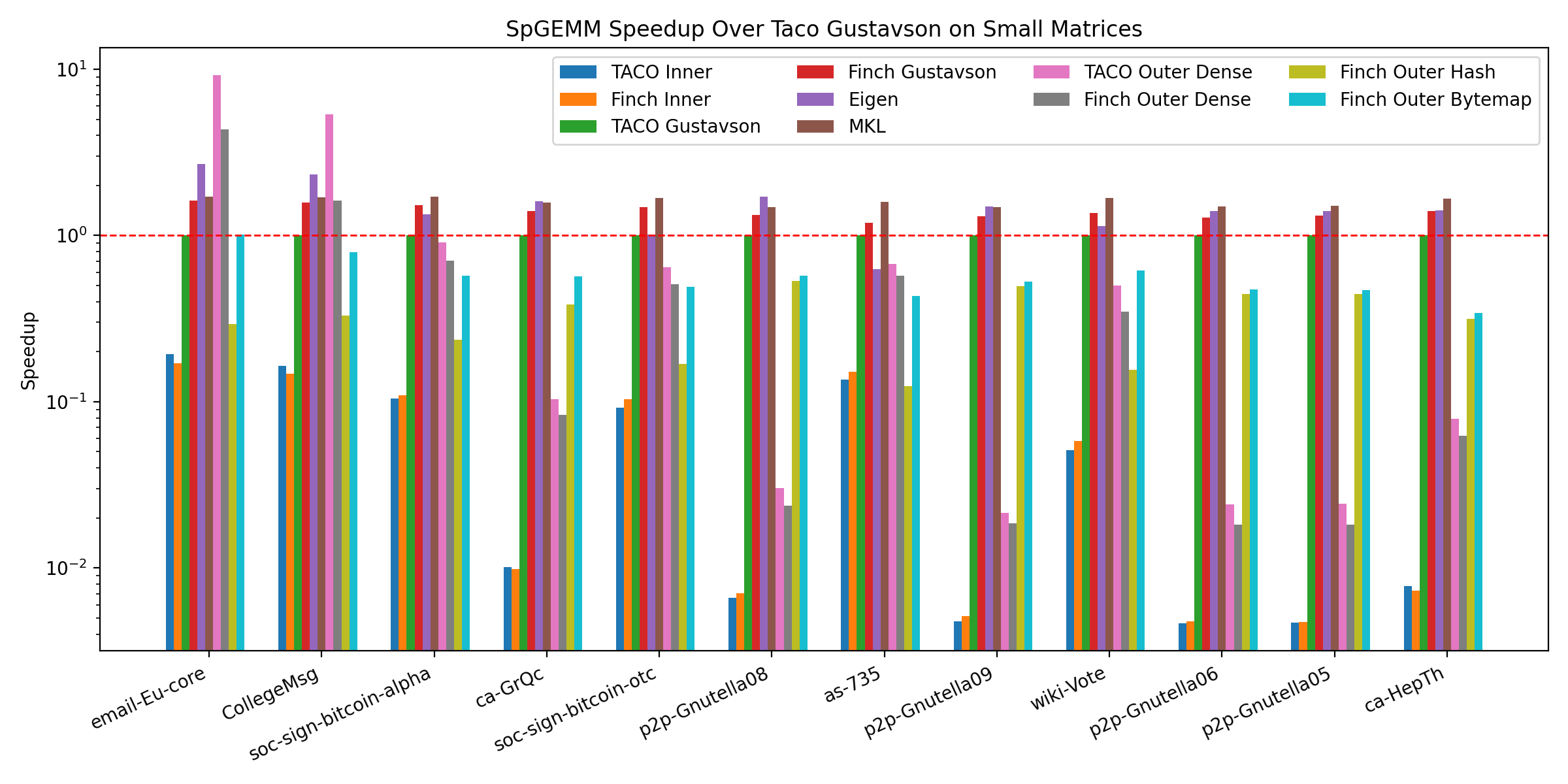}%
    \end{minipage}%
    \begin{minipage}[t]{0.5\linewidth}
	\includegraphics[width=\linewidth]{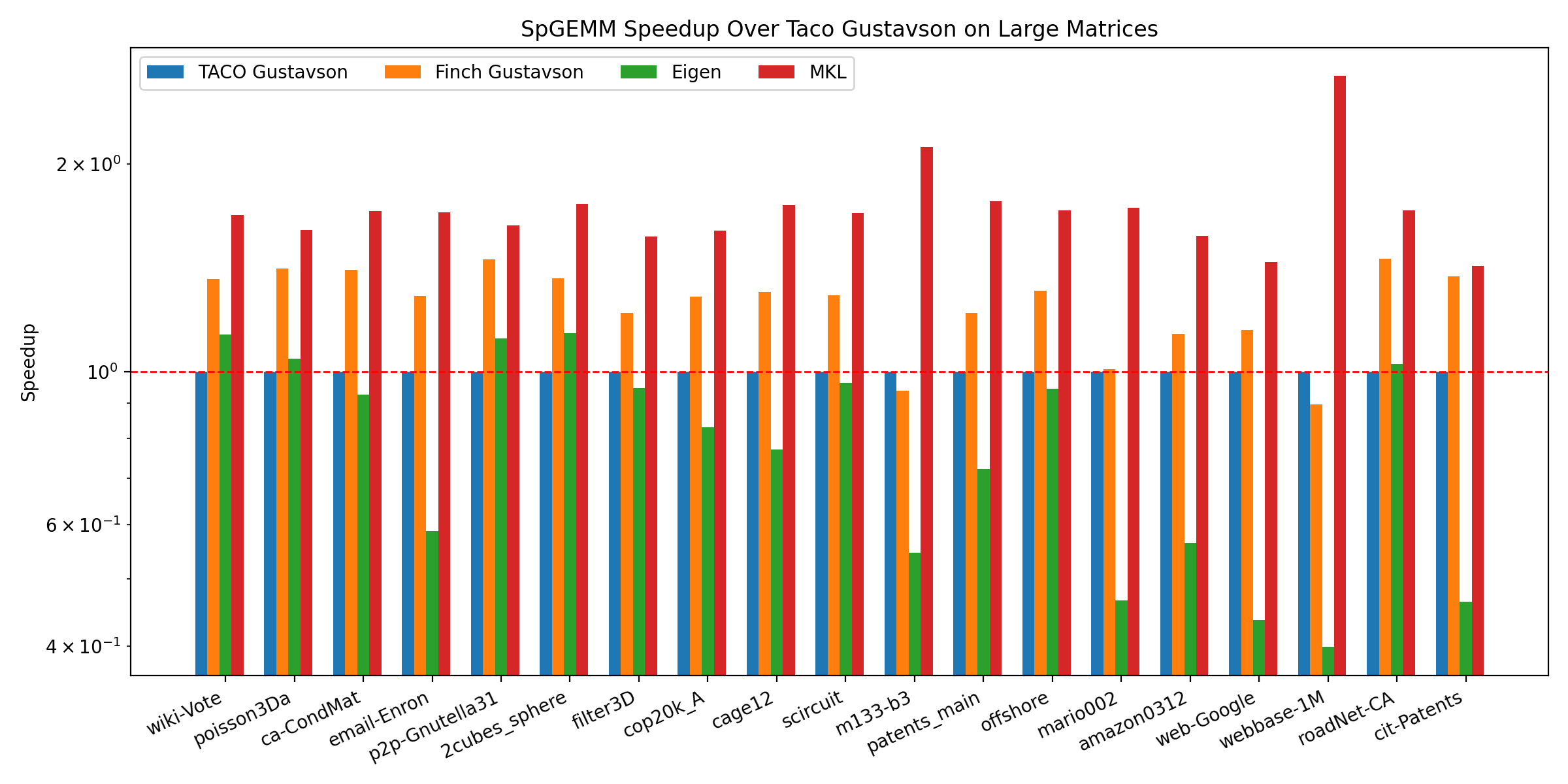}
    \end{minipage}%
    \vspace{-12pt}
    \begin{minipage}[t]{0.5\linewidth}
    \vspace{-12pt}
    \caption{A comparison of several matrix multiply algorithms between Finch,
    Taco, Eigen, and MKL. The top images show results on the same dataset as
    \cite{zhang_gamma_2021}.  We use only Gustavson's algorithm on larger
    matrices.  The image on bottom right show results on increasingly large
    Erd\H{o}s-Rényi matrices with an average of 4 nonzeros per row. Note that
    inner-products and outer products algorithms with a dense output have an
    asymptotic disadvantage, and that Gustavson's algorithm or an outer products
    with sparse format perform better as the problem size grows.}
    \label{fig:spgemm}
    \end{minipage}%
    \begin{minipage}[t]{0.5\linewidth}
      \vspace{6pt}
    \includegraphics[width=\linewidth]{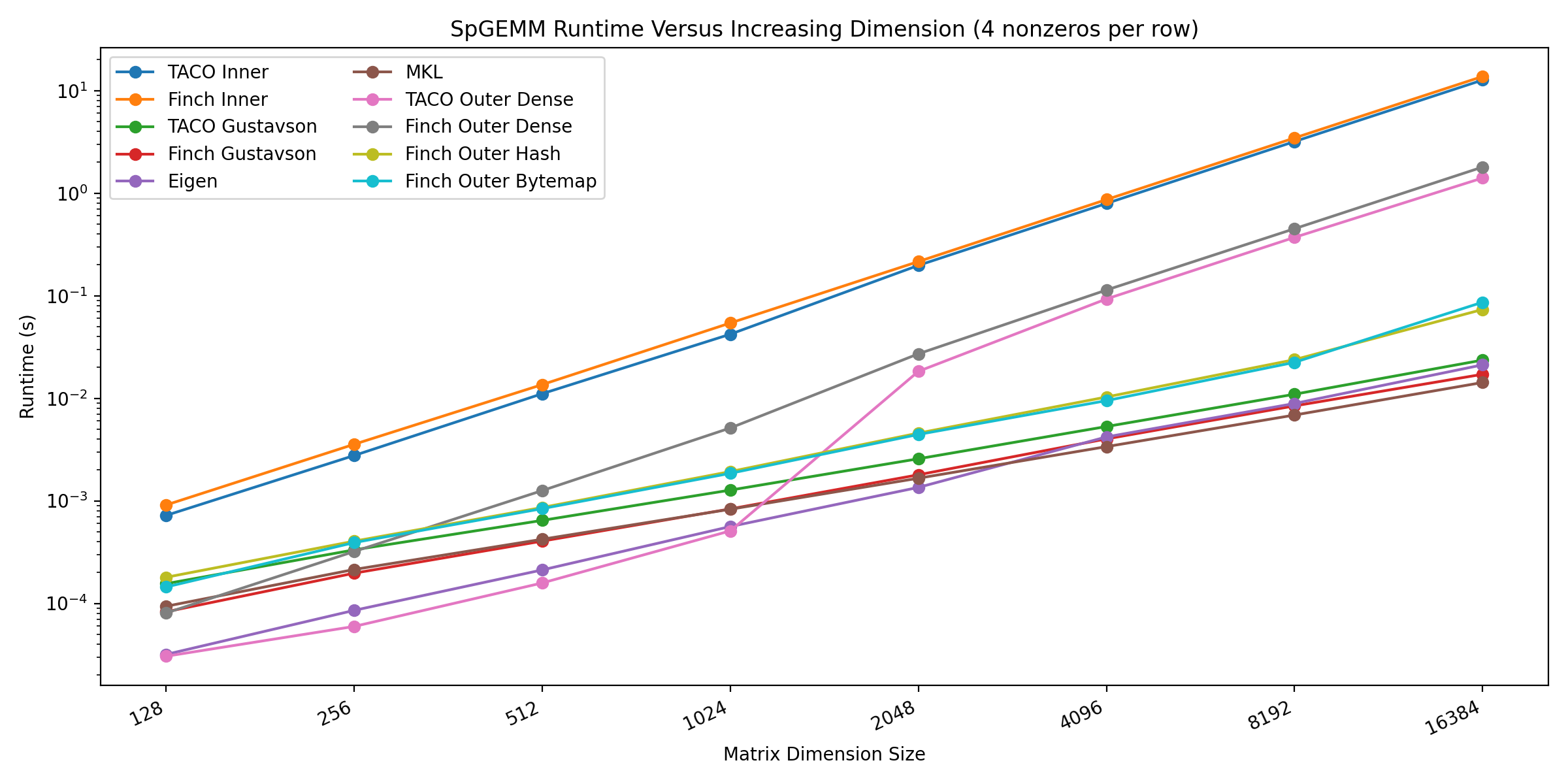}
    \end{minipage}%
      \vspace{-12pt}
\end{figure}

Figure~\ref{fig:spgemm_listing} implements all three approaches in Finch, and Figure~\ref{fig:spgemm} compares the performance of Finch to TACO, Eigen, and MKL on the matrices of Zhang et al. \cite{zhang_gamma_2021}.
Note that these algorithms mainly differ in their loop order, but that different data structures can be used to support the various access patterns induced.
In our Finch implementation of outer products, we use a sparse hash table, as it is fully-sparse and randomly accessible.
Since, TACO does not support multidimensional sparse workspaces, its outer products uses a dense intermediate, which leads to an asymptotic slow down shown in Figure~\ref{fig:spgemm}.
Similarly, although a sparse bytemap has a dense memory footprint, we use it in our Finch implementation of Gustavson's for the smaller $O(N)$ intermediate.
We note that the bytemap format in TACO's Gustavson's implementation is hard-wired, whereas Finch's programming model allows us to write algorithms with explicit temporaries and transpositions.
Without such hard wiring, TACO would have to use a dense intermediate to support random writes, which TACO would then propagate to the output, turning it dense and leading to the same asymptotic results as in the case of outer products. 
As depicted in Figure~\ref{fig:spgemm}, Finch achieves comparable performance with TACO on smaller matrices when we use the same datastructures, and significant improvements when we use better datastructures. 
Finch outperforms TACO overall, with a geomean speedup of 1.30.
Finch and TACO both outperform Eigen by a significant margin, as Eigen is designed for usability but is not heavily optimized to support a wide variety of matrix multiplication routines.
Finch is competitive with, but slightly slower than MKL. We cannot comment extensively on MKL’s good performance on SpGEMM workloads as we cannot access source code, but we suspect that MKL uses a Gustavson’s algorithm with a highly optimized sorting routine.

We also include a scaling study in Figure~\ref{fig:spgemm} to show the asymptotic impact of the output format as the SpGEMM problem size grows. We consider uniformly random $N \times N$ matrices with a fraction of $p = 4/N$ nonzeros. An inner-product approach runs in time $O(N^3p) = O(4N^2)$. An outer-products approach with a sparse output format runs in expected time $O(N^3p^2) = O(16N)$, which is an asymptotic improvement as the matrix gets sparser. An outer-products approach with a dense output format runs in expected time $O(N^3p^2 + N^2) = O(N^2)$, which is an asymptotic disadvantage when the number of nonzeros per row ($Np$) is small. Our plot shows that Finch's sparse outer products  routine outperforms TACO's outer products routine, since Finch supports a sparse output format but TACO does not. Finch is the first tensor compiler to support all three strategies with both sparse and dense output formats.


\subsection{Graph Analytics}

We used Finch to implement both Breadth-first search (BFS) and Bellman-Ford single-source shortest path.
Our BFS implementation and graphs datasets are taken from Yang et al.~\cite{yang_implementing_2018}, including both road networks and scale-free graphs (bounded node degree vs. power law node degree).


Direction-optimization~\cite{beamer_direction-optimizing_2012} is crucial for achieving high BFS performance in such scenarios, switching between push and pull traversals to efficiently explore graphs.
Push traversal visits the neighbors of each frontier node, while pull traversal visits every node and checks to see if it has a neighbor in the frontier. 
The advantage of pull traversal is that we may terminate our search once we find a node in the frontier, saving time in the event the push traversal were to visit most of the graph anyway. 
Early break is the critical part of control flow in this algorithm, though the algorithms also require different loop orders, multiple outputs, and custom operators.
Finch performs well because it can directly express algorithms comparable to competitive libraries.

\begin{figure}[h]
    \begin{minipage}{0.33\linewidth}
    \begin{minted}{julia}
    V = Tensor(Dense(Element(false)))
    P = Tensor(Dense(Element(0)))
    F = Tensor(SparseByteMap(Pattern()))
    _F = Tensor(SparseByteMap(Pattern()))
    A = Tensor(Dense(SparseList(Pattern())))
    AT = Tensor(Dense(SparseList(Pattern())))

    function bfs_push(_F, F, A, V, P)
      @finch begin
        _F .= false
        for j=_, k=_
          if F[j] && A[k, j] && !(V[k])
            _F[k] |= true
            P[k] <<choose(0)>>= j
        return _F

    \end{minted}
\end{minipage}%
\begin{minipage}{0.33\linewidth}
    \begin{minted}{julia}
    function bfs_pull(_F, F, AT, V, P)
      p = ShortCircuitScalar{0}()
      @finch begin
        _F .= false
        for k=_
          if !V[k]
            p .= 0
            for j=_
              if F[follow(j)] && AT[j, k]
                p[] <<choose(0)>>= j
            if p[] != 0
              _F[k] |= true
              P[k] = p[]
        return _F
    \end{minted}
\end{minipage}%
\begin{minipage}{0.33\linewidth}
  \begin{minted}{julia}
  _D = Tensor(Dense(Element(Inf)), n)
  D = Tensor(Dense(Element(Inf)), n)
  function bellmanford(A, _D, D, _F, F)
    @finch begin
    F .= false
    for j = _
      if _F[j]
        for i = _
          let d = _D[j] + A[i, j]
            D[i] <<min>>= d
            F[i] |= d < _D[i]
\end{minted}
\end{minipage}
\caption{Graph Applications written in Finch. Note that parents are calculated separately for Bellman-Ford. The $choose(z)$ operator is a GraphBLAS concept which returns any argument that is not $z$.}\label{fig:graph_listing}
\end{figure}

\begin{figure}[h]
    \vspace{-12pt}
	\includegraphics[width=0.5\linewidth]{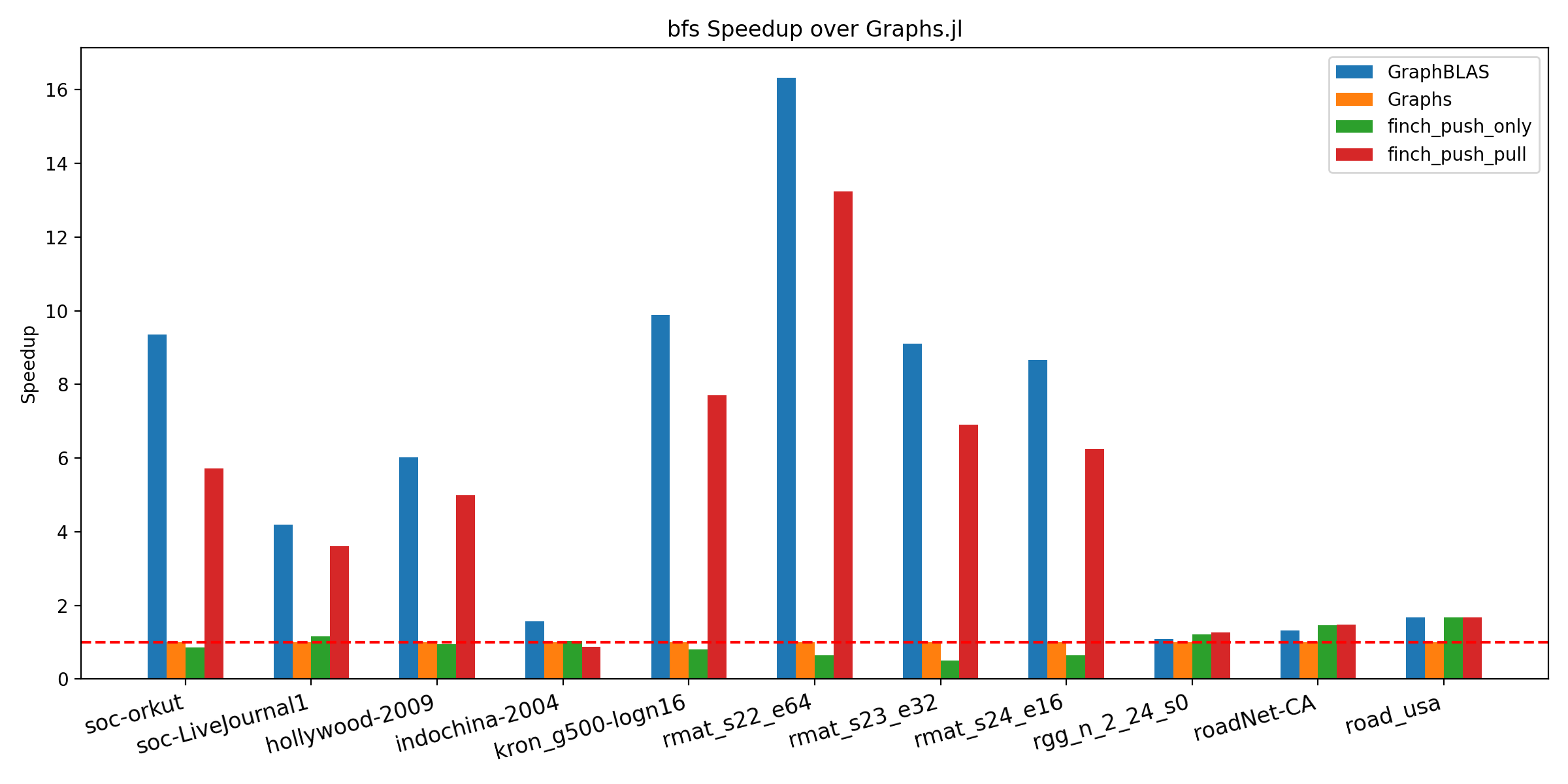}%
	\includegraphics[width=0.5\linewidth]{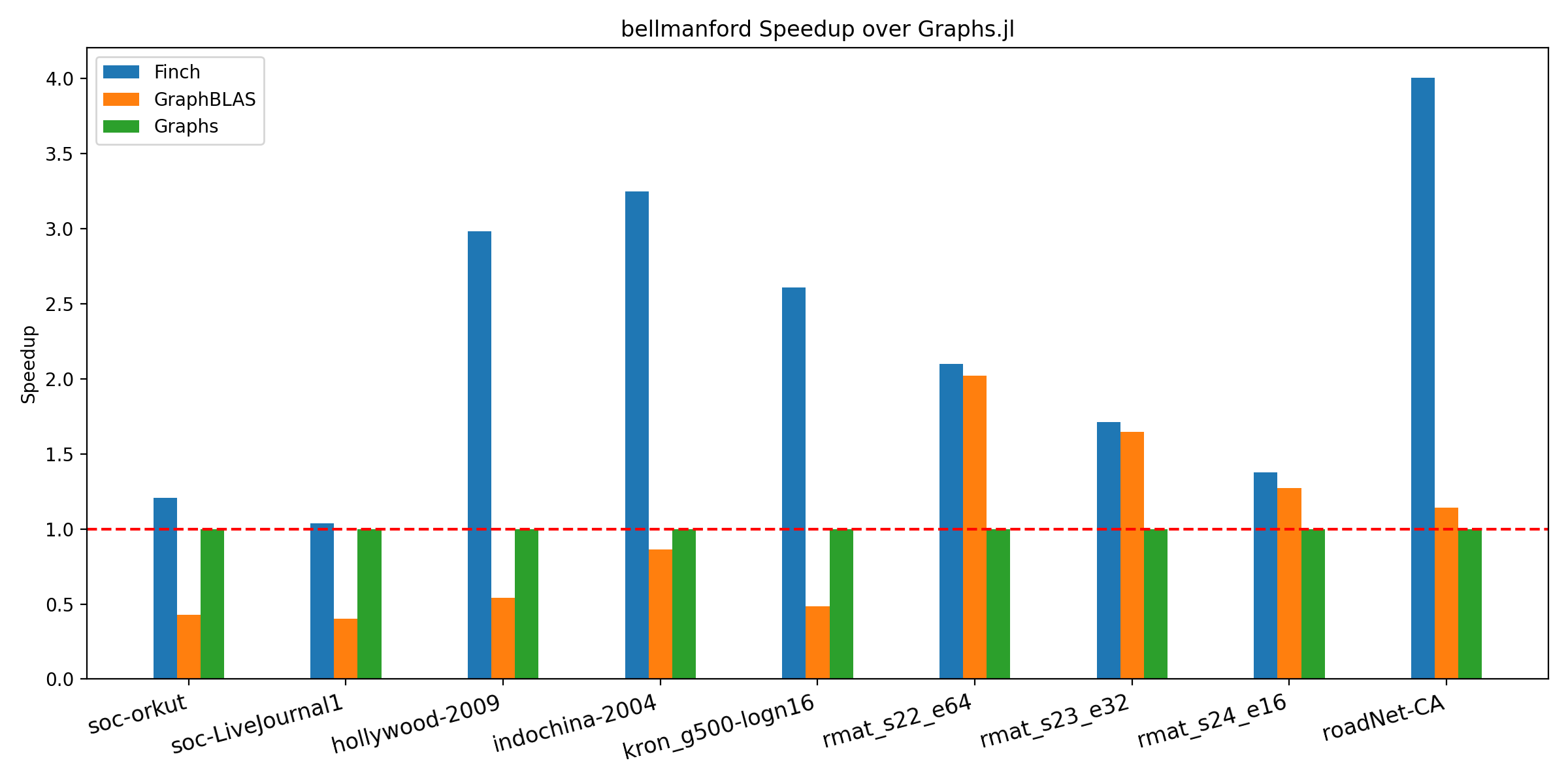}
    \vspace{-12pt}
    \caption{Performance of graph apps across various tools. finch\_push\_only exclusively utilizes push traversal, while finch\_push\_pull applies direction-optimization akin to GraphBLAS. Finch's support for push/pull traversal and early break facilitates direction-optimization. Among GraphBLAS's five variants for Bellman-Ford, we selected LAGraph\_BF\_full1a, consistently the fastest with our graphs. We did not include Bellman-Ford results for graphs with high diameter as they timed out (> 1 hour).}
     \label{fig:graph_result}
\end{figure}

Figure~\ref{fig:graph_result} compares performance to Graphs.jl, a Julia library, and the LAGraph Library, which implements graph algorithms with sparse linear algebra using GraphBLAS~\cite{mattson_lagraph_2019}.
On BFS, Finch is competitive even with the hardwired optimizations of GraphBLAS, a geomean slowdown of 1.22. Direction-optimization notably enhances performance for scale-free graphs. 
On Bellman-Ford (with path lengths and shortest-path tree), Finch's support for multiple outputs, sparse inputs, and masks leads to a geomean speedup of 2.47 over GraphBLAS. 
Appendix B displays code for BFS and Bellman-Ford in Finch (57 and 50 LOC) and LAGraph (215 and 227 LOC), and we invite readers to compare the clarity of the algorithms.
\subsection{Image Morphology}

Some image processing pipelines stand to benefit from structured data processing \cite{donenfeld_unified_2022}.
We focus on binary image morphology and the logical transformation of binary images and masks.
We consider two operations: binary erosion (computing a mask), and a masked histogram (using a mask to avoid work).
We use images that are all binary, either by design or having been thresholded.

Finch allows us choose our datastructure, so we may choose to use either a dense representation with bytes ($Dense(Element(0x00))$), a bit-packed representation ($Dense(Element(UInt64))$), or a run-length encoded representation that represents runs of true or false ($SparseRunList(Pattern())$).
All of these have their advantages.
The dense representation induces the least overhead, the bit-packed representation can take advantage of bitwise binary ops, and the run-length encoded version only uses memory and compute when the pattern changes. 

Similarly, since Finch lets us choose our algorithm we can implement erosion in a few ways.
The erosion operation turns off a pixel unless all of it's neighbors are on.
This can be used to shrink the boundaries of a mask, and remove point instances of noise \cite{fisher_hypermedia_1996}.
%
This introduces three instances of structure in the control flow: the mask, the padding of inputs, and the convolutional filter.
We focused on the filter.
%
We might understand the filter as a structured tensor of circular shifts, or we might understand each shifted view of the data in an unrolled stencil computation as a structured tensor, or a two part stencil where we compute the horizontal then vertical part of the stencil.
We experimented with these options and found that the last approach performed best, due to fitting the storage formats while reducing the amount of work with intermediate temporaries.
Figure \ref{fig:morphology_listing} displays example erosion algorithms for bitwise
or run-length-encoded algorithms.
%

We compared against OpenCV on four datasets. We randomly selected 100 images from the MNIST \cite{lecun_gradient-based_1998} and Omniglot \cite{lake_human-level_2015} character recognition datasets, as well as a dataset of human line drawings \cite{eitz_how_2012}. 
We also hand-selected a subset of mask images (these images were less homogeneous, so we listed them in Appendix C) from a digital image processing textbook \cite{gonzalez_digital_2006}. 
All images were thresholded, and we also include versions of the images that have been magnified before thresholding, to induce larger constant regions. 
In our erosion task, the SparseRunList format performs the best as it is asymptotically faster and uses less memory, leading to a 19.5X speedup over OpenCV on the sketches dataset, which becomes arbitrarily large as we magnify the images (here shown as 266X). 
Finch achieves these speedups by exploiting structured sparsity to straightforwardly do less work than OpenCV's more naive dense implementation, which must still read most of an image or mask even when it is unnecessary.
However, we believe the 51.6x on MNIST is due to calling overhead in OpenCV. 
The bitwise kernels were effective as well, and would be more effective on datasets with less structure. 
A strength of Finch is that it supports structured datasets, even over bitwise operations, allowing us to implement the bitwise kernel and then mask it.

We also implemented a histogram kernel.
We used an indirect access into the output to implement this (Figure \ref{fig:morphology_listing}), something not many sparse frameworks support.
We compare to OpenCV since the OpenCV histogram function also accepts a mask. 
If we use $SparseRunList(Pattern())$ for our mask, we can reduce the branching
in the masked kernel and get better performance.
The improvements with SparseRunList are seen in the histogram task too, as it allows us to mask off contiguous regions of computation, instead of individual pixels, reducing the branches and leading to a significant speedup (20.3x on Omniglot and 20.8x on sketches).
In a low compute task such as a histogram, skipping many reads for the mask via structured sparsity can lead to huge speedups. 
%

\begin{figure}
    \scriptsize
    \begin{minipage}{0.5\linewidth}
    Wordwise Erosion:
    \begin{minted}{julia}
        output .= false
        for y = _
          tmp .= false
          for x = _
            tmp[x] = coalesce(input[x, ~(y-1)], true) & input[x, y] & coalesce(input[x, ~(y+1)], true)
          for x = _
            output[x, y] = coalesce(tmp[~(x-1)], true) & tmp[x] & coalesce(tmp[~(x+1)], true)
    \end{minted}
    \vspace{12pt}
    Masked Histogram:
    \begin{minted}{julia}
        bins .= 0 
        for x=_
          for y=_
            if mask[y, x]
              bins[div(img[y, x], 16) + 1] += 1
    \end{minted}
    \end{minipage}%
    \begin{minipage}{0.5\linewidth}
    Bitwise Erosion:
    \begin{minted}{julia}
        output .= 0
        for y = _
          tmp .= 0
          for x = _
            if mask[x, y]
              tmp[x] = coalesce(input[x, ~(y-1)], 0xFFFFFFFF) & input[x, y] & coalesce(input[x, ~(y+1)], 0xFFFFFFFF)
          for x = _
            if mask[x, y]
              let tl = coalesce(tmp[~(x-1)], 0xFFFFFFFF), t = tmp[x], tr = coalesce(tmp[~(x+1)], 0xFFFFFFFF)
                let res = ((tr << (8 * sizeof(UInt) - 1)) | (t >> 1)) & t & ((t << 1) | (tl >> (8 * sizeof(UInt) - 1)))
                  output[x, y] = res
    \end{minted}
    \end{minipage}
    \vspace{-12pt}
    \caption{Two approaches to erosion in Finch. The $coalesce$ function defines the out of bounds value. On left, the naive approach. On $SparseRunList(Pattern())$ inputs, this only performs operations at the boundaries of constant regions. On right, a bitwise approach, using a mask to limit work to nonzero blocks of bits.}
    \label{fig:morphology_listing}
\end{figure}

\begin{figure}
	\includegraphics[width=0.5\linewidth]{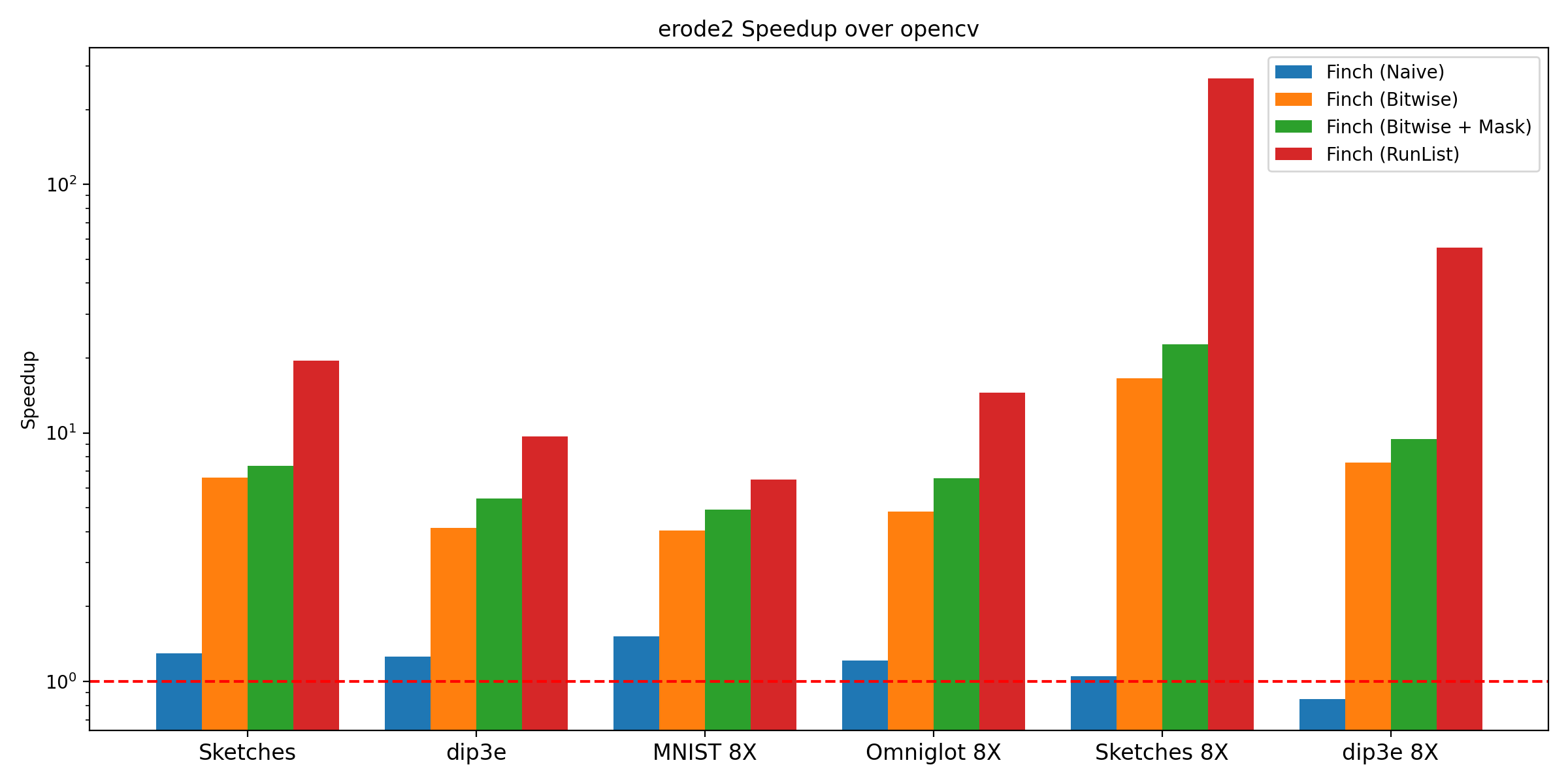}%
	\includegraphics[width=0.5\linewidth]{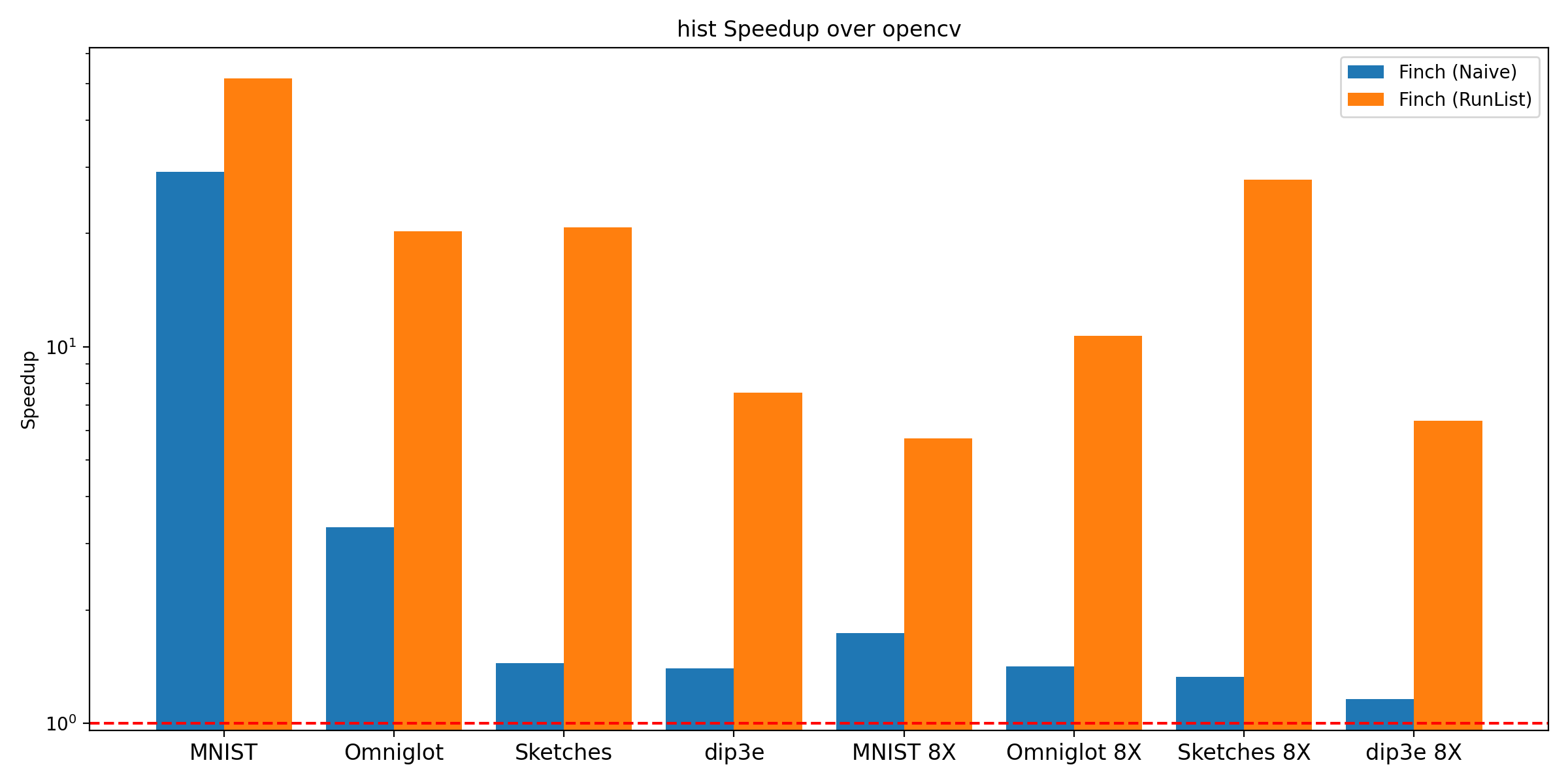}
 \vspace{-12pt}
    \caption{Performance of Finch on image morphology tasks. On left, we run 2 iterations of erosion. On right, we run a masked histogram. We display the geomean speedup within each dataset.}\label{fig:morphology}
\end{figure}

\section{Related Work}

The related work on tensor languages and libraries spans several areas, from libraries to languages, from dense to structured computation.

\paragraph{Libraries for Dense Data:}
Many libraries specialize in dense computations.
Perhaps the most well-known example is NumPy~\cite{harris_array_2020},
and a classic example is the BLAS, though several BLAS routines are specialized to symmetric, hermitian, and triangular matrices~\cite{anderson_lapack_1999}. 
Many research projects have advanced on BLAS, such as BatchedBlas and BLIS~\cite{dongarra_design_2017,van_zee_blis:_2015}.

\paragraph{Libraries for Structured Data:}

Many libraries support BLAS plus a few sparse tensor types, typically CSR, CSC, BCSR, Banded, and COO.
Examples include SciPy~\cite{virtanen_scipy_2020}, PETSc~\cite{abhyankar_petsc_nodate}, Armadillo~\cite{rumengan_pyarmadillo_2021}, OSKI~\cite{vuduc_oski:_2005}, Cyclops~\cite{solomonik_cyclops_2013}, MKL~\cite{noauthor_developer_2024}, and Eigen~\cite{guennebaud_eigen_2010}.
There are even libraries for very specific kernels and format combinations, such as SPLATT~\cite{smith_splatt:_2015} (MTTKRP on CSF).
Several of these libraries also feature some graph or mesh algorithms built on sparse matrices.
The GraphBLAS~\cite{kepner_mathematical_2016} supports primitive semiring operations (operations beyond $(+, *)$, such as $(min, +)$ multiplication) which can be composed to enable graph algorithms, some of which are collected in LAGraph~\cite{mattson_lagraph_2019}.
Similarly, the MapReduce and Hadoop platforms support operations on indexed collections~\cite{dean_mapreduce_2008}, and have been used to support graph
algorithms in the GBASE library\cite{kang_gbase_2011}.
Several machine learning frameworks support some sparse tensors and operations, most notably TorchSparse\cite{tang_torchsparse_2022,tang_torchsparse_2023}.
\paragraph{Compilers for Dense Data:}
Outside of general purpose compilers, many compilers have been developed for optimizing dense data on a variety of control flow.
Perhaps the most well known example is Halide~\cite{ragan-kelley_halide_2013} and its various descendant such as TVM~\cite{chen_tvm_2018}, Exo~\cite{ikarashi_exocompilation_2022}, Elevate~\cite{hagedorn_achieving_2020}, and ATL~\cite{liu_verified_2022}.
These languages typically support most control flow except for an early break though some don't support arbitrary reading/writing or even indirect accesses.
Several polyhedral languages, such as Polly~\cite{grosser_pollyperforming_2012}, Tiramisu~\cite{baghdadi_tiramisu_2019}, CHiLL~\cite{chen_framework_2008}, Pluto~\cite{bondhugula_pluto_2008}, and AlphaZ~\cite{yuki_alphaz_2012} offer similar capabilities in terms of control flow though they often support more irregular regions that the polyhedral framework supports.
These are based on ISL~\cite{verdoolaege_isl_2010}.
The density of this research represents the density of support for dense computation.

\paragraph{ Compilers for Structured Data:}
Several compilers exist for several types of structured data, often featuring separate languages for the storage of the structured data and the computation.
The TACO compiler originally supported just plain Einsum computations~\cite{kjolstad_tensor_2017}, but has been extended several times to support (single dimensional) local tensors \cite{kjolstad_tensor_2019}, imperfectly nested loops \cite{dias_sparselnr_2022}, breaks via semi-rings~\cite{henry_compilation_2021}, windowing and tiling \cite{senanayake_sparse_2020}, and convolution~\cite{won_unified_2023}, and compilation in MLIR \cite{bik_compiler_2022}, all as separate extensions.
Similarly, TACO originally support just dense and CSF like N dimensional structures, but was extended independently to support COO like structures~\cite{chou_format_2018}, and tree like structures~\cite{chou_compilation_2022}, as separate extensions. SparseTIR is a similar system supporting combined 
sparse formats (including block structures) \cite{ye_sparsetir_2023}.
The SDQL language offers a similar level of control flow~\cite{shaikhha_functional_2022}, but only on sparse hash tables.
Similarly, SDQL has been extended with a system that allows one to specify formats as queries on a set of base storage types~\cite{schleich_optimizing_2023} and separately by another system that describes static symmetries and other structures as predicates~\cite{ghorbani_compiling_2023}.
%
The Taichi language focuses on a single sparse data structure made from dense blocks, bit-masks, and pointers~\cite{hu_taichi_2019}.
%
%
The sparse polyhedral framework builds on CHiLL for the purpose of generating inspector/executor optimizations~\cite{strout_sparse_2018} though the branch of this work that specifies sparse formats separately from the computation (otherwise they are inlined into the computation manually) seems to apply mainly to Einsums~\cite{zhao_polyhedral_2022}.
Second to last, SQL's classical physical/logical distinction is the classic program/format distinction, and SQL supports a huge variety of control flow constructs~\cite{kotlyar_relational_1997,date_guide_1989}.
However, many SQL or dataframe systems rely on b-trees, columnar, or hash tables, with only a few systems, such as Vectorwise~\cite{boncz_vectorwise_2012}, LaraDB~\cite{hutchison_laradb_2017}, GMAP~\cite{tsatalos_gmap_1996}, or SciDB~\cite{stonebraker_scidb_2013} building physical layouts with other constructs based in tensor programming.
However, tensor based databases are a new focus given the rise of mixed ML/DB pipelines~\cite{baumann_array_2021,luo_scalable_2018}.
Lastly, SPIRAL focuses on recursively defined datastructures and recursively define linear algebra, and can therefore express a structure and computation that none of the systems mentioned above can: a Cooley–Tukey FFT ~\cite{franchetti_spiral_2018,franchetti_operator_2009}.
%

\paragraph{Other Architectures:} 

Sparse compilers have been extended to many architectures. An extension of TACO supports GPU~\cite{senanayake_sparse_2020}, Cyclops~\cite{solomonik_cyclops_2013,solomonik_sparse_2015} and SPDistal~\cite{yadav_spdistal_2022} support distributed memory, and the Sparse Abstract Machine~\cite{hsu_sparse_2023} supports custom hardware.
We believe that supporting control flow is the first step towards architectural support beyond unstructured sparsity.

%




\section{Conclusion}
Finch automatically specializes flexible control flow to diverse data structures,
facilitating productive algorithmic exploration, flexible tensor programming, and
efficient high-level interfaces for a wider variety of applications than ever before.
\begin{acks}
  Intel and NSF PPoSS Grant CCF-2217064; DARPA PROWESS Award HR0011-23-C-0101; NSF SHF Grant CCF-2107244; DoE PSAAP Center DE-NA0003965; DARPA SBIR HR001123C0139
\end{acks}

\clearpage
\section{Data Availability Statement}

The Finch compiler and the benchmarks used to construct this paper are both
currently available as open-source software, and we plan to make them available
as an artifact to the OOPSLA Artifact Evaluation Committee. We will provide
instructions to replicate all results of the paper.

\bibliographystyle{ACM-Reference-Format}
\bibliography{FinchOOPSLAWillow.bib}

\appendix
\section{Artifact Appendix}

\subsection{Abstract}

In this artifact, we provide an archive of the Finch compiler at the time of
writing and instructions to replicate all benchmarks in this paper.  We note
that the Finch compiler is separately available as open-source software.  We
claim that the results in this paper are reproducible with the provided artifact
on an identical machine. Some results, especially those regarding MKL, may be
architecture-specific.

\subsection{Artifact check-list (meta-information)}

{\em Obligatory. Use just a few informal keywords in all fields applicable to your artifacts
and remove the rest. This information is needed to find appropriate reviewers and gradually 
unify artifact meta information in Digital Libraries.}

{\small
\begin{itemize}
  \item {\bf Algorithm: } Sparse Tensors, Compilers, Image Processing, Scientific Computing, Graph Analytics
  \item {\bf Program: } Julia, C++, Python
  \item {\bf Compilation: } Makefile, CMake, gcc, Julia, LLVM
  \item {\bf Transformations: } Sparsity and Structural Specialization
  \item {\bf Binary: } x86-64, also ARM
  \item {\bf Data set: } Synthetic, SuiteSparse, MNIST, Omniglot, HumanSketches
  \item {\bf Run-time environment: } Ubuntu 22.04.5 LTS, Linux 5.15.0-119-generic, Root Access
  \item {\bf Hardware: } 12-core 2-socket Intel Xeon CPU E5-2680 v3 running at 2.50GHz with 128GB of memory.
  \item {\bf Run-time state: } Sensitive to memory bandwidth, cache size
  \item {\bf Execution: } Requires exclusive access to node for repeatable results
  \item {\bf Metrics: } Execution time, Speedup
  \item {\bf Output: } JSON table, PNG plot
  \item {\bf Experiments: } SpMV, SpGEMM, Erosion, Histogram, Breadth-First Search, Shortest Path
  \item {\bf How much disk space required (approximately)?: }
  \item {\bf How much time is needed to prepare workflow (approximately)?: }
  \item {\bf How much time is needed to complete experiments (approximately)?: }
  \item {\bf Publicly available?: } yes
  \item {\bf Code licenses (if publicly available)?: } MIT
  \item {\bf Data licenses (if publicly available)?: } MIT
  \item {\bf Workflow framework used?: }  SLURM, Shell
  \item {\bf Archived (provide DOI)?: } 10.5281/zenodo.14597754
\end{itemize}}

\subsection{Description}

\subsubsection{How delivered}
The artifact may be downloaded from zeonodo at \url{doi.org/10.5281/zenodo.14597754}, or cloned from the \texttt{oopsla-25-artifact} branch of the FinchBenchmarks repository on GitHub at
\url{https://github.com/finch-tensor/FinchBenchmarks}. The artifact contains a
copy of the Finch.jl compiler version v1.1.0, and all benchmarks used in the
paper. The artifact takes 1.6 MB to download, and 17 GB of disk space once it
has been extracted and built and datasets have been downloaded and generated.

The \verb|deps| subdirectory contains major dependencies required. The
\verb|spmv|, \verb|spgemm|, \verb|graphs|, and \verb|images| directories contain
the benchmarks corresponding to the SpMV, SpGEMM, Graphs, and Image Processing
sections of the paper, respectively.

\subsubsection{Hardware dependencies}

This artifact was run on a 12-core 2-socket Intel Xeon CPU E5-2680 v3 running at 2.50GHz with 128GB of memory.
The Intel MKL and CORA benchmarks require an x86-64 machine to build and run,
but we believe that the other benchmarks can be built on ARM hardware.

\subsubsection{Software dependencies}

The results in the paper concern the following software dependencies, and special notes for building are included. Although we include the sources of the artifact, the artifact itself contains these repositories as submodules.

\begin{enumerate}
  \item Julia \cite{bezanson_julia:_2017} v1.10.7
    Julia can be installed via `juliaup` at \url{https://github.com/JuliaLang/juliaup} or at \url{https://julialang.org/downloads/}.
  \item Finch 1.1.0
    Finch is a registered Julia package and will be installed automatically during setup.
    However, if for whatever reason you would like to use the copy included with the artifact, you may run
    {\tiny\begin{verbatim}
      julia --project=. -e 'using Pkg; Pkg.develop(PackageSpec(path="./deps/Finch.jl"))'\end{verbatim}}
    from the root of the artifact.
  \item TACO \cite{kjolstad_tensor_2017} at commit \texttt{1278503a1} from \url{https://github.com/tensor-compiler/taco}, corresponding to the ``benchmark'' branch.
    Taco requires CMake.
  \item Eigen 3.4.0 \cite{guennebaud_eigen_2010} from \url{https://gitlab.com/libeigen/eigen.git}
  \item GraphBLAS 9.4.2 from \url{https://github.com/DrTimothyAldenDavis/GraphBLAS}.
  \item LAGraph 1.1.4 from \url{https://github.com/GraphBLAS/LAGraph}.
  \item Graphs.jl 1.9 from \url{https://github.com/JuliaGraphs/Graphs.jl}.
  \item Intel MKL 2024.2 \cite{noauthor_developer_2024}, available from \url{https://www.intel.com/content/www/us/en/developer/tools/oneapi/onemkl-download.html}.
    This will need to be installed to the \verb|deps/intel| folder.
  \item The Cora tensor compiler \cite{fegade_cora_2022} at commit \texttt{8e7de1d7c} from \url{https://github.com/pratikfegade/cora.git}.
    An artifact is also available at \url{https://doi.org/10.5281/zenodo.6326456}.
    Cora requires MKL, LLVM 9.0.0, Z3 4.8.8, CMake, and Python 3.
    Instructions for building are available in the \verb|cora/ae_appendix_supplement.pdf| file.
\end{enumerate}

Our build process for all comparison frameworks is automated in our Makefile, included at the root of the directory.
The \verb|Project.toml| file contains all of the required Julia dependecies (as well as their versions).
The \verb|pyproject.toml| file contains all of the required python dependecies (as well as their versions).

We built our artifact on Ubuntu 22.04.5 LTS, Linux 5.15.0-119-generic, using the following dependencies:
\begin{enumerate}
  \item cmake 3.22.1
  \item gcc 11.4.0
  \item Python 3.10.12
  \item We used poetry 1.8.5 to manage python dependencies, which can be installed with \texttt{pip} or following \url{https://python-poetry.org/docs/#installation}.
  \item jq 1.6, git 2.34.1, curl 7.81.0, GNU tar 1.34, and UnZip 6.00
\end{enumerate}
  
\subsubsection{Data sets}

\begin{enumerate}
  \item SuiteSparse \\
    The SuiteSparse Matrix Collection \cite{davis_university_2011} is available at \url{https://sparse.tamu.edu/}.
    We use the MatrixDepot.jl package at \url{https://github.com/JuliaLinearAlgebra/MatrixDepot.jl} to download matrices from this collection.
    The precise datasets used for each benchmark are listed in the test harnesses.
  \item MNIST \\
    The MNIST dataset \cite{lecun_gradient-based_1998} is available at \url{http://yann.lecun.com/exdb/mnist/}.
    We use the MLDatasets.jl package at \url{https://github.com/JuliaML/MLDatasets.jl} to download this dataset.
  \item Omniglot \\
    The Omniglot dataset \cite{lake_human-level_2015} is available at \url{https://www.omniglot.com/}.
    We use the MLDatasets.jl package at \url{https://github.com/JuliaML/MLDatasets.jl} to download this dataset.
  \item HumanSketches \\
    We evaluate on a dataset of human line drawings \cite{eitz_how_2012}, available at \url{https://cybertron.cg.tu-berlin.de/eitz/projects/classifysketch/sketches_png.zip}. 
  \item Dip3masks \\
    We also hand-selected a subset of mask images from a digital image processing textbook \cite{gonzalez_digital_2006}.
    The precise set of images used is included with the artifact.
  \item Synthetic Data \\
    Scripts are included to generate synthetic data.
    For spmv, we generate banded, triangular, and a reverse permutation matrix.
    For SpGEMM, we generate a series of increasingly larger uniformly random sparse matrices.
    For the Graphs dataset, we generate a few RMAT \cite{chakrabarti_r-mat_2004} graphs to match the dataset used by Yang et. al. \cite{yang_implementing_2018}.
\end{enumerate}

\subsection{Installation}

\subsubsection{1. Install system dependecies}
  Install Julia, Python, CMake, and other system dependencies as described in the software dependencies section.

\subsubsection{2. Download the core dependencies}

Several dependencies are included as submodules in the \verb|deps/| folder,
referencing the precise commits we used to build the artifact.
Most of these can be installed via \begin{verbatim}
  git submodule update --init --recursive
\end{verbatim}

MKL must be manually installed to the \verb|deps/intel| folder. \url{Install Intel MKL version 2024.2, available from https://www.intel.com/content/www/us/en/developer/tools/oneapi/onemkl-download.html}. You'll need to request an academic license on the website, then download an offline installer. There should be instructions on the website for how to run the install script. When asked, you can install to the \verb|deps/intel| folder.

The makefile also contains instructions to clone the submodules.

\subsubsection{3. Build the dependencies and benchmarks}

The makefile contains targets to build all of the benchmarks and dependencies.
We refer to each individual dependency for more detailed instructions.
We expect that some system-specific adjustments to the makefile may be necessary to build CORA, as it has many complex subdependencies such as LLVM and Z3.

When all dependencies have been successfully installed, from the root of the artifact, run
\begin{verbatim}
  make
\end{verbatim}

\subsubsection{4. Instantiate Runtime Environments}

In this step, we will setup and install the Julia and Python environments. This can be achieved by running from the root of the artifact:
\begin{verbatim}
  bash instantiate_environments.sh
\end{verbatim}
which simply runs the following commands:
\begin{verbatim}
  julia --project=. -e 'using Pkg; Pkg.instantiate(); Pkg.precompile()'
  poetry install --no-root
\end{verbatim}

\subsection{Experiment workflow}

\subsubsection{1. Running the dataset generators}

To generate the synthetic data used in the benchmarks, run
\begin{verbatim}
  bash generate_data.sh
\end{verbatim}

\subsubsection{2. Running the benchmarks}

To run all benchmarks, run
\begin{verbatim}
  bash run_benchmarks.sh
\end{verbatim}

This script will run all benchmarks and generate the JSON tables and PNG plots used in the paper.

A more fine-grained approach may be taken to run the benchmarks individually.
Each experiment subdirectory contains a \verb|run_*.sh| script that will run that
particular benchmark, and the commands contained within may be modified to run
different subsets of experiments. We expect that running the whole set of
experiments may take in excess of 24 hours, so some experiments may be commented
out. Additionally, if evaluators have access to a SLURM cluster, we include
SLURM scripts that help accelerate the process, which may need adaptation to
your particular cluster. It is convenient to use \verb|jq| to combine json
outputs from parallel runs, for example,
\begin{verbatim}
  jq -s 'add' results_*.json > combined_results.json
\end{verbatim}

\subsubsection{3. Plotting the output}

To generate the plots used for each experiment, run the corresponding chart.py
script in each experiment directory. For example, to generate the plots for the
SpMV experiment, run from within the \verb|spmv| directory

\begin{verbatim}
  poetry run python chart.py
\end{verbatim}

Plots will be generated in the corresponding \verb|charts/| directory of each experiment.

\subsection{Evaluation and expected result}

Reference JSON and PNG plot results are stored for each experiement with the
prefix \verb|reference_|.  Evaluators can compare the reference plots with the
result plots. To generate our geomean speedup claims, evaluators may run
\begin{verbatim}
  poetry run get_geomean.py
\end{verbatim}
in the corresponding experiment directory and compare to the text.

\subsection{Reusability Guide}

All of the julia run scripts support a \verb|--help| flag describing parameters which
allowing one to customize the experiments.
Separately, the Finch compiler is a featureful Julia package, and the evaluators
are encouraged to experiment with the compiler using the
documentation at \url{https://finch-tensor.github.io/Finch.jl/stable/} and \url{https://finch-tensor.github.io/Finch.jl/stable/docs/language/calling_finch/}.
Reviewers may also build the documentation locally by running
\verb|julia docs/make.jl|
from the root of the \verb|Finch.jl| directory. Some example Finch
programs are given in the \verb|docs/examples| directory.
%
%

\section{Graph Algorithm Listings}\label{sec:graph_listings}

\subsection{Finch Breadth-First Search}
\begin{minted}{julia}
    function bfs_finch_kernel(edges, edgesT, source=5, alpha = 0.01)
    (n, m) = size(edges)
    edges = pattern!(edges)
    @assert n == m
    F = Tensor(SparseByteMap(Pattern()), n)
    _F = Tensor(SparseByteMap(Pattern()), n)
    @finch F[source] = true
    F_nnz = 1 
    V = Tensor(Dense(Element(false)), n)
    @finch V[source] = true
    P = Tensor(Dense(Element(0)), n)
    @finch P[source] = source
    while F_nnz > 0 
        if F_nnz/m > alpha # pull
            p = ShortCircuitScalar{0}()
            _F .= false
            for k=_ 
                if !V[k]
                    p .= 0
                    for j=_ 
                        if F[follow(j)] && AT[j, k]
                            p[] <<choose(0)>>= j
                        end 
                    end 
                    if p[] != 0
                        _F[k] |= true
                        P[k] = p[] 
                    end 
                end 
            end
        else # push
            _F .= false
            for j=_, k=_ 
                if F[j] && A[k, j] && !(V[k])
                    _F[k] |= true
                    P[k] <<choose(0)>>= j
                end 
            end 
        end 
        c = Scalar(0)
        @finch begin
            for k=_ 
                let _f = _F[k]
                    V[k] |= _f
                    c[] += _f
                end 
            end 
        end 
        (F, _F) = (_F, F)
        F_nnz = c[] 
    end 
    return P
end
\end{minted}
\subsection{GraphBLAS Breadth-First Search}
\begin{minted}{c}
//------------------------------------------------------------------------------
// LAGr_BreadthFirstSearch:  breadth-first search dispatch
//------------------------------------------------------------------------------

// LAGraph, (c) 2019-2022 by The LAGraph Contributors, All Rights Reserved.
// SPDX-License-Identifier: BSD-2-Clause
//
// For additional details (including references to third party source code and
// other files) see the LICENSE file or contact permission@sei.cmu.edu. See
// Contributors.txt for a full list of contributors. Created, in part, with
// funding and support from the U.S. Government (see Acknowledgments.txt file).
// DM22-0790

// Contributed by Scott McMillan, SEI Carnegie Mellon University

//------------------------------------------------------------------------------

// Breadth-first-search via push/pull method if using SuiteSparse:GraphBLAS
// and its GxB extensions, or a push-only method otherwise.  The former is
// much faster.

// This is an Advanced algorithm.  SuiteSparse can use a push/pull method if
// G->AT and G->out_degree are provided.  G->AT is not required if G is
// undirected.  The vanilla method is always push-only.

#include "LG_alg_internal.h"

int LAGr_BreadthFirstSearch
(
    // output:
    GrB_Vector *level,
    GrB_Vector *parent,
    // input:
    const LAGraph_Graph G,
    GrB_Index src,
    char *msg
)
{

#if LAGRAPH_SUITESPARSE
    return LG_BreadthFirstSearch_SSGrB   (level, parent, G, src, msg) ;
#else
    return LG_BreadthFirstSearch_vanilla (level, parent, G, src, msg) ;
#endif
}

//------------------------------------------------------------------------------
// LG_BreadthFirstSearch_SSGrB:  BFS using Suitesparse extensions
//------------------------------------------------------------------------------

// LAGraph, (c) 2019-2022 by The LAGraph Contributors, All Rights Reserved.
// SPDX-License-Identifier: BSD-2-Clause
//
// For additional details (including references to third party source code and
// other files) see the LICENSE file or contact permission@sei.cmu.edu. See
// Contributors.txt for a full list of contributors. Created, in part, with
// funding and support from the U.S. Government (see Acknowledgments.txt file).
// DM22-0790

// Contributed by Timothy A. Davis, Texas A&M University

//------------------------------------------------------------------------------

// This is an Advanced algorithm.  G->AT and G->out_degree are required for
// this method to use push-pull optimization.  If not provided, this method
// defaults to a push-only algorithm, which can be slower.  This is not
// user-callable (see LAGr_BreadthFirstSearch instead).  G->AT and
// G->out_degree are not computed if not present.

// References:
//
// Carl Yang, Aydin Buluc, and John D. Owens. 2018. Implementing Push-Pull
// Efficiently in GraphBLAS. In Proceedings of the 47th International
// Conference on Parallel Processing (ICPP 2018). ACM, New York, NY, USA,
// Article 89, 11 pages. DOI: https://doi.org/10.1145/3225058.3225122
//
// Scott Beamer, Krste Asanovic and David A. Patterson, The GAP Benchmark
// Suite, http://arxiv.org/abs/1508.03619, 2015.  http://gap.cs.berkeley.edu/

// revised by Tim Davis (davis@tamu.edu), Texas A&M University

#define LG_FREE_WORK        \
{                           \
    GrB_free (&w) ;         \
    GrB_free (&q) ;         \
}

#define LG_FREE_ALL         \
{                           \
    LG_FREE_WORK ;          \
    GrB_free (&pi) ;        \
    GrB_free (&v) ;         \
}

#include "LG_internal.h"

int LG_BreadthFirstSearch_SSGrB
(
    GrB_Vector *level,
    GrB_Vector *parent,
    const LAGraph_Graph G,
    GrB_Index src,
    char *msg
)
{

    //--------------------------------------------------------------------------
    // check inputs
    //--------------------------------------------------------------------------

    LG_CLEAR_MSG ;
    GrB_Vector q = NULL ;           // the current frontier
    GrB_Vector w = NULL ;           // to compute work remaining
    GrB_Vector pi = NULL ;          // parent vector
    GrB_Vector v = NULL ;           // level vector

#if !LAGRAPH_SUITESPARSE
    LG_ASSERT (false, GrB_NOT_IMPLEMENTED) ;
#else

    bool compute_level  = (level != NULL) ;
    bool compute_parent = (parent != NULL) ;
    if (compute_level ) (*level ) = NULL ;
    if (compute_parent) (*parent) = NULL ;
    LG_ASSERT_MSG (compute_level || compute_parent, GrB_NULL_POINTER,
        "either level or parent must be non-NULL") ;

    LG_TRY (LAGraph_CheckGraph (G, msg)) ;

    //--------------------------------------------------------------------------
    // get the problem size and cached properties
    //--------------------------------------------------------------------------

    GrB_Matrix A = G->A ;

    GrB_Index n, nvals ;
    GRB_TRY (GrB_Matrix_nrows (&n, A)) ;
    LG_ASSERT_MSG (src < n, GrB_INVALID_INDEX, "invalid source node") ;

    GRB_TRY (GrB_Matrix_nvals (&nvals, A)) ;

    GrB_Matrix AT = NULL ;
    GrB_Vector Degree = G->out_degree ;
    if (G->kind == LAGraph_ADJACENCY_UNDIRECTED ||
       (G->kind == LAGraph_ADJACENCY_DIRECTED &&
        G->is_symmetric_structure == LAGraph_TRUE))
    {
        // AT and A have the same structure and can be used in both directions
        AT = G->A ;
    }
    else
    {
        // AT = A' is different from A.  If G->AT is NULL, then a push-only
        // method is used.
        AT = G->AT ;
    }

    // direction-optimization requires G->AT (if G is directed) and
    // G->out_degree (for both undirected and directed cases)
    bool push_pull = (Degree != NULL && AT != NULL) ;

    // determine the semiring type
    GrB_Type int_type = (n > INT32_MAX) ? GrB_INT64 : GrB_INT32 ;
    GrB_Semiring semiring ;

    if (compute_parent)
    {
        // use the ANY_SECONDI_INT* semiring: either 32 or 64-bit depending on
        // the # of nodes in the graph.
        semiring = (n > INT32_MAX) ?
            GxB_ANY_SECONDI_INT64 : GxB_ANY_SECONDI_INT32 ;

        // create the parent vector.  pi(i) is the parent id of node i
        GRB_TRY (GrB_Vector_new (&pi, int_type, n)) ;
        GRB_TRY (GxB_set (pi, GxB_SPARSITY_CONTROL, GxB_BITMAP + GxB_FULL)) ;
        // pi (src) = src denotes the root of the BFS tree
        GRB_TRY (GrB_Vector_setElement (pi, src, src)) ;

        // create a sparse integer vector q, and set q(src) = src
        GRB_TRY (GrB_Vector_new (&q, int_type, n)) ;
        GRB_TRY (GrB_Vector_setElement (q, src, src)) ;
    }
    else
    {
        // only the level is needed, use the LAGraph_any_one_bool semiring
        semiring = LAGraph_any_one_bool ;

        // create a sparse boolean vector q, and set q(src) = true
        GRB_TRY (GrB_Vector_new (&q, GrB_BOOL, n)) ;
        GRB_TRY (GrB_Vector_setElement (q, true, src)) ;
    }

    if (compute_level)
    {
        // create the level vector. v(i) is the level of node i
        // v (src) = 0 denotes the source node
        GRB_TRY (GrB_Vector_new (&v, int_type, n)) ;
        GRB_TRY (GxB_set (v, GxB_SPARSITY_CONTROL, GxB_BITMAP + GxB_FULL)) ;
        GRB_TRY (GrB_Vector_setElement (v, 0, src)) ;
    }

    // workspace for computing work remaining
    GRB_TRY (GrB_Vector_new (&w, GrB_INT64, n)) ;

    GrB_Index nq = 1 ;          // number of nodes in the current level
    double alpha = 8.0 ;
    double beta1 = 8.0 ;
    double beta2 = 512.0 ;
    int64_t n_over_beta1 = (int64_t) (((double) n) / beta1) ;
    int64_t n_over_beta2 = (int64_t) (((double) n) / beta2) ;

    //--------------------------------------------------------------------------
    // BFS traversal and label the nodes
    //--------------------------------------------------------------------------

    bool do_push = true ;       // start with push
    GrB_Index last_nq = 0 ;
    int64_t edges_unexplored = nvals ;
    bool any_pull = false ;     // true if any pull phase has been done

    // {!mask} is the set of unvisited nodes
    GrB_Vector mask = (compute_parent) ? pi : v ;

    for (int64_t nvisited = 1, k = 1 ; nvisited < n ; nvisited += nq, k++)
    {

        //----------------------------------------------------------------------
        // select push vs pull
        //----------------------------------------------------------------------

        if (push_pull)
        {
            if (do_push)
            {
                // check for switch from push to pull
                bool growing = nq > last_nq ;
                bool switch_to_pull = false ;
                if (edges_unexplored < n)
                {
                    // very little of the graph is left; disable the pull
                    push_pull = false ;
                }
                else if (any_pull)
                {
                    // once any pull phase has been done, the # of edges in the
                    // frontier has no longer been tracked.  But now the BFS
                    // has switched back to push, and we're checking for yet
                    // another switch to pull.  This switch is unlikely, so
                    // just keep track of the size of the frontier, and switch
                    // if it starts growing again and is getting big.
                    switch_to_pull = (growing && nq > n_over_beta1) ;
                }
                else
                {
                    // update the # of unexplored edges
                    // w<q>=Degree
                    // w(i) = outdegree of node i if node i is in the queue
                    GRB_TRY (GrB_assign (w, q, NULL, Degree, GrB_ALL, n,
                        GrB_DESC_RS)) ;
                    // edges_in_frontier = sum (w) = # of edges incident on all
                    // nodes in the current frontier
                    int64_t edges_in_frontier = 0 ;
                    GRB_TRY (GrB_reduce (&edges_in_frontier, NULL,
                        GrB_PLUS_MONOID_INT64, w, NULL)) ;
                    edges_unexplored -= edges_in_frontier ;
                    switch_to_pull = growing &&
                        (edges_in_frontier > (edges_unexplored / alpha)) ;
                }
                if (switch_to_pull)
                {
                    // switch from push to pull
                    do_push = false ;
                }
            }
            else
            {
                // check for switch from pull to push
                bool shrinking = nq < last_nq ;
                if (shrinking && (nq <= n_over_beta2))
                {
                    // switch from pull to push
                    do_push = true ;
                }
            }
            any_pull = any_pull || (!do_push) ;
        }

        //----------------------------------------------------------------------
        // q = kth level of the BFS
        //----------------------------------------------------------------------

        int sparsity = do_push ? GxB_SPARSE : GxB_BITMAP ;
        GRB_TRY (GxB_set (q, GxB_SPARSITY_CONTROL, sparsity)) ;

        // mask is pi if computing parent, v if computing just level
        if (do_push)
        {
            // push (saxpy-based vxm):  q'{!mask} = q'*A
            GRB_TRY (GrB_vxm (q, mask, NULL, semiring, q, A, GrB_DESC_RSC)) ;
        }
        else
        {
            // pull (dot-product-based mxv):  q{!mask} = AT*q
            GRB_TRY (GrB_mxv (q, mask, NULL, semiring, AT, q, GrB_DESC_RSC)) ;
        }

        //----------------------------------------------------------------------
        // done if q is empty
        //----------------------------------------------------------------------

        last_nq = nq ;
        GRB_TRY (GrB_Vector_nvals (&nq, q)) ;
        if (nq == 0)
        {
            break ;
        }

        //----------------------------------------------------------------------
        // assign parents/levels
        //----------------------------------------------------------------------

        if (compute_parent)
        {
            // q(i) currently contains the parent id of node i in tree.
            // pi{q} = q
            GRB_TRY (GrB_assign (pi, q, NULL, q, GrB_ALL, n, GrB_DESC_S)) ;
        }
        if (compute_level)
        {
            // v{q} = k, the kth level of the BFS
            GRB_TRY (GrB_assign (v, q, NULL, k, GrB_ALL, n, GrB_DESC_S)) ;
        }
    }

    //--------------------------------------------------------------------------
    // free workspace and return result
    //--------------------------------------------------------------------------

    if (compute_parent) (*parent) = pi ;
    if (compute_level ) (*level ) = v ;
    LG_FREE_WORK ;
    return (GrB_SUCCESS) ;
#endif
}
\end{minted}
\subsection{Finch Bellman-Ford}
\begin{minted}{julia}
function bellmanford_finch_kernel(edges, source=1)
    (n, m) = size(edges)
    @assert n == m
    dists_prev = Tensor(Dense(Element(Inf)), n)
    dists_prev[source] = 0 
    dists = Tensor(Dense(Element(Inf)), n)
    active_prev = Tensor(SparseByteMap(Pattern()), n)
    active_prev[source] = true
    active = Tensor(SparseByteMap(Pattern()), n)
    parents = Tensor(Dense(Element(0)), n)
    for iter = 1:n  
        @finch begin
            for j=_ 
                if active_prev[j]
                    dists[j] <<min>>= dists_prev[j]
                end 
            end 
        end 
        @finch begin
            active .= false
            for j = _ 
                if active_prev[j]
                    for i = _ 
                        let d = dists_prev[j] + edges[i, j]
                            dists[i] <<min>>= d
                            active[i] |= d < dists_prev[i]
                        end 
                    end 
                end 
            end 
        end 
        if countstored(active) == 0
            break
        end 
        dists_prev, dists = dists, dists_prev
        active_prev, active = active, active_prev
    end 
    @finch begin
        for j = _ 
            for i = _ 
                let d = edges[i, j]
                    if d < Inf && dists[j] + d <= dists[i]
                        parents[i] <<choose(0)>>= j
                    end 
                end 
            end 
        end 
    end 
    return (dists=dists, parents=parents)
end
\end{minted}

\subsection{GraphBLAS Bellman-Ford}
\begin{minted}{c}
//------------------------------------------------------------------------------
// LAGraph_BF_full1a.c: Bellman-Ford single-source shortest paths, returns tree,
// while diagonal of input matrix A needs not to be explicit 0
//------------------------------------------------------------------------------

// LAGraph, (c) 2019-2022 by The LAGraph Contributors, All Rights Reserved.
// SPDX-License-Identifier: BSD-2-Clause
//
// For additional details (including references to third party source code and
// other files) see the LICENSE file or contact permission@sei.cmu.edu. See
// Contributors.txt for a full list of contributors. Created, in part, with
// funding and support from the U.S. Government (see Acknowledgments.txt file).
// DM22-0790

// Contributed by Jinhao Chen and Timothy A. Davis, Texas A&M University

//------------------------------------------------------------------------------

// This is the fastest variant that computes both the parent & the path length.

// LAGraph_BF_full1a: Bellman-Ford single source shortest paths, returning both
// the path lengths and the shortest-path tree.

// LAGraph_BF_full performs a Bellman-Ford to find out shortest path, parent
// nodes along the path and the hops (number of edges) in the path from given
// source vertex s in the range of [0, n) on graph given as matrix A with size
// n*n. The sparse matrix A has entry A(i, j) if there is an edge from vertex i
// to vertex j with weight w, then A(i, j) = w.

// LAGraph_BF_full1a returns GrB_SUCCESS if it succeeds.  In this case, there
// are no negative-weight cycles in the graph, and d, pi, and h are returned.
// The vector d has d(k) as the shortest distance from s to k. pi(k) = p+1,
// where p is the parent node of k-th node in the shortest path. In particular,
// pi(s) = 0. h(k) = hop(s, k), the number of edges from s to k in the shortest
// path.

// If the graph has a negative-weight cycle, GrB_NO_VALUE is returned, and the
// GrB_Vectors d(k), pi(k) and h(k)  (i.e., *pd_output, *ppi_output and
// *ph_output respectively) will be NULL when negative-weight cycle detected.

// Otherwise, other errors such as GrB_OUT_OF_MEMORY, GrB_INVALID_OBJECT, and
// so on, can be returned, if these errors are found by the underlying
// GrB_* functions.

//------------------------------------------------------------------------------

#define LG_FREE_WORK                   \
{                                      \
    GrB_free(&d);                      \
    GrB_free(&dmasked);                \
    GrB_free(&dless);                  \
    GrB_free(&Atmp);                   \
    GrB_free(&BF_Tuple3);              \
    GrB_free(&BF_lMIN_Tuple3);         \
    GrB_free(&BF_PLUSrhs_Tuple3);      \
    GrB_free(&BF_LT_Tuple3);           \
    GrB_free(&BF_lMIN_Tuple3_Monoid);  \
    GrB_free(&BF_lMIN_PLUSrhs_Tuple3); \
    LAGraph_Free ((void**)&I, NULL);   \
    LAGraph_Free ((void**)&J, NULL);   \
    LAGraph_Free ((void**)&w, NULL);   \
    LAGraph_Free ((void**)&W, NULL);   \
    LAGraph_Free ((void**)&h, NULL);   \
    LAGraph_Free ((void**)&pi, NULL);  \
}

#define LG_FREE_ALL                    \
{                                      \
    LG_FREE_WORK ;                     \
    GrB_free (pd_output);              \
    GrB_free (ppi_output);             \
    GrB_free (ph_output);              \
}

#include <LAGraph.h>
#include <LAGraphX.h>
#include <LG_internal.h>  // from src/utility

typedef void (*LAGraph_binary_function) (void *, const void *, const void *) ;

//------------------------------------------------------------------------------
// data type for each entry of the adjacent matrix A and "distance" vector d;
// <INFINITY,INFINITY,INFINITY> corresponds to nonexistence of a path, and
// the value  <0, 0, NULL> corresponds to a path from a vertex to itself
//------------------------------------------------------------------------------

typedef struct
{
    double w;    // w  corresponds to a path weight.
    GrB_Index h; // h  corresponds to a path size or number of hops.
    GrB_Index pi;// pi corresponds to the penultimate vertex along a path.
                 // vertex indexed as 1, 2, 3, ... , V, and pi = 0 (as nil)
                 // for u=v, and pi = UINT64_MAX (as inf) for (u,v) not in E
}
BF_Tuple3_struct;

//------------------------------------------------------------------------------
// binary functions, z=f(x,y), where Tuple3xTuple3 -> Tuple3
//------------------------------------------------------------------------------

void BF_lMIN3
(
    BF_Tuple3_struct *z,
    const BF_Tuple3_struct *x,
    const BF_Tuple3_struct *y
)
{
    if (x->w < y->w
        || (x->w == y->w && x->h < y->h)
        || (x->w == y->w && x->h == y->h && x->pi < y->pi))
    {
        if (z != x) { *z = *x; }
    }
    else
    {
        *z = *y;
    }
}

void BF_PLUSrhs3
(
    BF_Tuple3_struct *z,
    const BF_Tuple3_struct *x,
    const BF_Tuple3_struct *y
)
{
    z->w = x->w + y->w ;
    z->h = x->h + y->h ;
    z->pi = (x->pi != UINT64_MAX && y->pi != 0) ?  y->pi : x->pi ;
}

void BF_LT3
(
    bool *z,
    const BF_Tuple3_struct *x,
    const BF_Tuple3_struct *y
)
{
    (*z) = (x->w < y->w
        || (x->w == y->w && x->h < y->h)
        || (x->w == y->w && x->h == y->h && x->pi < y->pi)) ;
}

// Given a n-by-n adjacency matrix A and a source vertex s.
// If there is no negative-weight cycle reachable from s, return the distances
// of shortest paths from s and parents along the paths as vector d. Otherwise,
// returns d=NULL if there is a negtive-weight cycle.
// pd_output is pointer to a GrB_Vector, where the i-th entry is d(s,i), the
//   sum of edges length in the shortest path
// ppi_output is pointer to a GrB_Vector, where the i-th entry is pi(i), the
//   parent of i-th vertex in the shortest path
// ph_output is pointer to a GrB_Vector, where the i-th entry is h(s,i), the
//   number of edges from s to i in the shortest path
// A has weights on corresponding entries of edges
// s is given index for source vertex
GrB_Info LAGraph_BF_full1a
(
    GrB_Vector *pd_output,      //the pointer to the vector of distance
    GrB_Vector *ppi_output,     //the pointer to the vector of parent
    GrB_Vector *ph_output,      //the pointer to the vector of hops
    const GrB_Matrix A,         //matrix for the graph
    const GrB_Index s           //given index of the source
)
{
    GrB_Info info;
    char *msg = NULL ;
    // tmp vector to store distance vector after n (i.e., V) loops
    GrB_Vector d = NULL, dmasked = NULL, dless = NULL;
    GrB_Matrix Atmp = NULL;
    GrB_Type BF_Tuple3;

    GrB_BinaryOp BF_lMIN_Tuple3;
    GrB_BinaryOp BF_PLUSrhs_Tuple3;
    GrB_BinaryOp BF_LT_Tuple3;

    GrB_Monoid BF_lMIN_Tuple3_Monoid;
    GrB_Semiring BF_lMIN_PLUSrhs_Tuple3;

    GrB_Index nrows, ncols, n, nz;  // n = # of row/col, nz = # of nnz in graph
    GrB_Index *I = NULL, *J = NULL; // for col/row indices of entries from A
    GrB_Index *h = NULL, *pi = NULL;
    double *w = NULL;
    BF_Tuple3_struct *W = NULL;

    if (pd_output  != NULL) *pd_output  = NULL;
    if (ppi_output != NULL) *ppi_output = NULL;
    if (ph_output  != NULL) *ph_output  = NULL;

    LG_ASSERT (A != NULL && pd_output != NULL &&
        ppi_output != NULL && ph_output != NULL, GrB_NULL_POINTER) ;

    GRB_TRY (GrB_Matrix_nrows (&nrows, A)) ;
    GRB_TRY (GrB_Matrix_ncols (&ncols, A)) ;
    GRB_TRY (GrB_Matrix_nvals (&nz, A));
    LG_ASSERT_MSG (nrows == ncols, -1002, "A must be square") ;
    n = nrows;
    LG_ASSERT_MSG (s < n, GrB_INVALID_INDEX, "invalid source node") ;

    //--------------------------------------------------------------------------
    // create all GrB_Type GrB_BinaryOp GrB_Monoid and GrB_Semiring
    //--------------------------------------------------------------------------
    // GrB_Type
    GRB_TRY (GrB_Type_new(&BF_Tuple3, sizeof(BF_Tuple3_struct)));

    // GrB_BinaryOp
    GRB_TRY (GrB_BinaryOp_new(&BF_LT_Tuple3,
        (LAGraph_binary_function) (&BF_LT3), GrB_BOOL, BF_Tuple3, BF_Tuple3));
    GRB_TRY (GrB_BinaryOp_new(&BF_lMIN_Tuple3,
        (LAGraph_binary_function) (&BF_lMIN3), BF_Tuple3, BF_Tuple3,BF_Tuple3));
    GRB_TRY (GrB_BinaryOp_new(&BF_PLUSrhs_Tuple3,
        (LAGraph_binary_function)(&BF_PLUSrhs3),
        BF_Tuple3, BF_Tuple3, BF_Tuple3));

    // GrB_Monoid
    BF_Tuple3_struct BF_identity = (BF_Tuple3_struct) { .w = INFINITY,
        .h = UINT64_MAX, .pi = UINT64_MAX };
    GRB_TRY (GrB_Monoid_new_UDT(&BF_lMIN_Tuple3_Monoid, BF_lMIN_Tuple3,
        &BF_identity));

    //GrB_Semiring
    GRB_TRY (GrB_Semiring_new(&BF_lMIN_PLUSrhs_Tuple3,
        BF_lMIN_Tuple3_Monoid, BF_PLUSrhs_Tuple3));

    //--------------------------------------------------------------------------
    // allocate arrays used for tuplets
    //--------------------------------------------------------------------------
#if 1
    LAGRAPH_TRY (LAGraph_Malloc ((void **) &I, nz, sizeof(GrB_Index), msg)) ;
    LAGRAPH_TRY (LAGraph_Malloc ((void **) &J, nz, sizeof(GrB_Index), msg)) ;
    LAGRAPH_TRY (LAGraph_Malloc ((void **) &w, nz, sizeof(double), msg)) ;
    LAGRAPH_TRY (LAGraph_Malloc ((void **) &W, nz, sizeof(BF_Tuple3_struct),
        msg)) ;

    //--------------------------------------------------------------------------
    // create matrix Atmp based on A, while its entries become BF_Tuple3 type
    //--------------------------------------------------------------------------

    GRB_TRY (GrB_Matrix_extractTuples_FP64(I, J, w, &nz, A));
    int nthreads, nthreads_outer, nthreads_inner ;
    LG_TRY (LAGraph_GetNumThreads (&nthreads_outer, &nthreads_inner, msg)) ;
    nthreads = nthreads_outer * nthreads_inner ;
    printf ("nthreads %d\n", nthreads) ;
    int64_t k;
    #pragma omp parallel for num_threads(nthreads) schedule(static)
    for (k = 0; k < nz; k++)
    {
        W[k] = (BF_Tuple3_struct) { .w = w[k], .h = 1, .pi = I[k] + 1 };
    }
    GRB_TRY (GrB_Matrix_new(&Atmp, BF_Tuple3, n, n));
    GRB_TRY (GrB_Matrix_build_UDT(Atmp, I, J, W, nz, BF_lMIN_Tuple3));
    LAGraph_Free ((void**)&I, NULL);
    LAGraph_Free ((void**)&J, NULL);
    LAGraph_Free ((void**)&W, NULL);
    LAGraph_Free ((void**)&w, NULL);

#else

    todo: GraphBLAS could use a new kind of unary operator, not z=f(x), but

    [z,flag] = f (aij, i, j, k, nrows, ncols, nvals, etc, ...)
    flag: keep or discard.  Combines GrB_apply and GxB_select.

    builtins:
        f(...) =
            i, bool is true
            j, bool is true
            i+j*nrows, etc.
            k
            tril, triu (like GxB_select): return aij, and true/false boolean

        z=f(x,i).  x: double, z:tuple3, i:GrB_Index with the row index of x
        // z = (BF_Tuple3_struct) { .w = x, .h = 1, .pi = i + 1 };

    GrB_apply (Atmp, op, A, ...)

    in the BFS, this is used:
        op:  z = f ( .... ) = i
        to replace x(i) with i

#endif

    //--------------------------------------------------------------------------
    // create and initialize "distance" vector d, dmasked and dless
    //--------------------------------------------------------------------------
    GRB_TRY (GrB_Vector_new(&d, BF_Tuple3, n));
    // make d dense
    GRB_TRY (GrB_Vector_assign_UDT(d, NULL, NULL, (void*)&BF_identity,
        GrB_ALL, n, NULL));
    // initial distance from s to itself
    BF_Tuple3_struct d0 = (BF_Tuple3_struct) { .w = 0, .h = 0, .pi = 0 };
    GRB_TRY (GrB_Vector_setElement_UDT(d, &d0, s));

    // creat dmasked as a sparse vector with only one entry at s
    GRB_TRY (GrB_Vector_new(&dmasked, BF_Tuple3, n));
    GRB_TRY (GrB_Vector_setElement_UDT(dmasked, &d0, s));

    // create dless
    GRB_TRY (GrB_Vector_new(&dless, GrB_BOOL, n));

    //--------------------------------------------------------------------------
    // start the Bellman Ford process
    //--------------------------------------------------------------------------
    bool any_dless= true;      // if there is any newly found shortest path
    int64_t iter = 0;          // number of iterations

    // terminate when no new path is found or more than V-1 loops
    while (any_dless && iter < n - 1)
    {
        // execute semiring on dmasked and A, and save the result to dmasked
        GRB_TRY (GrB_vxm(dmasked, GrB_NULL, GrB_NULL,
            BF_lMIN_PLUSrhs_Tuple3, dmasked, Atmp, GrB_NULL));

        // dless = d .< dtmp
        GRB_TRY (GrB_eWiseMult(dless, NULL, NULL, BF_LT_Tuple3, dmasked, d,
            NULL));

        // if there is no entry with smaller distance then all shortest paths
        // are found
        GRB_TRY (GrB_reduce (&any_dless, NULL, GrB_LOR_MONOID_BOOL, dless,
            NULL)) ;
        if(any_dless)
        {
            // update all entries with smaller distances
            //GRB_TRY (GrB_apply(d, dless, NULL, BF_Identity_Tuple3,
            //    dmasked, NULL));
            GRB_TRY (GrB_assign(d, dless, NULL, dmasked, GrB_ALL, n, NULL));

            // only use entries that were just updated
            //GRB_TRY (GrB_Vector_clear(dmasked));
            //GRB_TRY (GrB_apply(dmasked, dless, NULL, BF_Identity_Tuple3,
            //    d, NULL));
            //try:
            GRB_TRY (GrB_assign(dmasked, dless, NULL, d, GrB_ALL, n, GrB_DESC_R));
        }
        iter ++;
    }

    // check for negative-weight cycle only when there was a new path in the
    // last loop, otherwise, there can't be a negative-weight cycle.
    if (any_dless)
    {
        // execute semiring again to check for negative-weight cycle
        GRB_TRY (GrB_vxm(dmasked, GrB_NULL, GrB_NULL,
            BF_lMIN_PLUSrhs_Tuple3, dmasked, Atmp, GrB_NULL));

        // dless = d .< dtmp
        GRB_TRY (GrB_eWiseMult(dless, NULL, NULL, BF_LT_Tuple3, dmasked, d,
            NULL));

        // if there is no entry with smaller distance then all shortest paths
        // are found
        GRB_TRY (GrB_reduce (&any_dless, NULL, GrB_LOR_MONOID_BOOL, dless,
            NULL)) ;
        if(any_dless)
        {
            // printf("A negative-weight cycle found. \n");
            LG_FREE_ALL;
            return (GrB_NO_VALUE) ;
        }
    }

    //--------------------------------------------------------------------------
    // extract tuple from "distance" vector d and create GrB_Vectors for output
    //--------------------------------------------------------------------------

    LAGRAPH_TRY (LAGraph_Malloc ((void **) &I, n, sizeof(GrB_Index), msg)) ;
    LAGRAPH_TRY (LAGraph_Malloc ((void **) &W, n, sizeof(BF_Tuple3_struct),
        msg)) ;
    LAGRAPH_TRY (LAGraph_Malloc ((void **) &w, n, sizeof(double), msg)) ;
    LAGRAPH_TRY (LAGraph_Malloc ((void **) &h, n, sizeof(GrB_Index), msg)) ;
    LAGRAPH_TRY (LAGraph_Malloc ((void **) &pi, n, sizeof(GrB_Index), msg)) ;

    // todo: create 3 unary ops, and use GrB_apply?

    GRB_TRY (GrB_Vector_extractTuples_UDT (I, (void *) W, &n, d));

    for (k = 0; k < n; k++)
    {
        w [k] = W[k].w ;
        h [k] = W[k].h ;
        pi[k] = W[k].pi;
    }
    GRB_TRY (GrB_Vector_new(pd_output,  GrB_FP64,   n));
    GRB_TRY (GrB_Vector_new(ppi_output, GrB_UINT64, n));
    GRB_TRY (GrB_Vector_new(ph_output,  GrB_UINT64, n));
    GRB_TRY (GrB_Vector_build (*pd_output , I, w , n, GrB_MIN_FP64  ));
    GRB_TRY (GrB_Vector_build (*ppi_output, I, pi, n, GrB_MIN_UINT64));
    GRB_TRY (GrB_Vector_build (*ph_output , I, h , n, GrB_MIN_UINT64));
    LG_FREE_WORK;
    return (GrB_SUCCESS) ;
}

\end{minted}

\section{Mask Images}
We interpreted the following images from ``Digital Image Processing'' \cite{gonzalez_digital_2006} as masks:

\textbf{FigP1012.png}

\includegraphics[scale=0.20]{{./dip3e\_masks/FigP1012.png}}

\textbf{Fig1008\_a\_\_step\_edge\_.png}

\includegraphics[scale=0.20]{{./dip3e\_masks/Fig1008\_a\_\_step\_edge\_.png}}

\textbf{FigP0905\_d\_.png}

\includegraphics[scale=0.20]{{./dip3e\_masks/FigP0905\_d\_.png}}

\textbf{FigP0528\_c\_\_doughnut\_.png}

\includegraphics[scale=0.20]{{./dip3e\_masks/FigP0528\_c\_\_doughnut\_.png}}

\textbf{FigP0616\_b\_.png}

\includegraphics[scale=0.20]{{./dip3e\_masks/FigP0616\_b\_.png}}

\textbf{Fig0114\_c\_\_bottles\_.png}

\includegraphics[scale=0.20]{{./dip3e\_masks/Fig0114\_c\_\_bottles\_.png}}

\textbf{FigP0616\_c\_.png}

\includegraphics[scale=0.20]{{./dip3e\_masks/FigP0616\_c\_.png}}

\textbf{FigP0433\_b\_.png}

\includegraphics[scale=0.20]{{./dip3e\_masks/FigP0433\_b\_.png}}

\textbf{Figp0917.png}

\includegraphics[scale=0.20]{{./dip3e\_masks/Figp0917.png}}

\textbf{FigP0905\_b\_.png}

\includegraphics[scale=0.20]{{./dip3e\_masks/FigP0905\_b\_.png}}

\textbf{Fig1059\_c\_\_NegADI\_.png}

\includegraphics[scale=0.20]{{./dip3e\_masks/Fig1059\_c\_\_NegADI\_.png}}

\textbf{Fig1111\_a\_\_triangle\_.png}

\includegraphics[scale=0.20]{{./dip3e\_masks/Fig1111\_a\_\_triangle\_.png}}

\textbf{Fig1111\_b\_\_square\_.png}

\includegraphics[scale=0.20]{{./dip3e\_masks/Fig1111\_b\_\_square\_.png}}

\textbf{FigP0905\_top\_.png}

\includegraphics[scale=0.20]{{./dip3e\_masks/FigP0905\_top\_.png}}

\textbf{FigP1110.png}

\includegraphics[scale=0.20]{{./dip3e\_masks/FigP1110.png}}

\textbf{Fig0533\_a\_\_circle\_.png}

\includegraphics[scale=0.20]{{./dip3e\_masks/Fig0533\_a\_\_circle\_.png}}

\textbf{FigP0917\_noisy\_rectangle\_.png}

\includegraphics[scale=0.20]{{./dip3e\_masks/FigP0917\_noisy\_rectangle\_.png}}

\textbf{Fig0230\_b\_\_dental\_xray\_mask\_.png}

\includegraphics[scale=0.20]{{./dip3e\_masks/Fig0230\_b\_\_dental\_xray\_mask\_.png}}

\textbf{FigP0528\_b\_\_two\_dots\_.png}

\includegraphics[scale=0.20]{{./dip3e\_masks/FigP0528\_b\_\_two\_dots\_.png}}

\textbf{Fig1059\_a\_\_AbsADI\_.png}

\includegraphics[scale=0.20]{{./dip3e\_masks/Fig1059\_a\_\_AbsADI\_.png}}

\textbf{Fig1059\_b\_\_PosADI\_.png}

\includegraphics[scale=0.20]{{./dip3e\_masks/Fig1059\_b\_\_PosADI\_.png}}

\textbf{Fig0539\_c\_\_shepp-logan\_phantom\_.png}

\includegraphics[scale=0.20]{{./dip3e\_masks/Fig0539\_c\_\_shepp-logan\_phantom\_.png}}

\textbf{FigP0905\_c\_.png}

\includegraphics[scale=0.20]{{./dip3e\_masks/FigP0905\_c\_.png}}

\textbf{Fig1043\_a\_\_yeast\_USC\_.png}

\includegraphics[scale=0.20]{{./dip3e\_masks/Fig1043\_a\_\_yeast\_USC\_.png}}

\textbf{FigP0905\_U\_.png}

\includegraphics[scale=0.20]{{./dip3e\_masks/FigP0905\_U\_.png}}

\textbf{Fig0524\_b\_\_blurred-impulse\_.png}

\includegraphics[scale=0.20]{{./dip3e\_masks/Fig0524\_b\_\_blurred-impulse\_.png}}

\textbf{Fig0424\_a\_\_rectangle\_.png}

\includegraphics[scale=0.20]{{./dip3e\_masks/Fig0424\_a\_\_rectangle\_.png}}

\textbf{Fig1008\_c\_\_roof\_edge\_.png}

\includegraphics[scale=0.20]{{./dip3e\_masks/Fig1008\_c\_\_roof\_edge\_.png}}

\textbf{Fig0539\_a\_\_vertical\_rectangle\_.png}

\includegraphics[scale=0.20]{{./dip3e\_masks/Fig0539\_a\_\_vertical\_rectangle\_.png}}

\textbf{FigP0905\_a\_.png}

\includegraphics[scale=0.20]{{./dip3e\_masks/FigP0905\_a\_.png}}

\textbf{FigP0433\_a\_.png}

\includegraphics[scale=0.20]{{./dip3e\_masks/FigP0433\_a\_.png}}

\textbf{Fig0.15\_a\_\_translated\_rectangle\_.png}

\includegraphics[scale=0.20]{{./dip3e\_masks/Fig0425\_a\_\_translated\_rectangle\_.png}}

\textbf{FigP0918\_c\_.png}

\includegraphics[scale=0.20]{{./dip3e\_masks/FigP0918\_c\_.png}}

\textbf{Fig0524\_a\_\_impulse\_.png}

\includegraphics[scale=0.20]{{./dip3e\_masks/Fig0524\_a\_\_impulse\_.png}}

\textbf{Fig0236\_a\_\_letter\_T\_.png}

\includegraphics[scale=0.20]{{./dip3e\_masks/Fig0236\_a\_\_letter\_T\_.png}}

\textbf{Fig0503\_\_original\_pattern\_.png}

\includegraphics[scale=0.20]{{./dip3e\_masks/Fig0503\_\_original\_pattern\_.png}}

\textbf{FigP0501.png}

\includegraphics[scale=0.20]{{./dip3e\_masks/FigP0501.png}}

\textbf{Fig1218\_airplanes\_.png}

\includegraphics[scale=0.20]{{./dip3e\_masks/Fig1218\_airplanes\_.png}}

\textbf{Fig0534\_a\_\_ellipse\_and\_circle\_.png}

\includegraphics[scale=0.20]{{./dip3e\_masks/Fig0534\_a\_\_ellipse\_and\_circle\_.png}}

\textbf{FigP0616\_a\_.png}

\includegraphics[scale=0.20]{{./dip3e\_masks/FigP0616\_a\_.png}}

\textbf{FigP0918\_b\_.png}

\includegraphics[scale=0.20]{{./dip3e\_masks/FigP0918\_b\_.png}}

\textbf{FigP0528\_c\_.png}

\includegraphics[scale=0.20]{{./dip3e\_masks/FigP0528\_c\_.png}}

\textbf{FigP0528\_a\_\_single\_dot\_.png}

\includegraphics[scale=0.20]{{./dip3e\_masks/FigP0528\_a\_\_single\_dot\_.png}}

\end{document}